\documentclass[format=acmsmall, review=false, screen=true]{acmart}

\usepackage{booktabs} 
\usepackage{enumitem} 
\usepackage{longtable}
\usepackage{multirow}
\usepackage{multicol}
\usepackage{tikz}
\setcitestyle{numbers,sort}
\usepackage{url}
\usepackage[utf8]{inputenc}

\acmDOI{0000001.0000001}

\begin{document}
	\title[Microservice Transition and its Granularity Problem: A Systematic Mapping Study]{Microservice Transition and its Granularity Problem:\\ A Systematic Mapping Study}
	
	\author{Sara Hassan}
	\orcid{https://orcid.org/0000-0001-7481-0434}
	\affiliation{%
		\institution{University of Birmingham}
		\city{Birmingham}
		\state{West Midlands}
		\postcode{B15 2TT}
		\country{UK}}
	\email{ssh195@cs.bham.ac.uk}
	\author{Rami Bahsoon}
	\affiliation{%
		\institution{University of Birmingham}
		\city{Birmingham}
		\country{UK}
	}
	\email{r.bahsoon@cs.bham.ac.uk}
	\author{Rick Kazman}
	\affiliation{%
		\institution{Software Engineering Institute (SEI)/CMU}
		\city{Pittsburgh}
		\country{USA}
	}
	\affiliation{%
		\institution{University of Hawaii}
		\city{Honolulu}
		\country{USA}
	}
	\email{kazman@hawaii.edu}
	
	\begin{abstract}
		Microservices have gained wide recognition and acceptance in software industries as an emerging architectural style for autonomic, scalable, and more reliable computing. The transition to microservices has been highly motivated by the need for better alignment of technical design decisions with improving value potentials of architectures. Despite microservices' popularity, research still lacks disciplined understanding of transition and consensus on the principles and activities underlying "micro-ing" architectures. In this paper, we report on a systematic mapping study that consolidates various views, approaches and activities that commonly assist in the transition to microservices. The study aims to provide a better understanding of the transition; it also contributes a working definition of the transition and technical activities underlying it. We term the transition and technical activities leading to microservice architectures as microservitization. We then shed light on a fundamental problem of microservitization: microservice granularity and reasoning about its adaptation as first-class entities. This study reviews state-of-the-art and -practice related to reasoning about microservice granularity; it reviews modelling approaches, aspects considered, guidelines and processes used to reason about microservice granularity. This study identifies opportunities for future research and development related to reasoning about microservice granularity.
	\end{abstract}

	\ccsdesc[500]{General and reference~Surveys and overviews}
	\ccsdesc[500]{Software and its engineering~Extra-functional properties}
	\ccsdesc[500]{Software and its engineering~Designing software}
	\ccsdesc[500]{Software and its engineering~Software design tradeoffs}
	\ccsdesc[300]{Software and its engineering~Abstraction, modeling and modularity}
	\ccsdesc[300]{Software and its engineering~Software design engineering}
	\ccsdesc[300]{Software and its engineering~Agile software development}
	\ccsdesc[100]{Software and its engineering~Publish-subscribe / event-based architectures}
	\ccsdesc[100]{Software and its engineering~Model-driven software engineering}
	\ccsdesc[100]{Software and its engineering~Ultra-large-scale systems}
	\ccsdesc[100]{Software and its engineering~Rapid application development}
	\ccsdesc[100]{Software and its engineering~Design patterns}
	%
	%

	\keywords{Microservices, software economics, systematic mapping study, design decision support, granularity}

	\maketitle
	
	
\section{Introduction}\label{mapover}
Several industries have migrated their applications (or actively considering migrating) to microservices \cite{front,ottodev,bbc,uber,buzzfeed,netflixeng,zalando,aws,freelunch,netflixeng,scc}. The transition to microservices has not been purely driven by technical objectives; the transition requires careful alignment of the technical design decisions with the business ones. The ultimate objective of this alignment is to enhance utilities of the application's software architecture and to improve its value potentials. For example, among the technical design decisions is isolating business functionalities into microservices that interact through standardised interfaces. Isolation motivated only by technical objectives can lead to aggressive decomposition of functionalities favouring service autonomy without considering its impact on value potentials. However, isolation motivated by technical and business objectives can be more informed. It can enhance utilities such as autonomy, replaceability and decentralised governance (among other utilities) to improve the microservice architecture's ability in coping with operation, maintenance and evolution uncertainties. Ultimately, this can also  relate to improved maintenance costs and cost-effective quality of service (QoS) provision to end users; these are examples of improved value potentials in the architecture.

Due to the recency of microservices, they have a multitude of definitions; each definition captures different properties of microservices. Definitions mostly agree that the fundamental properties of microservices include enabling facilitated improvement of component characteristics --- autonomy, replaceability, independent deployability --- and of architectural characteristics --- improved reliability, scalability, resilience to failure, availability, and evolvability \cite{microfow,solid,refarch,scc,martin,7943959,smartbear,differencesoa,samn,practical2,7300793,7856721}. In essence, these definitions capture some drivers of the transition to microservices aimed at enhancing utility in the application's software architecture. 
The utility enhanced through the transition can render benefits that can cross-cut architectural design, testing, maintenance and service management \cite{bluemix}. For example, microservice autonomy allows the architects to easily locate, implement and test necessary service amendments \cite{scc}. Microservice replaceability allows architects to confidently and independently add and/or manage new business functionalities over the system lifetime \cite{scc}. 

The transition to microservices can help the application in better meeting its quality of services (QoS) requirements; this may consequently result into improved compliance with service level agreements for QoS, potential economics gains, and better alignment with the business objectives of microservice adopters \cite{scc}. Because of their ``micro'' character, microservices can be mobilised to the benefit of several service-oriented applications that can require ``lighter weight'' processing (e.g., mobile services and Internet-of-Things (IoT)) \cite{scc}. 

Despite the industrial push towards microservices, there is no disciplined understanding of their transition nor consensus on the principles and activities underlying the transition \cite{chal}. A disciplined understanding of the transition is of paramount importance to inform and/or to justify its technical activities by aligning them with their added value and cost. Currently however, the state-of-the-practice in microservice adoption lacks appropriate methods and techniques that can justify value added of technical design decisions. For example, the software architect can be equipped with mechanisms and tools that can enhance replaceability by standardising the communication paradigms across microservices \cite{amqp,aws}. Reasoning about the added value and possible cost becomes essential to justify this technical design decision regarding communication paradigms.   

\textit{This paper is an attempt for a better understanding of the transition to microservices. It conducts a systematic study to consolidate various views about microservices; it then uses the study results to contribute to a well-rounded working definition describing the transition and technical activities of the transition to microservice architectures. We term the transition and technical activities leading to microservice architectures as microservitization.}

This working definition has explicitly considered a fundamental problem of microservitization: reasoning about the granularity of a microservice (i.e. whether a microservice should be decomposed/merged further or not). A granularity level determines "the service size and the scope of functionality a service exposes \cite[p.426]{4578356}." Granularity adaptation entails merging or decomposing microservices thereby moving to a finer or more coarse grained granularity level. 

Determining the granularity level too early in the software architecture's lifetime can lead to problems in reasoning about microservices \cite{10.1007/978-3-319-74433-9_3,nuha2016}. This problem is of significance both in brownfield and greenfield development \cite{front}. In both fields, a suitable granularity level is paramount to inform choosing concrete services from a plethora of COTS microservices. 

"A systematic mapping study allows the evidence in a domain to be plotted at a high level of granularity. This allows for the identification of evidence clusters and evidence deserts to direct the focus of future systematic reviews and to identify areas for more primary studies to be conducted \cite[p.5]{slr}." Directing the focus of future systematic reviews aligns with our aforementioned objectives. Ultimately,  our attempt at defining the transition to microservices (Objective 1 above) will pave the way for future development and research related to microservice transition. Furthermore, understanding the problem of reasoning about microservice granularity (Objective 2 above) allows identifying areas for future primary studies. Since the examined literature regarding microservices spanned a broad variety of aspects, we found a systematic mapping study to be suitable given the amount of reviewed literature \cite{slr}. 

In Section \ref{process} we describe the steps we followed in the systematic mapping study. In Section \ref{litanalysis} we report and briefly analyse our mapping study results. In Section \ref{gaps} we use this analysis to: (1) present our working definition for the transition to microservices (Section \ref{obj1gap}) and (2) identify gaps in the state-of-the-art and -practice related to reasoning about microservice granularity (Section \ref{obj2gap}). Overall, the identified gaps motivate the need for:
\begin{itemize}
	\item Microservice-specific modelling support potentially using an architecture definition language (ADL) that "treats microservice boundaries as adaptable first-class entities \cite[p.2]{icsa2017}" thereby facilitating runtime analysis of microservice granularity in a systematic architecture-oriented manner \cite{icsa2017}.
	\item A dynamic architectural evaluation approach that captures two dimensions under uncertainty at runtime: added value to be introduced and cost to be incurred if the granularity of microservice architectures is adapted. 
	\item Effective decision support that should systematically guide the software architects towards suitable granularity adaptation strategies at runtime or suggest re-visiting their expectations of the architecture's runtime environment. 
\end{itemize}
In Section \ref{pubrelated} we compare and contrast related literature reviews, studies and surveys in the field of microservices against our systematic mapping study. In Section \ref{threats} we reflect on threats to the validity of our study. In Section \ref{relevant} we summarise contributions that are not directly linked to microservices but can be relevant to their emergence and development. In Section \ref{conclusion} we conclude by summarising the results of our systematic mapping study.
\section{Systematic Mapping Study Process}\label{process}
The process we follow in our mapping study is inspired by guidelines from \cite{slr}. In the subsections below, we describe our application of each stage of this process.
\subsection{Research Questions}\label{mappingresquest}
Overall, this paper conducts a systematic mapping study to address the following objectives:
\begin{itemize} 
	\item \textit{Objective 1}: providing a better understanding of the transition to microservices --- we consolidate various views (industrial, research/academic) of the principles, methods and techniques that are commonly adopted to assist the transition to microservices. This consolidation allows us to reach a working definition for the transition to microservices; we term this transition \textit{microservitization}. 
	\item \textit{Objective 2}: understanding a fundamental problem of the transition to microservices related to reasoning about their granularity --- we review state-of-the-art and -practice related to reasoning about microservice granularity. This review allows us to understand the state-of-the-art and -practice in the modelling approaches, aspects considered, guidelines and processes used to reason about microservice granularity. 
\end{itemize}
Given these objectives, we reify them into the following research questions. Along with each question we outline the rationale behind it. 

\textit{Objective 1: Providing a better understanding of the transition to microservices}
\begin{itemize}
	\item What are the activities undertaken to adopt microservices? This question helps understand the principles, methods and techniques of the transition to microservices by digesting experiences of microservice adopters in industry and in academic research.
\end{itemize}
\textit{Objective 2: Understanding a fundamental problem of the transition to microservices related to reasoning about their granularity}
\begin{itemize}
	\item What are the modelling approaches used to define the granularity of a microservice? The aim of this question is to identify the support provided by microservice-specific models for reasoning about granularity and to investigate how systematic (i.e. standardised and methodological) these models are.
	\item Which quality attributes are considered when reasoning about microservice granularity and how are they captured? The aim of this question is to elicit the possible trade-offs which software architects need to balance when reasoning about microservice granularity and/or how these trade-offs can be captured objectively.
	\item How is reasoning about microservice granularity described? The aim of this question is to explore the state of the art regarding triggers and steps of microservice granularity adaptation and their suitability to the dynamic microservice environment. 
\end{itemize}
\subsection{Search Strategy}\label{searchstrat}
The terms used when searching for English publications were "microservice" and "micro-service"; Google Scholar, ACM Digital Library, IEEE Xplore, ScienceDirect, SpringerLink and Wiley  InterScience were used during this search. Our scope of publications includes journals, theses, books, conferences, workshops, blog articles, presentations, and videos. For academic publications, snowballing was applied \cite{PETERSEN20151} to further extract relevant literature. Non-academic publications were included since most industrial experiences regarding microservices were published in these forms. We made our best effort however to only include articles that either transcribe the view of adopters or ones where the author is the opinion holder. We believe answering the research questions above comprehensively requires examining both academic and non-academic experiences with microservices. This broad scope also aligns with a property of systematic mapping studies --- aiming for broad coverage rather than narrow focus \cite{slr}. Our search was restricted to publications between 2013 and 2018, since the microservice trend had not emerged prior to that; non-academic literature started to appear in 2013 \cite{googletrend} while peer-reviewed publications started to appear in 2014 \cite{7930195}. Meta-data of the search results was maintained using a tool called "Publish or Perish" \cite{perish}.
\subsection{Selection of Primary Studies}\label{selection}
Initially, the search results were examined for relevance according to inclusion and exclusion criteria below. Each research question elicited in Section \ref{mappingresquest} has corresponding inclusion/exclusion criteria (their structure is inspired by \cite{Petersen:2008:SMS:2227115.2227123}).  Along with each criterion is the rationale justifying it italicised.

\textit{What activities are undertaken to adopt microservices?}\\
\textbf{Inclusion Criteria:}\\
\begin{itemize}[leftmargin=*]
	\item Publications generically presenting the challenges of adopting microservices \textit{since they can be used to infer the activities comprising the transition to microservices}.
	\item Case studies of adopting microservices \textit{are used to complement and verify the activities in generic publications.}
	\item Publications comparing specific development tools in the microservice industry \textit{because they can be used to infer activities comprising the transition.}
\end{itemize}
\textbf{Exclusion Criteria:}\\
\begin{itemize}[leftmargin=*]
	\item Publications without any reference to microservices. \textit{Including such publications would confuse rather than clarify understanding the transition to microservices. This is against Objective 1 of our study.}
	\item Publications that refer to servitization in the business not the software context \textit{since we are only concerned about the activities of shifting a software system from another architectural style to microservices.}
\end{itemize}
\textit{What modelling approaches are used to define the granularity of a microservice?}\\
\textbf{Inclusion Criteria:}\\
\begin{itemize}[leftmargin=*]
	\item Publications defining formal notations/diagrams for modelling microservices. \textit{Such publications can be used to assess how systematic the state-of-the-art is for modelling microservice granularity.}
	\item Proposals of modelling concepts for microservices. \textit{Even when unverified, a proposed modelling concept can provide an insight for the building units of reasoning about microservice granularity.}
	\item Publications presenting industrial case studies for modelling microservices. \textit{Such publications would verify and illustrate the expressiveness of proposed modelling concepts to microservice granularity.}
\end{itemize}
\textbf{Exclusion Criteria:}\\
\begin{itemize}[leftmargin=*]
	\item Papers which provide binding and re-configuration solutions for SOAs/web-services/mobile services only \textit{are excluded since they do not capture the properties specific to microservices, hence they are not suited for reasoning about microservice-specific decision problems (in this case microservice granularity).}
	\item Papers which provide modelling approaches for SOAs/web-services/mobile services are \textit{excluded since they do not capture properties specific to microservices, hence they are not suited for reasoning about microservice-specific decision problems (in this case microservice granularity).}
\end{itemize}
\textit{Which quality attributes are considered when reasoning about microservice granularity and how are they captured?}\\
\textbf{Inclusion Criteria:}\\
\begin{itemize}[leftmargin=*]
	\item Publications presenting metrics used when reasoning about microservice granularity. \textit{Such publications would help assess how objectively the trade-offs affecting microservice granularity are captured in academia and/or industry.} 
	\item Case studies involving the quality drivers considered when reasoning about granularity. \textit{Such publications would realistically capture the significance of specific trade-offs when reasoning about granularity adaptation.}
	\item Publications focused on vendor-specific comparisons between platforms supporting reasoning about microservice granularity. \textit{This can help derive the quality attributes and metrics considered when reasoning about granularity.} 
\end{itemize}
\textbf{Exclusion Criteria:}\\
\begin{itemize}[leftmargin=*]
	\item Case studies that do not explicitly relate a challenge to its impact on microservice granularity. \textit{Since case studies report concrete challenges and trade-offs impacting them, it is unreasonable to claim an impact of a reported trade-off on granularity if that is not reported explicitly in a case study.}
\end{itemize}
\textit{How is reasoning about microservice granularity described?}\\
\textbf{Inclusion Criteria:}\\
\begin{itemize}[leftmargin=*]
	\item Publications including guidelines for reasoning about microservice granularity. \textit{This helps to identify the state-of-the-art regarding triggers and/or steps for granularity adaptation.}
	\item Publications showing a sequence of when and how granularity is reasoned about in the application's lifecycle. \textit{These publications help assess how much the state-of-the-practice considers dynamicity in microservice environments when reasoning about granularity adaptation.}
\end{itemize}
\textbf{Exclusion Criteria:}\\
\begin{itemize}[leftmargin=*]
	\item Publications that provide generic best practices for the granularity of applications with no reference to microservices (e.g. related to web services, SOA, mobile services). \textit{Such best practices are not targeted specifically at microservices, so it would not be reasonable to use them in the context of microservice granularity.}
\end{itemize} 
\subsection{Keywords and Classification}\label{keywords}
"The purpose of this stage is to classify papers with sufficient detail to answer the broad research questions and identify papers for later reviews without being a time consuming task \cite[p.44]{slr}." Here we classify the included publications according to two classification frameworks. Initially, we classify them according to the following research approaches, elicited from \cite{Wieringa:2005:REP:1107677.1107683}: 
\begin{itemize}
	\item Evaluation research: these investigate the significance of a problem and/or investigate the feasibility of a contribution in practice.
	\item Opinion papers: "these contain the author's opinion about what is wrong or good about something, how we should do something, etc \cite[p.105]{Wieringa:2005:REP:1107677.1107683}".
	\item Solution proposals: these "propose a solution technique and argue for its relevance, without a full-blown validation. The technique must be novel, or at least a significant improvement of an existing technique. A proof-of-concept may be offered by means of a small example, a sound argument, or by some other means \cite[p.86]{10.1007/978-3-319-19593-3_7}."
	\item Experience paper: in these publications "the emphasis is on what and not on why. The experience may concern one project or more, but it must be the author's personal experience  \cite[p.105]{Wieringa:2005:REP:1107677.1107683}." Such papers should "contain a list of lessons learned by the author from his or her experience. Papers in this category will often come from industry practitioners or from researchers who have used their tools in practice, and the experience will be reported without a discussion of research methods. The evidence presented in the paper can be anecdotal   \cite[p.106]{Wieringa:2005:REP:1107677.1107683}."
	\item Validation research: such papers validate solution proposals which "may have been proposed elsewhere, by the author or by someone else. The investigation uses a thorough, methodologically sound research setup. Possible research methods are experiments, simulation, prototyping, mathematical analysis, mathematical proof of properties, etc  \cite[p.105]{Wieringa:2005:REP:1107677.1107683}."
	\item Philosophical paper: "these papers sketch a new way of looking at things, a new conceptual framework, etc \cite[p.105]{Wieringa:2005:REP:1107677.1107683}."
\end{itemize}
The second classification framework entails categorising the publications under categories derived from our research questions of concern. A publication belongs in a category if it contains any of the corresponding keywords identified in Table \ref{categories}. 
\begin{footnotesize}
	\begin{longtable}{p{.35\textwidth}p{.20\textwidth}p{.43\textwidth}}
		\hline
		Research Question&Category&Keywords\\
		\hline
		\endhead
		\multirow{7}{.35\textwidth}{What activities are undertaken to adopt microservices?}&Architectural design&	architectural style, communication mechanism, boundaries, orchestration, service choreography, service registration, service discovery, design patterns, proxy, bulkhead, circuit breaker, router, routing\\
		&Organisation&Conway's law, decentralised governance, cross-functional teams, hierarchical teams\\
		&Operation&Devops, NoOps, configuration settings,  operation\\
		&Deployment&Continuous integration, CICD, continuous deployment, deployment pipeline, automated deployment, virtualisation, hypervisors, containerisation, configuration provider\\
		&Development&Heterogenous tools, agile development, extreme programming\\
		&Monitoring&Regression unit testing, health monitoring, cluster monitoring, troubleshooting, debugging, failure\\
		&Logging&central logging, decentralised logging, profiling, tracing\\
		\hline
		\multirow{2}{.35\textwidth}{What modelling approaches are used to define the granularity of a microservice?}&Structural&Domain-driven, classes, instances, resources, components, message format, data item, topology, service dependency, nodes, type definition\\
		&Behavioural&Event flow, message stream, activity flow, communication flow, event triggers, execution timeline, use cases\\
		\hline
		\multirow{5}{.35\textwidth}{Which quality attributes are considered when reasoning about microservice granularity and how are they captured?}&Performance&QoS, efficiency, service contracts, SLA, performance, response time, throughput, performance bottleneck, invocation duration, transaction duration, customer value\\
		&Reliability&Fault tolerance, disaster recovery, single point of failure, resilience, robustness, failure rate, error rate\\
		&Scalability&Auto-scale, scalable, scaling, load balancing, load distribution, completed transactions per second\\
		&Maintainability&Maintainable, changeable, maintenance cost, maintenance overhead, adaptability, changeability, effort cost, expandability, dynamicity\\
		&Complexity&Communication overhead, complexity cost, development cost, tight coupling, low cohesion\\
		\hline
		\multirow{2}{.35\textwidth}{How is reasoning about microservice granularity described?}&Guidelines&Two-pizza team, lines of code, half-life, agile manifesto, one task, single responsibility, fine-grained functionality, separation of business concerns, high cohesion, loose coupling\\
		&Processes&Iterative, strangler pattern\\
		\hline
		\caption{Inferred classification framework used to classify the included publications in our study}
		\label{categories}
	\end{longtable}
\end{footnotesize}
The identification of the keywords is inspired by a microservice-specific systematic mapping study \cite{nuha2016} and refined iteratively as more publications were examined. It is worth noting that for all the included publications, we manually categorised them to ensure that synonyms or partial matches of these keywords are accurately handled. We justify the keywords (italicised below) under each category as follows:
\textit{What activities were undertaken to adopt microservices?} 
\begin{itemize}
	\item Architectural design: publications in this category are concerned with all the technical activities which comprise adopting microservices. We use this category to verify whether or not microservice adopters call microservices an \textit{architectural style} and/or \textit{design pattern}. Choosing the generic \textit{communication paradigm} of the architecture (e.g., \textit{orchestration, choreography}) and the more concrete message exchange pattern (e.g., using a \textit{router/proxy}) are also among the technical activities of microservice adoption. In addition, the \textit{boundaries} of the system and individual components of the system is a technical design decision when moving to microservices. The fault tolerance mechanisms of the system also need to be identified (e.g. \textit{bulkhead} and \textit{circuit breaker} patterns). Because microservice applications are highly distributed and scalable, 2 additional technical decisions are critical in microservitization: \textit{service registration} and \textit{service discovery}.
	\item Organisation: in this category we are concerned with the organisational impact of adopting microservices. The state-of-the-practice in the microservice industry is to motivate \textit{decentralised governance} (i.e. holding the responsibility of building and running \cite{govern}) through microservitization \cite{microfow}. The three most common means of achieving that is through following \textit{Conway's law \cite{microfow}, cross-functional teams, or hierarchical teams \cite{smartbear}}. Conway's law states "organisations which design systems ... are constrained to produce designs which are copies of the communication structures of these organisations \cite{conway1968committees}." 
	\item Operation: publications in this category are concerned with identifying how the system will be governed post-deployment. This includes defining who is responsible for this governance to begin with. The state-of-the-practice is in the microservice industry is 2 alternative approaches: \textit{DevOps} where the governance is shared across a development team and an operations team; and \textit{NoOps}, where the governance is fully the responsibility of the development team.  In addition, operational activities include defining the \textit{configuration settings} and \textit{configuration provider} which the governor of the microservice application needs to follow or  in different runtime situations. 
	\item Deployment: this category includes publications that define the activities constituting the \textit{deployment pipeline} of the microservice application. The state-of-the-practice in the microservice industry is 2 alternative approaches \cite{cicdmicro,CIUFFOLETTI2015163}: \textit{continuous integration} and \textit{continuous delivery} (abbreviated as CICD), and \textit{automated deployment}. "Continuous integration is a coding philosophy and set of practices that drive development teams to implement small changes and check in code to version control repositories frequently \cite{cicdterms}." "Continuous delivery picks up where continuous integration ends. CD automates the delivery of applications to selected infrastructure environments \cite{cicdterms}." "Deployment automation allows applications to be deployed across the various environments used in the development process, as well as the final production environments \cite{deploy}." Regardless of the deployment pipeline, the host on which this pipeline is enforced is another critical decision in this category. The 3 most common approaches for deploying microservices are \textit{virtualisation}, using \textit{hypervisors} and/or \textit{containerisation}.
	\item Development: publications in this category describe how the transition to microservices impacts software development practices. The state-of-the-practice in microservice development is adopting \textit{extreme programming and agile} practices using \textit{heterogenous tools} (e.g., Kubernetes, Istio, Springboot, fabric8).
	\item Monitoring: publications in this category are concerned with identifying the alternative rationales of runtime monitoring (e.g. \textit{health, cluster}) of microservice application and the alternative approaches which support monitoring (e.g. \textit{regression unit testing, troubleshooting, debugging} and \textit{failure} identification). 
	\item Logging: in this category we are concerned with where the monitoring results are to be stored (alternatively called \textit{profiling} and \textit{tracing}). The 2 alternative approaches to this in the microservice industry as we have examined are \textit{central} or \textit{decentralised logging}.
\end{itemize}
\textit{What modelling approaches are used to define the granularity of a microservice?}
\begin{itemize}
	\item Structural: publications in this category are concerned about contributions that capture the structure (alternatively called \textit{topology}) of the microservice architecture. Depending on the nature of the contribution (e.g., \textit{domain-driven} \cite{domdriv}), the units of this structure differ (e.g. \textit{classes, instances, resources, components, data items, messages}, and/or \textit{nodes}). Modelling these units includes capturing the \textit{dependencies} across those units and their \textit{types}.
	\item Behavioural: alternative to the approach above, publications in this category capture the sequence of actions in the microservice application rather the structure of the units constituting the application. This sequence can be in the form of \textit{events, messages, activities, communication, execution steps} and/or \textit{use cases}.
\end{itemize}
\textit{Which quality attributes are considered when reasoning about microservice granularity and how are they captured?}
\begin{itemize}
	\item Performance: wherever the main driving force of reasoning about granularity is performance (alternatively called \textit{QoS, long-term value, efficiency or customer value}), the publication is put under this category. There means of capturing this objectively include \textit{response time, throughput, invocation duration}, identifying the \textit{performance bottlenecks} and/or \textit{transaction duration}. Thresholds on these metrics are captured in \textit{service contracts, service level agreements} (abbreviated as SLAs). We subsume service contracts as "an agreement between the a consumer and provider service about the format of data that they transfer between each other. Normally, the format of the contract is defined by the consumer and shared with the corresponding provider. Afterwards, tests are being implemented in order to verify that the contract is being kept \cite{cdc}."
	\item Reliability: publications that imply or explicitly focus on enhancing reliability as the primary driving force when reasoning about granularity are included in this category. Reliability entails exhibiting \textit{fault tolerance, disaster recovery, resilience}, eliminating \textit{single points of failure} and/or \textit{robustness}. Reliability is captured objectively in terms of \textit{failure/error rate}. 
	\item Scalability: publications that reason about granularity in terms of how it enhances the scalability of the architecture are included in this category. Exhibiting scalability entails employing strategies such as \textit{auto-scaling, load balancing} and/or \textit{load distribution}. Measuring scalability objectively can be done by looking at the \textit{number of completed transactions per second} (or per unit of time more generally). 
	\item Maintainability: exhibiting maintainability entails exhibiting \textit{adaptability, changeability, expandability} and/or \textit{dynamicity}. These properties are objectively captured in terms of \textit{maintenance cost, maintenance overhead}, and/or \textit{effort cost}. Publications concerned with reasoning about granularity in terms of enhancing maintainability are included in this category.
	\item Complexity: minimising complexity entails following 2 crucial design principles: \textit{tight coupling} and/or \textit{low cohesion}. This is measured in terms of \textit{communication overhead, complexity cost}, and/or \textit{development cost}. Publications where reasoning about granularity considers and/or measures its complexity are included in this category.
\end{itemize}
\textit{How is reasoning about microservice granularity described?}
\begin{itemize}
	\item Guidelines: publications under this category provide decision-making strategies for reasoning about granularity regardless of the steps of applying these strategies. The keywords under this category capture the state-of-the-practice guidelines.
	\item Processes: publications under this category are more elaborate in the sense that they enrich the granularity adaptation strategies with a sequence for applying them. The keywords under this category capture the alternative state-of-the-practice processes encountered for reasoning about microservice granularity.
\end{itemize}
\subsection{Data Synthesis}
RSS feeds and manual search were used to obtain publications complying with the strategy defined in Section \ref{searchstrat}. The results were then manually examined for inclusion and categorised according to the criteria and frameworks described above (Sections \ref{selection} and \ref{keywords} respectively). 
\section{Result Reporting and Analysis}\label{litanalysis}
In this section we present graphs summarising distributions of the included publications along the categories described in Table \ref{categories}. For each graph, we discuss how it helps serve the objectives outlined in Section \ref{mapover}.
\subsection{Publication Distribution Overview}
\begin{figure}[h]\centering\includegraphics[width=0.7\linewidth]{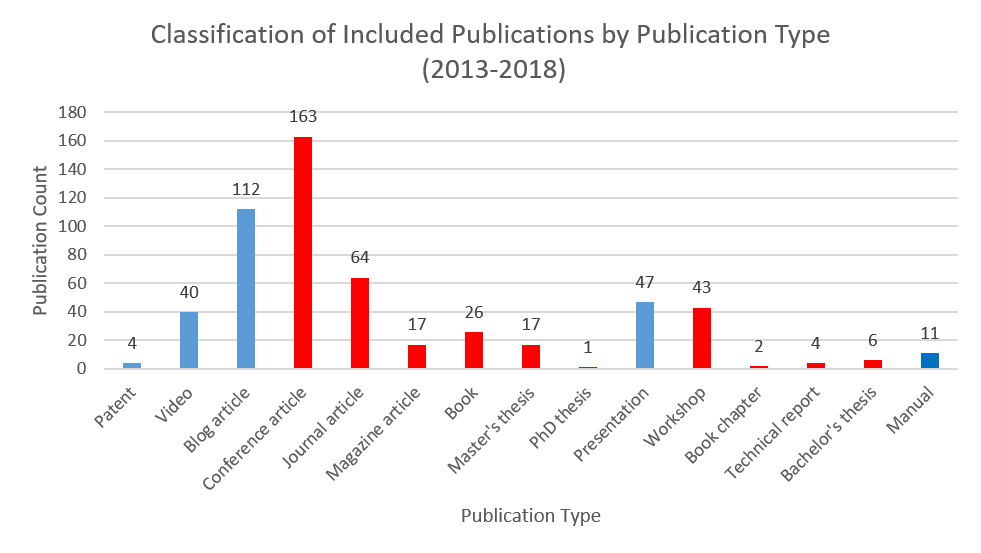}\caption{Publications between 2013 and 2018 as per the search strategy defined in Section \ref{searchstrat} included as per criteria from Section \ref{selection} classified according to their publication type; the red bars are the publications we consider as academic and the blue bars are those we consider non-academic}\label{pubtype}\end{figure}

A total of 560 publications met the inclusion criteria in Section \ref{selection}. Table \ref{examples} lists representative examples of the included publications categorised according to Table \ref{categories}. Figure \ref{pubtype} shows the overall distribution of the publications according to the publication type. 

As justified in Section \ref{process}, we widened the scope of our study to include both academic and industrial publications. Broadly, we consider a publication type to be academic if it has gone through editing or peer-revision (the red bars in Figure \ref{pubtype}); about 62\% of the included publications. On the other hand, non-academic publications account for about 38\% of the total. Although the majority of publications are academic, non-academic literature still comprises a significant percentage. Had we excluded these publications, attempting a definition for microservitization (Objective 1 in Section \ref{mapover}) would have been biased. The exclusion of non-academic sources would lead to missing relevant keywords related to each category and the coverage of industrial opinions and experiences related to microservice adoption would be narrower.
\begin{footnotesize}
	\begin{longtable}{p{.35\textwidth}p{.20\textwidth}p{.43\textwidth}}
		\hline
		Research Question&Category&Representative Examples\\
		\hline
		\endhead
		\multirow{7}{.35\textwidth}{What activities are undertaken to adopt microservices?}&Architectural design&\cite{article,krause2015microservices,7930199,infoqmag,antifrag,wolff2016microservices,Pahl:2016:MSM:3021834.3021846,release,ibmredbooks,rodger2017tao,fairbanks2010just,datamanage}\\
		&Organisation&\cite{Zimmermann2017,Killalea:2016:HDM:2975594.2948985,7819415,7436659,nginx,doi:10.1002/smr.1866,benchmicro,DBLP:journals/corr/DragoniDLM17,12fact,7965739}\\
		&Operation&\cite{qna,enterpriselandscape,workinprog,shrinking,ibmdevops,tame,netflixipc,guts,contdevops,7573165,7515686,7436659}\\
		&Deployment&\cite{7830692,DBLP:journals/corr/MontesiW16,8026911,Brilhante:2017:AQB:3126858.3126873,devopstoolkit,Cerny:2018:CUM:3183628.3183631,Wizenty:2017:MBM:3129790.3129821,8004304,Toffetti:2015:ASM:2747470.2747474,Zeiner2016}\\
		&Development&\cite{8216583,pop00011,sharma2017mastering,RODS,redhat,microbuilder,Esposte2017InterSCityAS,giallorenzo2017programming,pop00081,andrawos2017cloud,10.1007/978-3-319-71255-0_37}\\
		&Monitoring&\cite{pop00049,205506,pop00133,Flygare1119785,observe}\\
		&Logging&\cite{Salvadori:2017:OAF:3151759.3151793}\\
		\hline
		\multirow{2}{.35\textwidth}{What modelling approaches are used to define the granularity of a microservice?}&Structural&\cite{domdriv,sla,8093034,depmodel,10.1007/978-3-319-74781-1_17,8005348,bounded,future,intercept}\\
		&Behavioural&\cite{richardson2018microservice,eventcol,chal,conduct}\\
		\hline
		\multirow{5}{.35\textwidth}{Which quality attributes are considered when reasoning about microservice granularity and how are they captured?}&Performance&\cite{infocom,lessonslearneddeploy,autoscaleindia,econmic,8281846}\\
		&Reliability&\cite{10.1007/978-3-319-74781-1_18,derakhshanmanesh2016model,190613,10.1007/978-3-319-74781-1_16}\\
		&Scalability&\cite{7568389,micromatter,Hasselbring:2016:MSK:2851553.2858659,Klinaku:2018:CEC:3185768.3186296,8258201}\\
		&Maintainability&\cite{pop00190,8169955}\\
		&Complexity&\cite{iceis17,pop00150,gluecon,Kleindienst2017}\\
		\hline
		\multirow{2}{.35\textwidth}{How is reasoning about microservice granularity described?}&Guidelines&\cite{peopleimpact,netflixaws,uber,decoupled,chris,Sampaio2017SupportingME,7742277,7958428,stratdecomp}\\
		&Processes&\cite{strang,intercept,panel,kaiser,hardest,micropart2,randy,balazs}\\
		\hline
		\caption{Representative examples of publications included in the systematic mapping study}
		\label{examples}
	\end{longtable}
\end{footnotesize}
\begin{figure}[h]\centering\includegraphics[width=0.7\linewidth]{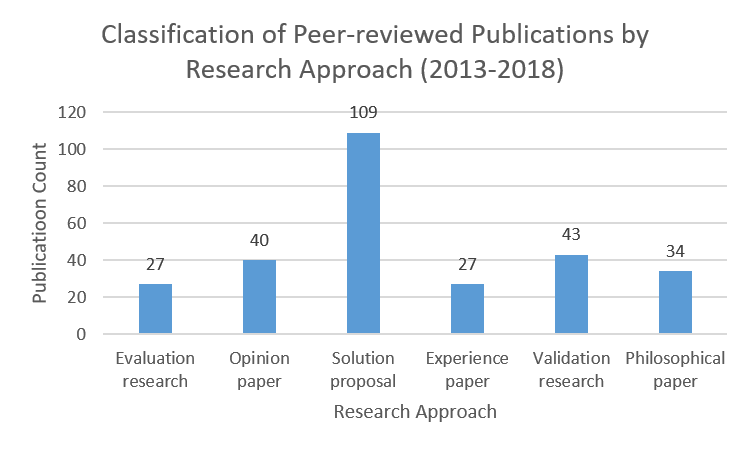}\caption{Academic (i.e. peer-reviewed) included publications between 2013 and 2018 as per the search strategy defined in Section \ref{searchstrat} classified according to their research approach; the classification criteria are derived from \cite{Wieringa:2005:REP:1107677.1107683}; }\label{pubres}\end{figure}
We further classify peer-reviewed publications according to a highly-cited paper classification framework \cite{Wieringa:2005:REP:1107677.1107683} which targets IEEE papers. This framework classifies papers according to their research approach; a brief description of each approach is presented in Section \ref{process}. This framework has been applied before in the context of microservices \cite{nuha2016}, so we consider it a neat fit for our study. To match the target context of the framework, we only apply it to peer-reviewed publications (Figure \ref{pubres}). 

Solution proposals by far comprise the largest number of peer-reviewed publications. Solution proposals present novel, significant techniques without a full-blown validation. A proof-of-concept may be offered in solution proposals by means of a small example, a sound argument, or by some other means \cite{Wieringa:2005:REP:1107677.1107683}. Therefore, the microservices trend is a thriving field for novelty but it is still lacking maturity. Validation research publications --- which thoroughly investigate solution proposals \cite{Wieringa:2005:REP:1107677.1107683} --- only amounted to 43 publications, which further proves the lack of maturity in the field. The large difference between the solutions proposals and validation research publications proves the need for disciplining the transition to microservices. The following subsections further discuss this need then focus on one of the fundamental problems of the transition --- reasoning about microservice granularity.
\subsection{Objective 1: Providing a Better Understanding of the Transition to Microservices}
\begin{figure}[h]\centering\includegraphics[width=0.7\linewidth]{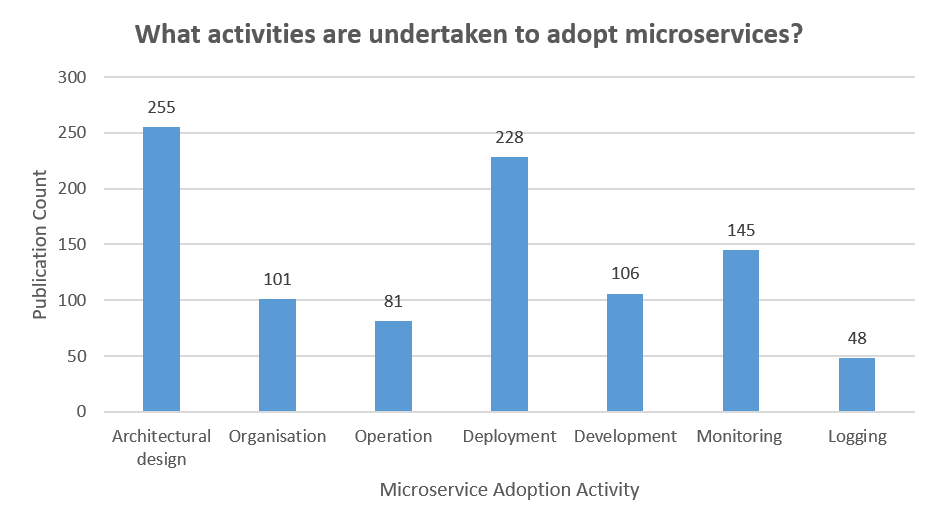}\caption{Included publications between 2013 and 2018 as per the search strategy defined in Section \ref{searchstrat} that have keywords related to the first research question in  Table \ref{categories}}\label{activities}\end{figure}

Figure \ref{activities} classifies the publications which include keywords related to: \textit{what are the activities undertaken to adopt microservices?} Architectural design and managing deployment are the most popular activities undertaken when adopting microservices. Therefore, we infer they are crucial activities in the transition to microservices. Nevertheless, there is a significant number of publications in the other categories of Figure \ref{activities}. The variation in number of publications across categories of Figure \ref{activities} implies there is no consensus in describing the transition to microservices. This leaves room for us to contribute the microservitization term which attempts to provide a better understanding of the transition to microservices.
\subsection{Objective 2: Understanding the Microservice Granularity Problem}
\begin{figure}[h]\centering\includegraphics[width=0.7\linewidth]{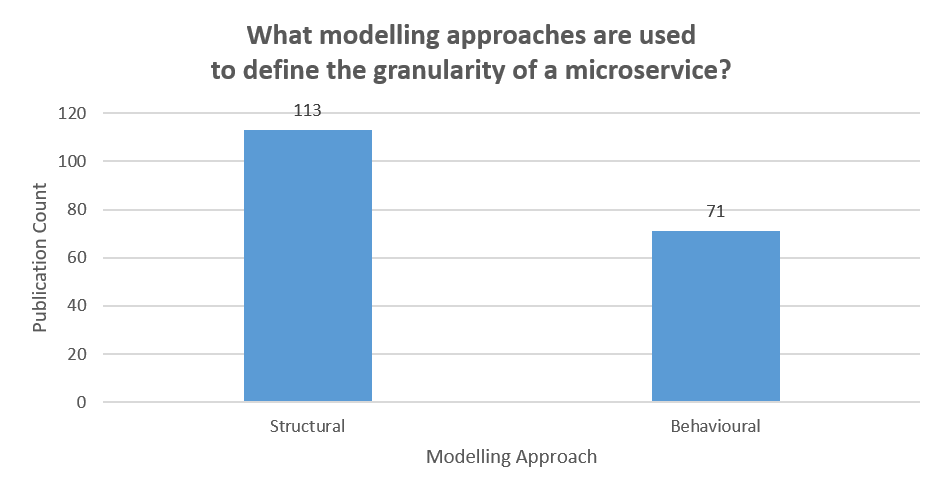}\caption{Included publications between 2013 and 2018 as per the search strategy defined in Section \ref{searchstrat} which have keywords related to the second research question in Table \ref{categories}}\label{pubmodel}\end{figure}

One of the fundamental problems of the transition to microservices is finalising  their level of granularity \cite{stefanpresent,nealford,infoqmag,antifrag,release,front}.  
Architecture definition language (ADL) classification frameworks \cite{adl} indicate that structural (or "topological \cite[p.26]{adl}") as well behavioural aspects of an architecture need to be modelled. Figure \ref{pubmodel} helps to clearly identify the state-of-the-practice in modelling microservices; to answer \textit{what are the modelling approaches used to define the granularity of a microservice?} The structural modelling approaches proposed for microservices are almost double the behavioural approaches. Structural approaches capture the topology and/or dependencies across building units of the microservice architecture. On the other hand, behavioural approaches capture the actions of these units in several runtime scenarios in which the modelled microservice operates.  Therefore, a systematic, architecture-oriented modelling approach for microservice architectures which facilitates reasoning about granularity needs to capture the architecture's structural and behavioural aspects. The difference in numbers between structural- and behavioural-oriented publications in Figure \ref{pubmodel} indicates that there is a lack in modelling approaches that capture both aspects of a microservice architecture.

\begin{figure}[h]\centering\includegraphics[width=0.7\linewidth]{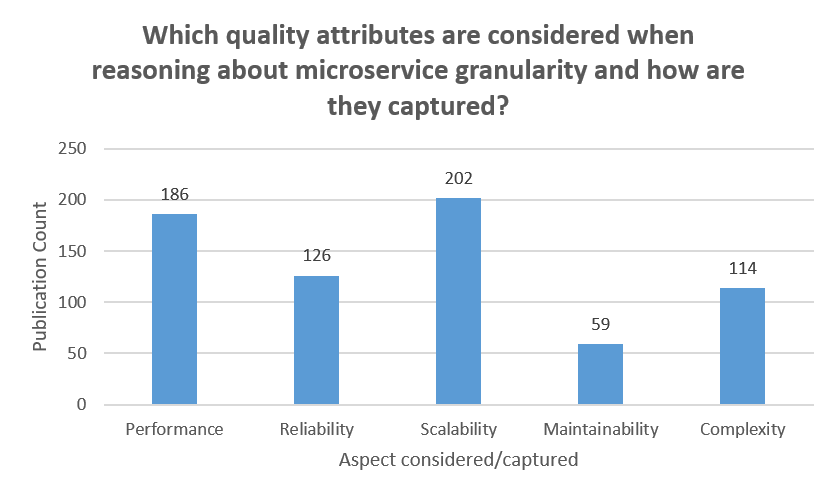}\caption{Included publications between 2013 and 2018 as per the search strategy defined in Section \ref{searchstrat} that have keywords related the third research question in Table \ref{categories}}\label{pubaspect}\end{figure}

Figure \ref{pubaspect} classifies publications according to the quality attributes they aim to optimise when reasoning about microservice granularity; to answer \textit{which quality attributes are considered when reasoning about microservice granularity and how are they captured?} In other words, they can be the most common means to introduce utility through cost-effective microservitization. Scalability is the most common quality considered in the examined literature. This is reasonable given the dynamic, large-scale environment in which microservices operate \cite{bbc,aws,smartbear,freelunch,uber,chal,deadly,zalando,adrianstate,ottodev}. Therefore, we infer that scalability can introduce added value to most microservice architectures. Relatively few publications have considered complexity/cost when reasoning about microservice granularity. Therefore, there is room for contributing to dynamic decision support which objectively considers both the potential value and cost of decisions related to microservice granularity (i.e. adapting granularity by decomposing or merging microservices).

\begin{figure}[h]\centering\includegraphics[width=0.7\linewidth]{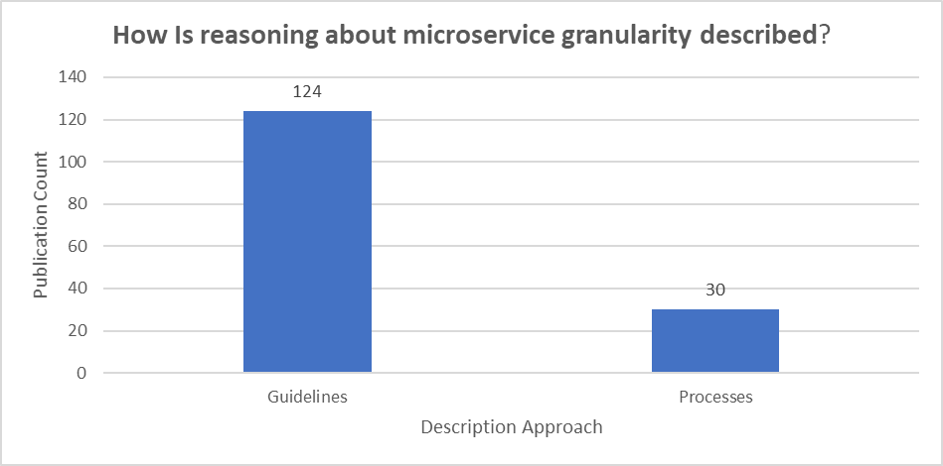}\caption{Included publications between 2013 and 2018 as per the search strategy defined in Section \ref{searchstrat} that have keywords related the fourth research question in Table \ref{categories}}\label{pubstep}\end{figure}
Figure \ref{pubstep} classifies proposed approaches for reasoning about microservice granularity according to how they are described; this answers \textit{how is reasoning about microservice granularity described?} Most of the proposed approaches are ad-hoc guidelines that could be applied differently at various points of the microservice application's lifecycle. However, effective decision support for microservice granularity should comprise a clear sequence of distinct steps to be triggered under clear conditions. We have not found such effective support even in the 30 publications proposing a process to reason about microservice granularity. Therefore, we infer that there is a lack in effective decision support for reasoning about microservice granularity in terms of clear adaptation steps and triggers.

\section{Addressing Systematic Mapping Study Objectives}\label{gaps}
In Section \ref{obj1gap} we contribute to an empirically grounded working definition for the transition to microservices; we call this transition microservitization. In Section \ref{obj2gap}, we identify the gaps in state-of-the-art and -practice related to reasoning about microservice granularity (inferred from Section \ref{litanalysis}) and discuss how they can be addressed. 

\subsection{Objective 1: Providing a Better Understanding of the Transition to Microservices}\label{obj1gap}
We have not found in the surveyed publications an empirically grounded definition which characterises the transition to microservices, but rather several conflicting, informal attempts coming from industry and academia. Table \ref{definitions} analyses these attempts in terms of whether or not they explicitly include the activities derived from the microservice-specific systematic mapping study \cite{nuha2016} and described in Section \ref{keywords}. It is worth noting that the attempts analysed in this table are just a fraction of the publications summarised in Figure \ref{activities}. In Table \ref{definitions} we focus on the explicit attempts to define the transition to microservices, while Figure \ref{activities} includes both explicit attempts to define the transition and case studies of microservice adoption which do not explicitly attempt to define the transition.
\begin{footnotesize}
	\begin{longtable}{|c|c|c|c|c|c|c|c|}
		\hline
		\multicolumn{8}{|c|}{Microservice Adoption Activity}\\
		\cline{2-8}
		Publication&Architectural design&Organisation&Operation&Deployment&Development&Monitoring&Logging\\
		\hline
		\cite{stefanpresent}&\checkmark&\checkmark&&\checkmark&&&\\
		\hline
		\cite{differencesoa}&&&&\checkmark&&&\\
		\hline
		\cite{7958454}&\checkmark&\checkmark&\checkmark&&&&\\
		\hline
		\cite{7867539}&\checkmark&&&&&\checkmark&\checkmark\\
		\hline
		\cite{nealford}&\checkmark&\checkmark&&\checkmark&&\checkmark&\checkmark\\
		\hline
		\cite{Koutsouras:2015:ASI:2801948.2802030}&&&&\checkmark&&&\\
		\hline
		\cite{DBLP:journals/corr/DragoniLLMMS17}&&&&&\checkmark&&\\
		\hline
		\cite{7856721}&\checkmark&\checkmark&&&&\checkmark&\\
		\hline
		\cite{relpatterns}&\checkmark&\checkmark&&&\checkmark&&\\
		\hline
		\cite{7943959}&\checkmark&&\checkmark&\checkmark&\checkmark&\checkmark&\checkmark\\
		\hline
		\cite{stateofpractice}&\checkmark&&&&&&\\
		\hline
		\cite{7930195}&\checkmark&&&&\checkmark&\checkmark&\\
		\hline
		\cite{10.1007/978-3-319-74433-9_3}&&\checkmark&&\checkmark&&&\\
		\hline
		\cite{joselyneframework}&\checkmark&&\checkmark&&\checkmark&&\\
		\hline
		\cite{almeida2017survey}&&&&\checkmark&\checkmark&&\\
		\hline
		\cite{stratdecomp}&\checkmark&&&&&&\\
		\hline
		\caption{Analysing publications that attempt to define the transition to microservices included in the systematic mapping study; a check means the publication includes the activity in the corresponding column}
		\label{definitions}
	\end{longtable}
\end{footnotesize}	
Observing Table \ref{definitions}, we introduce a definition for the transition to microservices (adapted from the activities included in \cite{nuha2016}) which we call microservitization to cover all the relevant activities of the transition to microservices. This definition is adapted from activities included in \cite{nuha2016} and inspired by a trending concept in business and manufacturing domains \cite{b6b81df2-e27d-4bfe-81a4-9e44c7c0c6a3} --- servitization. 

In the manufacturing and business domains, servitization is seen as a paradigm shift entailing ``manufacturers growing their revenues and profits through services \cite{aston}" rather than tangible functional products. A service in this context is any feature that helps the business to (1) ``really make money" and (2) deliver new outcomes to customers. Examples of services in servitization include software applications, customer support, and self-service capabilities \cite{VANDERMERWE1988314}. 

Servitization ``embraces business model innovation, organisational change, and new technology adoption. Services exist in various forms, and represent differing values to both the customer and provider \cite{aston}". To benefit from these values, servitization involves developing new relationships with customers, innovating customer value propositions, forming new value chain relationships, and adapting business models \cite{aston,SLACK2005,doi:10.1108/17410380910960984,Davies01102004}. The key to successful servitization is ``choosing the right technology, picking the appropriate moment to invest, and ensuring successful implementation \cite{aston2}" which aligns with the business objectives \cite{VANDERMERWE1988314}.

We liken the transition to microservices to servitization because of the following resemblances:
\begin{itemize}
	\item Servitization is driven by delivering new outcomes to customers which can translate into revenues and profits. Similarly, the transition to microservices is driven by improving QoS provision to end users and translating that into an economic gain. 
	\item Servitization entails embracing innovation in the manufacturing activities (e.g., building business models and defining customer value propositions) are carried out. Similarly, the transition to microservices entails a dramatic change (motivated by value creation) to the way technical activities manifested in the software architecture are carried out. 
\end{itemize}

Therefore, we define \textit{microservitization as a form of servitization where "services/components are transformed into  microservices --- a more fine-grained and autonomic form of services \cite[p.1]{icsa2017}" --- to introduce added value to the architecture \cite{scc}. "Microservitization is also an example of a paradigm shift \cite[p.1]{icsa2017}" since it involves dramatically changing how the following technical activities are carried out to align them with a microservice adopter's business objectives:}
\begin{itemize}
	\item \textit{Architectural design}: microservitization introduces the following critical architectural design activities:
	\begin{itemize}
		\item Choosing light-weight communication mechanisms: microservitization can increase the distribution of functionalities across the architecture thereby introducing extra communication calls between microservices \cite{chal}. Therefore, rather than relying on enterprise service buses (which are the state-of-the-practice in SOA architectures), more light-weight mechanisms such as pipes and filters, event-based queues, and correlation identifiers are critical to avoid very high communication costs in microservice architectures. The large number of microservices can lead to a large volume of message exchange and hence high communication costs.  
		\item Reasoning about microservice granularity levels: a suitable granularity level is paramount to inform buying commercial-off-the-shelf (COTS) concrete services or developing them in-house \cite{scc}. Choosing these services correctly is critical to introducing added value to the microservice architecture.
		\item Adopting fault tolerance design patterns: although striving for fault tolerance is a best practice in any architecture, investing in fault tolerance design patterns is all the more critical to microservitization. The criticality is due to the scale of industries adopting microservices (e.g., retail \cite{otto}, entertainment \cite{netflixipc,bbc}) where microservices span different continents with a wide variety of end users. Microservice-specific fault tolerance design patterns include circuit breakers and bulkheads.
		\item Incorporating microservice registration and discovery mechanisms: microservices are typically developed, deployed and replaced at a very quick rate \cite{chal}. Therefore, it is critical to incorporate robust registry and discovery mechanisms in microservice architectures to ensure an up to date record of the currently ``alive" microservices. 
	\end{itemize}
	\item \textit{Managing the organisational hierarchy}: microservitization has a direct impact on the organisational hierarchy \cite{practical1}. In particular, the autonomy and independent deployability enhanced in the architecture through microservitization facilitate decentralised governance by breaking ``silos" (based around strict separation of job roles) in the organisation. 
	\item \textit{Operation}: microservitization aligns operation management with breaking organisational ``silos". DevOps and NoOps are among the state-of-the-practice operation management approaches in microservitization; they involve "a set of practices intended to reduce the time between committing a change to a system and the change being placed into normal production, while ensuring high quality \cite{bass2015devops}". Decentralised operation management can reduce the risk of bottlenecks that can materialise into economic losses to the microservice adopter. Reducing this risk is conditional upon development operation teams adhering to service level agreements between them.
	\item \textit{Deployment}: microservitization introduces a critical challenge of determining the hosts on which a deployment pipeline is implemented \cite{practical1}. This is critical to balancing between the added value of microservitization and cost that can be introduced by a deployment pipeline implementation choice (e.g., physical link installation, server rental and maintenance costs). Virtualisation and containerisation are among the common deployment pipeline implementation choices. They can enable swift auto-scaling the microservice architecture in response to changes in its runtime environment; this can materialise into economic gains for competitive microservice adopters. 
	\item \textit{Development}: unlike other software development paradigms, microservitization enables freedom in choosing development tools which can in turn introduce more added value to the architecture \cite{nuha2016}. This freedom is a bi-product of the decentralised governance enabled by microservitization. It is worth noting that communication and knowledge sharing across teams using different development tools needs to be carefully managed to ensure this freedom actually introduces added value. 
	\item \textit{Monitoring}: microservitization requires much more robust, decentralised and customisable monitoring than that of classical SOAs due to the heterogeneity of tools used to develop microservices and scale at which they typically operate \cite{chal}. These requirements are critical to cater for the heterogeneity, scale and dynamism of microservice architectures. 
	\item \textit{Logging}: maintaining logs of the monitoring data needs to be more customisable and distributable than logging classical SOAs due to the heterogeneity of tools used to develop microservices and scale at which they typically operate \cite{chal}. This requirements are aligned with the aforementioned monitoring challenges introduced by microservitization. 
\end{itemize}
\subsection{Objective 2: Understanding the Microservice Granularity Problem}\label{obj2gap}
Based on our definition above, microservitization introduces the challenge of reasoning about the suitable granularity level of a microservice. 

To formalise the microservice granularity problem, the gap we identified is the lack of an architecture-oriented modelling approach that captures a microservice's granularity behaviour, thereby supporting runtime analysis of this behaviour. The approach should treat microservice boundaries as the primitives for formulating the microservice granularity decision problem and actuators of microservice granularity adaptation decisions. These decisions include merging multiple microservices into a single boundary and decomposing a microservice into multiple ones encapsulated by multiple boundaries. In other words, this approach should treat microservice boundaries as adaptable first-class entities to ensure that both the structural and behavioural aspects of the microservice architecture are captured; we contribute to this approach in \cite{icsa2017}. While conducting our study, we have seen architecture-oriented modelling approaches that treat the notion of boundaries statically (e.g., \cite{hexagon}), or provide support for adaptability but without explicitly capturing the notion of adaptable boundaries (e.g., \cite{7930194}). Because the role of microservices is to encapsulate functionality, it is intuitive to use boundaries as adaptable first-class entities in the modelling approach. By contrast, if the decision problem was to determine the optimal physical infrastructure to host the microservice for example, the adaptable first-class entity would be different (e.g., microservice configuration variables).

To reason about microservice granularity objectively and dynamically, the gap we identified is the lack for an architectural evaluation approach that captures two aspects explicitly: added value to be introduced and cost to be incurred by pursuing granularity adaptation. 

To effectively support reasoning about granularity adaptation, the gap we identified is the need for a decision support tool for reasoning about this problem at runtime. It is seen as a runtime problem since the suitable granularity level highly depends on the current scenario in which the microservice architecture is operating \cite{scc,proser}. For example, if a certain functionality in a microservice-based application is continuously receiving a large volume of requests at runtime, it makes sense to decompose this functionality in a separate microservice to manage its load separately. On the other hand, if two microservices are continuously communicating across a network at runtime causing latency, then merging these microservices is sensible to help reduce such latency. Overall, uncertainties related to the expected environment and behaviour \cite{layers,roadmap} of the microservice architecture can not be fully captured at design time. Therefore, a runtime decision support tool is necessary to track and analyse this uncertainty. The tool should systematically guide the software architects towards suitable granularity adaptation strategies at runtime or suggest re-visiting their expectations of the microservice runtime environment. Each candidate granularity adaptation strategy must be systematically described as a sequence of merging/decomposition steps accompanied by triggers on them. Moreover, the tool's suggestions need to be justified objectively while leaving the final decision to the architects for adopting the suggested strategies; approaches such as \cite{10.1007/978-3-319-65831-5_11} can inspire the design of this tool. 
\section{Threats to Validity}\label{threats}
In this subsection we acknowledge the threats to validity in the process we use in our systematic mapping study as well as the application of each stage. Threats to validity are "influences that may limit our ability to interpret or draw conclusions from the study's data \cite[p.351]{Perry:2000:ESS:336512.336586}".

When defining our search strategy in Section \ref{searchstrat}, we considered blog articles, presentations, and videos as the means of reporting first-hand industrial experiences with microservice adoption. We acknowledge that an alternative means could be interviews with microservice adopters in the industry. However, published blog articles, presentations and videos are arguably more trusted since they present a more responsible and objective view than interviews. Though they can enrich the study with diverse opinions, interviews tend to suffer from bias, subjectivity and irresponsible answers \cite{Kitchenham:2004:ESE:998675.999432}. Moreover, our search strategy yields publications that explicitly mention microservices. Nevertheless, we acknowledge that there are publications prior without direct mention of microservices that could be relevant to their emergence and thereby to our research (e.g., web service composition and agile development). We attempt to address this threat briefly in Section \ref{relevant}.

We acknowledge that the inclusion and exclusion criteria might have led to missing contributions that can inspire the microservices field. However, since our study was motivated by studying the state-of-the-art and -practice in the microservices trend we based the criteria on publications which already have a link to microservices. Nevertheless, we briefly outline the research areas that can inspire the microservice trend in Section \ref{relevant}. On a more technical level, we excluded publications that could not be translated to English which might have affected the study results. 

We iteratively built Table \ref{categories} to include all the relevant keywords and made our best effort to justify them (Section \ref{process}). However, we acknowledge that some keywords might have been missed related to each research question.

When extracting videos and presentation for inclusion in the study results, we made our best effort to include videos whose content contained keywords from each category. The keywords are either mentioned by the speaker in the video or in the slides presented in the presentation. We acknowledge however that this might have biased the study results and that in the future transcription of the videos/presentations would be a more accurate means of determining their relevance. 

When categorising publications according to Table \ref{categories}, we made our best effort to considers synonyms of the keywords in each category. However, since we categorised the publications by manual examination, we acknowledge that their distributions might have been skewed. In particular, we acknowledge that subjective interpretation of keywords might need to be complemented with more systematic approaches of categorising the yielded publications. For example, the context in which a keyword or a fragment is mentioned needs to be considered before putting each publication under a certain category. A more systematic categorisation of the publications can ensure the reproducibility of our results. 

We acknowledge that skewed distributions might lead to biassed inferences regarding gaps in the literature related to each objective of our study. For example, the variation in numbers of publications across categories in Figure \ref{activities} might be due to the inadequacies in keywords under each category. It might also indicate interest in architectural design across practitioners (i.e. in non-academic publications); this might not be as accurate for academic publications. Nevertheless, we argue that our mapping study results give a strong insight into the microservice trend and opens directions for more detailed research down each direction. For example, a systematic mapping study can be conducted which focusses specifically on microservice architectural design and/or development --- the most dense categories in Figure \ref{activities}.
\section{Related Studies}\label{pubrelated}
Motivated by disciplining the understanding of microservices, several studies have been conducted to examine the existing literature in this young yet trending field. They analyse existing literature with a variety of focuses. In this section, we compare and contrast the examined studies against the systematic study we conducted.

The closest to our study are \cite{nuha2016,Pahl:2016:MSM:3021834.3021846} since they both adopt a systematic mapping study process when examining the literature. However, the motivations in these studies are different from ours. In \cite{nuha2016} for example, the research questions motivating the study are related to challenges, modelling approaches and quality attributes considered when adopting microservices. These questions do overlap partially with both objectives of our study but they do not focus on reasoning about microservice granularity as we do in our study. In \cite{10.1007/978-3-319-74781-1_15}, the activities comprising the transition to microservices are inferred through a literature review (aligned to Objective 1 of our study). However, it does not focus on microservice granularity so our study is significant to understanding this microservitization challenge. 

Several systematic literature reviews and surveys focus on defining the fundamental properties of microservices and the challenges of adopting them. They range in their rigour: some follow a rigorous search protocol \cite{Cerny:2018:CUM:3183628.3183631,Cerny:2017:DCS:3129676.3129682,Zimmermann2017,stateofpractice,10.1007/978-3-319-62407-5_14,10.1007/978-3-319-74433-9_3,7930195,SOLDANI2018215} while others are less formal \cite{DBLP:journals/corr/DragoniGLMMMS16}. These studies overlap partially with Objective 1 of our study; they present activities related to microservices thereby contributing to a better understanding of the transition. However, they do not define on the transition to microservices nor can they be used to understand the microservice granularity problem. In \cite{relpatterns}, the transition to microservices is partially described in terms of design patterns that can be applied to microservices. However, the scope of activities comprising the transition is not clearly defined. 

Some studies mainly focus on modelling microservices  \cite{archrecnour,doi:10.1080/17517575.2018.1462406}. Their focus partially overlaps with our following research question: \textit{what are the modelling approaches used to define the granularity of a microservice?}

On the other hand, \cite{8312516} aims to "construct knowledge of quality attributes in architecture through a Systematic Literature Review (SLR), (an) exploratory case study and (an) explanatory survey \cite[p.1]{8312516}." In essence, this study can be used to partially address our question: \textit{what are the quality attributes considered when reasoning about microservice granularity and how are they captured?} Nevertheless, the other research questions of our study were not answered in this study. 

Overall, the examined studies can complement this paper to discipline the understanding of the transition to microservices. In this paper, our research questions are formulated with focus on a specific problem of this transition --- reasoning about microservice granularity.
\section{Relevant Research Fields}\label{relevant}
Although we focus on publications that have direct links to microservices in our study, we acknowledge that there are publications prior to the time period considered in this mapping study that could be relevant to the emergence of microservices and thereby to our research. Such publications do not fit our search strategy because they do not make direct reference to microservices. Nevertheless, in this section we briefly summarise the examined literature in these areas indicating how they can be relevant to our systematic mapping study objectives.
\subsection{Research Fields Relevant to Understanding the Microservice Transition}
Even among microservice adopters, there are still debates over distinguishing service-oriented architectures (SOA) from microservice architectures \cite{relpatterns,almeida2017survey,whataremicro,heritage,microfow,ibmredbooks,amazonmicro,release}. Therefore, we infer that contributions defining the properties and challenges of adopting SOA architectures can be relevant to understanding the transition to microservices. 

Seminal work defines SOA as an architectural style which can guide business process definition \cite{erl2007soa,erl2005service,mackenzie2006reference,rosen2012applied,1210138,Zimmermann:2005:SAB:1094855.1094965} and support "rapid, low-cost composition of distributed applications \cite[p.1]{1254461,Papazoglou2006ServiceOrientedCR,Papazoglou:2003:ISC:944217.944233}." The SOA style was introduced to address architectural complexity, redundant programming and inconsistent interfaces \cite{channabasavaiah2003migrating,evalsoa}. In SOAs a service is a self-describing unit that "consists of a contract, one or more interfaces and an implementation \cite[p.57]{krafzig2005enterprise}." SOAs in turn comprise  an application frontend, services, a service repository and enterprise service bus. 

Despite the resemblance between microservices and SOAs, there is a subtle distinction we infer from our microservitization definition. The distinction comes from \cite{scc}: 1) the potential of microservices as autonomous fine-grained computational units with lightweight communication mechanisms rather than service buses and, 2) the operational and organisational flexibility enhanced by microservitization. Further elaboration of these points is presented in \cite{markrich}. 

\subsection{Research Fields Relevant to Understanding the Microservice Granularity Problem}
Modelling microservice granularity can be inspired by architectural modelling approaches --- a wide research field, where contributions capture different notions of the architecture at varying levels of abstraction \cite{shark}. Domain-driven modelling is the most relevant to the modelling approach we describe in Section \ref{obj2gap} because it strives for logical isolation of business functionalities \cite{domdriv,microbuild}. However, domain-driven modelling is more concerned about the relationships between functional boundaries rather than their scope. In essence, domain-driven modelling can inform the structural rather than behavioural aspect of modelling microservices. 

The Zachman framework \cite{zachman} provides a comprehensive guide to the different dimensions and perspectives for architectural modelling. Two dimensions in this framework are aligned with the modelling approach we call in Section \ref{obj2gap}: \textit{what} the model units are and \textit{where} these units are located relative to each other. We call for a modelling approach that explicitly captures \textit{what} each microservice is concerned about and helps define \textit{where} the business functionalities encapsulated by the microservice are located relative to each other. 

A more dynamic architectural modelling approach is feature modelling \cite{white08creatingself-healing}, where an architecture is defined as a set of variability points, the candidates for each variability point and the rules constricting the dependencies across variability points. We appreciate this modelling technique is useful for formulating architectural decision-making problems. However, we would need to leverage such concept to focus on microservice boundaries being the variability point. While dependency rules in a feature model can give an insight about microservice granularity, they do not explicitly model it. EUREMA \cite{Vogel:2014:MES:2578044.2555612} provides a yet more powerful dynamic modelling technique. Here an adaptation engine contains a runtime model which represents the evolution of the system as well as adaptation activities to be executed on that model. The link between the adaptation activities and the runtime model is expressed using runtime mega-models. Similar to feature modelling, EUREMA needs to capture the notion of microservice boundaries more  explicitly. 

Boundaries are modelled more explicitly in design structure matrices (DSMs) \cite{dsm}. Modularity metrics \cite{decouple} can be used to assess the degree of interdependence across these boundaries. We have not seen a dynamic application of DSMs that captures the changes in these dependencies across a time unit (e.g., release cycles). 

Since we call for objective reasoning about microservice granularity adaptation, design metrics can inspire this requirement since they provide an objective way to capturing attributes of a design decision. Effort-based metrics \cite{42969} evaluate software development and maintenance efforts when a transition is made from centralised to distributed system architectures \cite{1702220}. By analogy, an objective way is needed to evaluate development and maintenance costs when microservice granularity is adapted. Metrics related to cohesion, coupling and visibility of system components are presented and visualised in \cite{1702220,Briand:1993:MAM:645542.658018}, which can be used to assess the impact of granularity adaptation on the microservice architecture's modularity. 

Since reasoning about microservice granularity is in essence a dynamic architectural design decision problem, several software engineering fields can be relevant to it. Architectural analysis methods, architectural design patterns, service composition and orchestration approaches, runtime architectural adaptation, architectural refactoring and feedback control loops are only some of the relevant fields. In the following subsections we categorise contributions in these fields according to their level of autonomy:
\begin{itemize}
	\item \textit{Manual} contributions with full reliance on the software architect and/or stakeholders (e.g., architectural design patterns)
	\item \textit{Partially autonomous} contributions where there is an autonomous agent but the software architect still makes the final decision regarding the optimal architecture (e.g., interactive service orchestration and/or composition)
	\item \textit{Fully autonomous} contributions where the decision-making process and executing the decision are fully handled by an autonomous agent (e.g., feedback control loops, online architectural refactoring)
\end{itemize}
\subsubsection{Manual Contributions}
Out of the approaches in this subsection, the most cost- and value-aware approaches we examined are \cite{ipek,kevin,module}. Refactoring the architecture according to patterns \cite{ipek} or to introduce modularity \cite{module} are regarded as value-bearing investments \cite{kevin}. However, these approaches are only applied statically at design time. In this paper, we call for a similar view for reasoning about microservice granularity. 

In \cite{cbam}, a cost benefit analysis method (CBAM) is proposed as a generic architecture evaluation method which utilises techniques in decision analysis, optimisation, and statistics to evaluate architectural design decisions. 
However, CBAM does not dynamically track and update the added value of architectural decisions. These dynamic updates are critical to the nature of the microservice granularity problem.

In \cite{costeff}, the net benefit of a software is calculated by deducting its total costs from the total benefit. These are manually elicited and monetised from software architects through a series of questions (e.g., ``what is the status of the environment without the system?''). To our knowledge, this approach however does not consider the uncertainty in the answers to these questions nor does it update them at runtime. In \cite{shaw} the predictive analysis of design captures the value-driven impact of architectural decisions as multi-dimensional normalised, weighted cost and value vectors. Therefore, it can only provide static objective decision support for reasoning about microservice granularity. 

The techniques in \cite{1019479} present different architectural evaluation methods and their motivations. Software Architecture Analysis for Evolution and Reusability (SAAMER), Scenario-Based Architecture Re-engineering (SBAR) and Architecture Level Prediction of Software Maintenance (ALPSM) in particular take an objective approach to architectural evaluation. 

SBAR captures the runtime nature of decision-making by providing different quality attribute evaluation techniques depending on whether the quality attribute is concerned with the ``development'' of the system (i.e. design time, such as reusability, which is handled by scenario-based evaluation) or the ``operation'' of the system (i.e. runtime, such as performance, which is handled by simulation-based evaluation). SAAMER on the other hand partially addresses granularity problem by analysing the level of interaction between different scenarios of the system as a means of assessing the level of functionality isolation in the system. Nevertheless, both methods have not been explicitly applied in a dynamic environment to our knowledge. 

ALPSM takes a more value-driven approach to the evaluation, similar to the Cost Benefit Analysis Method (CBAM) \cite{cbam}, which makes them more systematic architectural evaluation approaches than SAAMER and SBAR above. Furthermore, ALPSM uses probabilities to capture the likelihood of the impact of scenarios and CBAM captures the uncertainty the architectural analysis. ALPSM and CBAM therefore partially capture uncertainty, although they do not operate at runtime and thereby they suffer from the limitation of design-time analysis. 

Classical design patterns presented in \cite{gamma} extensively study creational (concerned with object instantiation), structural (concerned with relationships between objects) and behavioural patterns (concerned with coordination between objects). These patterns are further categorised according to their static or runtime nature. In that context, reasoning about  microservice granularity can benefit from runtime creational design patterns. However, the design patterns of that category in \cite{gamma} do not capture the scope --- boundary --- where a pattern can be enforced. Service workflow patterns have been presented in a seminal work  \cite{workflow} which implicitly discussed the issue of granularity is SOAs. However, we envision that the distinction between microservices and SOAs calls for explicitly addressing granularity adaptation decisions in the context of microservice constraints. 
\subsubsection{Partially Autonomous Contributions}
In \cite{Casati:2000:ADS:646088.679914,caaa}, pattern-based engines are proposed to synthesize a composition of atomic and composite services. However, we envision that reasoning about microservice granularity needs to be grounded on objective rather than pattern-based evaluation. 
In \cite{fuzzycluster,sustainecon}, a microservice-specific approach for addressing the microservice granularity problem is proposed which relies on microservice web application log mining to extract the usage pattern and then making adaptive decisions regarding microservice granularity to ensure an economically sustainable architecture. We acknowledge this work is very closely related to the decision support called for in this paper. Nevertheless, it does not explicitly analyse the value-driven implications of such adaptive decisions as we call for in Section \ref{obj2gap}.

\subsubsection{Fully Autonomous Contributions}
The closest fully autonomous contribution to the effective decision support we call for in Section \ref{obj2gap} is the ASTRO-CAptEvo orchestration framework \cite{astro}. It is a runtime framework that allows partial definition of business processes for service-based systems at design-time. Subsequently, the framework orchestrates "automatically composing the currently available services, provided by other actors and systems, according to the execution context and the goal of the process to be refined" using state transition systems. This framework takes a runtime approach to decision-making. However, the decision problem targeted by ASTRO-CAptEvo is service composition rather than microservice granularity. Moreover, ASTRO-CAptEvo does not consider objectively reason about composing the available services.  

The Self-Serv framework presented in \cite{p2p} facilitates composite web service execution through peer-to-peer message exchange between coordination agents, which manage the service composition according to a static state chart. This framework can be utilised to address microservice granularity adaptation, but the knowledge that drives the service composition is static, meaning the runtime nature of the granularity problem is not captured in this framework.

Reputation-based dynamic service configuration techniques such as \cite{trust} use a policy language to capture service consumers' and providers' profiles and then utilise these profiles to dynamically configure an optimal concrete service architecture. The work in \cite{reput} leverages on the concept of reputation by capturing trust in the feedback given regarding the services. In particular, a model is proposed to aggregate feedback from several consumers of a service to reduce the effect of biased feedback. In both cases, a subjective user profile is used to drive the composition rather than an objective value- and cost-driven approach.

Case-based reasoning about concrete service composition is presented in \cite{case} where a solution space of composite services  is formalised using recursive tuples of services. An agent then synthesises the optimal service decomposition given a request for services from the user which are then bound at runtime to concrete services fetched from a registry. A distinction is therefore made here between concrete service selection and the higher level composition of services; this can be utilised to address the granularity problem. However, this solution does not capture the dynamic nature of the microservice granularity problem. 

Service composition techniques based on model checking \cite{assur1,assur2,assur3} dynamically adapt probabilistic models of the system according to runtime changes in the scenarios surrounding the system or runtime changes in the requirements of the system. In \cite{assur3} Bayesian learning improves service composition synthesis process through runtime knowledge updates about the system's behaviour over its lifetime. In \cite{assur2} an abstract service composition is mapped to a concrete service composition at runtime. The field of model-checking therefore is an attractive one for runtime decision-making. Such contributions however need to be leveraged for the specific problem we are concerned with (i.e. the granularity of microservices). 

Similar to the field of model checking is runtime architecture modelling. Several contributions in this field manifest runtime changes to the architecture \cite{Bencomo:2013:DDN:2663546.2663565,Blair:2009:MR:1638585.1638645,Zhang:2006:MDD:1134285.1134337,Amann2014,sysreconfig,4216407,1541182,Vogel:2014:MES:2578044.2555612,Fleurey:2009:MVD:1537875.1537890,5070514}. Other contributions are catered for systems which exhibit a similar level of dynamism to microservices \cite{Floch:2006:UAM:1128592.1128711,Mongiello2016,7958486}. However, these contributions would need to be leveraged with objective reasoning that considers both the added value and cost of granularity adaptation.

Another approach of an autonomous solution is dynamic service formation rather than dynamic service composition. Frameworks such as \cite{gsoap,cloudref} provide means to dynamically produce web service specifications conforming to a service composition. These approaches can be utilised to complement the decision support we call for in this paper. 

The runtime, uncertain context of microservice granularity adaptation calls for support similar to that provided by engineering self-adaptivity \cite{scc,roadmap} into an architecture. The role of a self-adaptive solution is to refine and update at runtime the architects' design-time expectations about the architecture's behaviour. There  are  several mechanisms which can be adopted in this solution \cite{ddn,explicit,rainbow,ibm,secroadmap}. Underlying most of them is the concept of feedback control loops \cite{loop,cont1} which can be used "to monitor, analyse, plan and execute adaptations \cite[p.16]{mapek}" in a system regarding trade-offs of concern; the knowledge learnt about the system needs to be maintained in a knowledge base (MAPE-K loop \cite{rainbow}). 

Control loops can be composed in a centralized, hierarchical, master-slave, or fully decentralized pattern \cite{secroadmap}. Each pattern varies in which components of the architecture carry out which phase(s) of the control loop. The centralized pattern is more suitable for monolithic architectures. In a hierarchical pattern, the full MAPE loop is effected at individual services, with higher level services having a more general view of the architecture. The individual service MAPE loops pass information to the higher level loops at short time intervals.  Although this pattern is well-suited to addressing the trade-offs of concern here, its only shortcoming is deciding what the higher level component with the global view of the architecture should comprise and the possibility of this service creating a bottleneck. A variant of the hierarchical pattern is the master-slave pattern where the individual services only monitor and execute while a higher level service comprises the analysis and planning phases. Although more lightweight than the hierarchical pattern, the master/slave pattern suffers from the same shortcomings as the hierarchical pattern. The decentralized pattern \cite{dec2,dec3,dec4,dec5} on the other hand takes away the need for a service with a global view of the control loop. The MAPE loop is implemented in each service and information is passed across the services for decentralized management. The major challenge of this pattern is guaranteeing a consistent view of the system and its environment across all the control loops \cite{secroadmap}. However, this pattern is the most aligned with the autonomy of microservice architectures and the scale at which microservices operate.

Runtime architectural adaptation approaches have been proposed before in the Rainbow framework \cite{rainbow} which is the most aligned with the concept of feedback loops. The Rainbow framework provides a reusable solution to induce self-adaptivity into a system in a cost-aware manner. However, it is debatable whether dynamically adapting the level of granularity of a microservice can be captured using the Rainbow framework. This is because the solution space here varies regarding the number of services used and the interaction patterns between them. To our knowledge, the Rainbow framework has not been applied to such a setting before. Another prominent approach is the architectural refactoring approach \cite{barrier} where a set of anti-patterns is proposed which can be detected dynamically and used to trigger refactoring an architecture. This approach has as its main motive enhancing the modularity of the architecture rather than reasoning about modularity in a objective manner. 

Realising microservice granularity adaptation involves changes to a deployed architecture. These activities are similar to those underlying the field of online architectural refactoring. Several contributions in this field pave the way to automated online architectural refactoring \cite{archzimmer,refactoring,Stal201463,Schmidt2012,10.1007/978-3-319-22885-3_2,Zimmermann:2012:ADI:2361999.2362021,Oreizy:1998:ARS:302163.302181} and modelling transformations \cite{dsm,1602365}. These contributions are driven by meeting a specific architectural design pattern, but they do not objectively reason about granularity adaptation.

In the field of AI planning, an ontology-based approach to architectural adaptation is proposed \cite{mci-son-kr02}. A shared ontology of generic ``procedures'' (or templates) is produced which the stakeholder can choose from at run-time. An agent then executes this procedure customizing it depending on the scenario in which the system will operate. It is appreciated that the use of ontologies promotes sharing knowledge across across architects. However, we envision that the decision support we are calling for in this paper can promote knowledge sharing and profiling to inform reasoning about microservice granularity.
\section{Conclusion}\label{conclusion}
In this paper we report on a systematic mapping study to consolidate various views, principles, methods and techniques that are commonly adopted to assist the transition to microservices. We systematically describe the study's process and report its results. We contribute a working definition capturing the fundamentals of the transition; we term it as microservitization. Microservitization is a form of servitization \cite{aston,aston2} where services/components are transformed into  microservices --- a more fine-grained and autonomic form of services --- to introduce added value to the architecture \cite{scc}. Microservitization is also an example of a paradigm shift since it involves a dramatic change to the way technical activities are carried out and aligned with a microservice adopter's business objectives. We then shed light on a fundamental problem of microservitization: microservice granularity and reasoning about its adaptation as first-class entities. This study has reviewed and identified gaps in the state-of-the-art and -practice that relate to the modelling approaches, aspects considered, guidelines and processes used to reason about microservice granularity. The identified gaps pave the way to opportunities for future research and development related to reasoning about microservice granularity. In particular, we identify there is room for: (1) systematic architecture-oriented modelling support for microservice granularity, (2) a dynamic architectural evaluation approach to reason about the cost and added value of granularity adaptation and (3) effective decision support to inform reasoning about microservice granularity at runtime.
	\bibliographystyle{ACM-Reference-Format}
	\bibliography{thesisref}


\begin{thebibliography}{272}


\ifx \showCODEN    \undefined \def \showCODEN     #1{\unskip}     \fi
\ifx \showDOI      \undefined \def \showDOI       #1{#1}\fi
\ifx \showISBNx    \undefined \def \showISBNx     #1{\unskip}     \fi
\ifx \showISBNxiii \undefined \def \showISBNxiii  #1{\unskip}     \fi
\ifx \showISSN     \undefined \def \showISSN      #1{\unskip}     \fi
\ifx \showLCCN     \undefined \def \showLCCN      #1{\unskip}     \fi
\ifx \shownote     \undefined \def \shownote      #1{#1}          \fi
\ifx \showarticletitle \undefined \def \showarticletitle #1{#1}   \fi
\ifx \showURL      \undefined \def \showURL       {\relax}        \fi
\providecommand\bibfield[2]{#2}
\providecommand\bibinfo[2]{#2}
\providecommand\natexlab[1]{#1}
\providecommand\showeprint[2][]{arXiv:#2}

\bibitem[\protect\citeauthoryear{Acevedo, y~Jorge, and Patino}{Acevedo
  et~al\mbox{.}}{2017}]%
        {8169955}
\bibfield{author}{\bibinfo{person}{C.~A.~J. Acevedo},
  \bibinfo{person}{J.~P.~Gomez y Jorge}, {and} \bibinfo{person}{I.~R. Patino}.}
  \bibinfo{year}{2017}\natexlab{}.
\newblock \showarticletitle{Methodology to transform a monolithic software into
  a microservice architecture}. In \bibinfo{booktitle}{\emph{2017 6th
  International Conference on Software Process Improvement (CIMPS)}}.
  \bibinfo{pages}{1--6}.
\newblock
\urldef\tempurl%
\url{https://doi.org/10.1109/CIMPS.2017.8169955}
\showDOI{\tempurl}


\bibitem[\protect\citeauthoryear{Alipour and Liu}{Alipour and Liu}{2017}]%
        {8258201}
\bibfield{author}{\bibinfo{person}{H. Alipour} {and} \bibinfo{person}{Y. Liu}.}
  \bibinfo{year}{2017}\natexlab{}.
\newblock \showarticletitle{Online machine learning for cloud resource
  provisioning of microservice backend systems}. In
  \bibinfo{booktitle}{\emph{2017 IEEE International Conference on Big Data (Big
  Data)}}. \bibinfo{pages}{2433--2441}.
\newblock
\urldef\tempurl%
\url{https://doi.org/10.1109/BigData.2017.8258201}
\showDOI{\tempurl}


\bibitem[\protect\citeauthoryear{Almeida, de~Aguiar~Monteiro, Hazin, de~Lima,
  and Ferraz}{Almeida et~al\mbox{.}}{2017}]%
        {almeida2017survey}
\bibfield{author}{\bibinfo{person}{Washington Henrique~Carvalho Almeida},
  \bibinfo{person}{Luciano de Aguiar~Monteiro},
  \bibinfo{person}{Raphael~Rodrigues Hazin},
  \bibinfo{person}{Anderson~Cavalcanti de Lima}, {and}
  \bibinfo{person}{Felipe~Silva Ferraz}.} \bibinfo{year}{2017}\natexlab{}.
\newblock \showarticletitle{Survey on Microservice Architecture-Security,
  Privacy and Standardization on Cloud Computing Environment}. In
  \bibinfo{booktitle}{\emph{The Twelfth International Conference on Software
  Engineering Advances (ICSEA 2017)}}. \bibinfo{pages}{210}.
\newblock


\bibitem[\protect\citeauthoryear{Alshuqayran, Ali, and Evans}{Alshuqayran
  et~al\mbox{.}}{2016}]%
        {nuha2016}
\bibfield{author}{\bibinfo{person}{Nuha Alshuqayran}, \bibinfo{person}{Nour
  Ali}, {and} \bibinfo{person}{Roger Evans}.} \bibinfo{year}{2016}\natexlab{}.
\newblock \showarticletitle{A Systematic Mapping Study in Microservice
  Architecture}. In \bibinfo{booktitle}{\emph{2016 IEEE 9th International
  Conference on Service-Oriented Computing and Applications}}.
\newblock


\bibitem[\protect\citeauthoryear{Alshuqayran, Ali, and Evans}{Alshuqayran
  et~al\mbox{.}}{2018}]%
        {archrecnour}
\bibfield{author}{\bibinfo{person}{Nuha Alshuqayran}, \bibinfo{person}{Nour
  Ali}, {and} \bibinfo{person}{Roger Evans}.} \bibinfo{year}{2018}\natexlab{}.
\newblock \showarticletitle{Towards Micro Service Architecture Recovery: An
  Empirical Study}. In \bibinfo{booktitle}{\emph{IEEE INTERNATIONAL CONFERENCE
  ON SOFTWARE ARCHITECTURE 2018}}.
\newblock


\bibitem[\protect\citeauthoryear{Andrawos and Helmich}{Andrawos and
  Helmich}{2017}]%
        {andrawos2017cloud}
\bibfield{author}{\bibinfo{person}{M. Andrawos} {and} \bibinfo{person}{M.
  Helmich}.} \bibinfo{year}{2017}\natexlab{}.
\newblock \bibinfo{booktitle}{\emph{Cloud Native programming with Golang:
  Develop microservice-based high performance web apps for the cloud with Go}}.
\newblock \bibinfo{publisher}{Packt Publishing}.
\newblock
\showISBNx{9781787127968}
\urldef\tempurl%
\url{https://books.google.co.uk/books?id=NvNFDwAAQBAJ}
\showURL{%
\tempurl}


\bibitem[\protect\citeauthoryear{Arcot~Rajasekar}{Arcot~Rajasekar}{2012}]%
        {datamanage}
\bibfield{author}{\bibinfo{person}{Wayne~Schroeder Arcot~Rajasekar, Mike~Wan}.}
  \bibinfo{year}{2012}\natexlab{}.
\newblock \bibinfo{booktitle}{\emph{Micro-Services: A Service-Oriented Paradigm
  for Scalable, Distributed Data Management}}.
\newblock \bibinfo{publisher}{Data Intensive Distributed Computing: Challenges
  and Solutions for Large-scale Information Management}.
\newblock


\bibitem[\protect\citeauthoryear{Ashikhmin, Radchenko, and Tchernykh}{Ashikhmin
  et~al\mbox{.}}{2017}]%
        {10.1007/978-3-319-71255-0_37}
\bibfield{author}{\bibinfo{person}{Nikita Ashikhmin}, \bibinfo{person}{Gleb
  Radchenko}, {and} \bibinfo{person}{Andrei Tchernykh}.}
  \bibinfo{year}{2017}\natexlab{}.
\newblock \showarticletitle{RAML-Based Mock Service Generator for Microservice
  Applications Testing}. In \bibinfo{booktitle}{\emph{Supercomputing}},
  \bibfield{editor}{\bibinfo{person}{Vladimir Voevodin} {and}
  \bibinfo{person}{Sergey Sobolev}} (Eds.). \bibinfo{publisher}{Springer
  International Publishing}, \bibinfo{address}{Cham},
  \bibinfo{pages}{456--467}.
\newblock
\showISBNx{978-3-319-71255-0}


\bibitem[\protect\citeauthoryear{Asik and Selcuk}{Asik and Selcuk}{2017}]%
        {7965739}
\bibfield{author}{\bibinfo{person}{T. Asik} {and} \bibinfo{person}{Y.~E.
  Selcuk}.} \bibinfo{year}{2017}\natexlab{}.
\newblock \showarticletitle{Policy enforcement upon software based on
  microservice architecture}. In \bibinfo{booktitle}{\emph{2017 IEEE 15th
  International Conference on Software Engineering Research, Management and
  Applications (SERA)}}. \bibinfo{pages}{283--287}.
\newblock
\urldef\tempurl%
\url{https://doi.org/10.1109/SERA.2017.7965739}
\showDOI{\tempurl}


\bibitem[\protect\citeauthoryear{A{\ss}mann, G{\"o}tz, J{\'e}z{\'e}quel, Morin,
  and Trapp}{A{\ss}mann et~al\mbox{.}}{2014}]%
        {Amann2014}
\bibfield{author}{\bibinfo{person}{Uwe A{\ss}mann}, \bibinfo{person}{Sebastian
  G{\"o}tz}, \bibinfo{person}{Jean-Marc J{\'e}z{\'e}quel},
  \bibinfo{person}{Brice Morin}, {and} \bibinfo{person}{Mario Trapp}.}
  \bibinfo{year}{2014}\natexlab{}.
\newblock \bibinfo{booktitle}{\emph{A Reference Architecture and Roadmap for
  Models@run.time Systems}}.
\newblock \bibinfo{publisher}{Springer International Publishing},
  \bibinfo{address}{Cham}, \bibinfo{pages}{1--18}.
\newblock
\showISBNx{978-3-319-08915-7}
\urldef\tempurl%
\url{https://doi.org/10.1007/978-3-319-08915-7_1}
\showDOI{\tempurl}


\bibitem[\protect\citeauthoryear{Asundi, Kazman, and Klein}{Asundi
  et~al\mbox{.}}{2001}]%
        {cbam}
\bibfield{author}{\bibinfo{person}{Jayatirtha Asundi}, \bibinfo{person}{Rick
  Kazman}, {and} \bibinfo{person}{Mark Klein}.}
  \bibinfo{year}{2001}\natexlab{}.
\newblock \bibinfo{booktitle}{\emph{Using Economic Considerations to Choose
  Among Architecture Design Alternatives}}.
\newblock \bibinfo{type}{{T}echnical {R}eport} CMU/SEI-2001-TR-035.
  \bibinfo{institution}{Software Engineering Institute, Carnegie Mellon
  University}, \bibinfo{address}{Pittsburgh, PA}.
\newblock
\urldef\tempurl%
\url{http://resources.sei.cmu.edu/library/asset-view.cfm?AssetID=5785}
\showURL{%
\tempurl}


\bibitem[\protect\citeauthoryear{Babar, Dings{\o}yr, Lago, and vander
  Vilet}{Babar et~al\mbox{.}}{2009}]%
        {shark}
\bibfield{author}{\bibinfo{person}{Muhammad~Ali Babar},
  \bibinfo{person}{Torgeir Dings{\o}yr}, \bibinfo{person}{Patricia Lago}, {and}
  \bibinfo{person}{Hans vander Vilet}.} \bibinfo{year}{2009}\natexlab{}.
\newblock \bibinfo{booktitle}{\emph{Software Architecture Knowledge Management
  - Theory and Practice} (\bibinfo{edition}{first} ed.)}.
\newblock \bibinfo{publisher}{Springer-Verlag Berlin Heidelberg}.
\newblock


\bibitem[\protect\citeauthoryear{Baines}{Baines}{[n. d.]a}]%
        {aston2}
\bibfield{author}{\bibinfo{person}{Tim Baines}.} \bibinfo{year}{[n.
  d.]}\natexlab{a}.
\newblock \bibinfo{title}{Digitalisation and servitization: the competitive
  advantage?}
\newblock
  \bibinfo{howpublished}{https://www.advancedservicesgroup.co.uk/single-post/2018/04/04/Digitalisation-and-servitization-the-competitive-advantage}.
\newblock


\bibitem[\protect\citeauthoryear{Baines}{Baines}{[n. d.]b}]%
        {aston}
\bibfield{author}{\bibinfo{person}{Tim Baines}.} \bibinfo{year}{[n.
  d.]}\natexlab{b}.
\newblock \bibinfo{title}{Servitization: From understanding to implementation}.
\newblock
  \bibinfo{howpublished}{https://www.advancedservicesgroup.co.uk/single-post/2016/10/12/Servitization-From-understanding-to-implementation}.
\newblock


\bibitem[\protect\citeauthoryear{Baines, Lightfoot, Benedettini, and
  Kay}{Baines et~al\mbox{.}}{2009}]%
        {doi:10.1108/17410380910960984}
\bibfield{author}{\bibinfo{person}{T.S. Baines}, \bibinfo{person}{H.W.
  Lightfoot}, \bibinfo{person}{O. Benedettini}, {and} \bibinfo{person}{J.M.
  Kay}.} \bibinfo{year}{2009}\natexlab{}.
\newblock \showarticletitle{The servitization of manufacturing: A review of
  literature and reflection on future challenges}.
\newblock \bibinfo{journal}{\emph{Journal of Manufacturing Technology
  Management}} \bibinfo{volume}{20}, \bibinfo{number}{5}
  (\bibinfo{year}{2009}), \bibinfo{pages}{547--567}.
\newblock
\urldef\tempurl%
\url{https://doi.org/10.1108/17410380910960984}
\showDOI{\tempurl}
\showeprint{http://dx.doi.org/10.1108/17410380910960984}


\bibitem[\protect\citeauthoryear{Bakshi}{Bakshi}{2017}]%
        {7943959}
\bibfield{author}{\bibinfo{person}{K. Bakshi}.}
  \bibinfo{year}{2017}\natexlab{}.
\newblock \showarticletitle{Microservices-based software architecture and
  approaches}. In \bibinfo{booktitle}{\emph{2017 IEEE Aerospace Conference}}.
  \bibinfo{pages}{1--8}.
\newblock
\urldef\tempurl%
\url{https://doi.org/10.1109/AERO.2017.7943959}
\showDOI{\tempurl}


\bibitem[\protect\citeauthoryear{Balalaie, Heydarnoori, and Jamshidi}{Balalaie
  et~al\mbox{.}}{2016}]%
        {7436659}
\bibfield{author}{\bibinfo{person}{A. Balalaie}, \bibinfo{person}{A.
  Heydarnoori}, {and} \bibinfo{person}{P. Jamshidi}.}
  \bibinfo{year}{2016}\natexlab{}.
\newblock \showarticletitle{Microservices Architecture Enables DevOps:
  Migration to a Cloud-Native Architecture}.
\newblock \bibinfo{journal}{\emph{IEEE Software}} \bibinfo{volume}{33},
  \bibinfo{number}{3} (\bibinfo{date}{May} \bibinfo{year}{2016}),
  \bibinfo{pages}{42--52}.
\newblock
\showISSN{0740-7459}
\urldef\tempurl%
\url{https://doi.org/10.1109/MS.2016.64}
\showDOI{\tempurl}


\bibitem[\protect\citeauthoryear{Baraiya and Singh}{Baraiya and Singh}{2016}]%
        {conduct}
\bibfield{author}{\bibinfo{person}{Viren Baraiya} {and} \bibinfo{person}{Vikram
  Singh}.} \bibinfo{year}{2016}\natexlab{}.
\newblock \bibinfo{title}{Netflix Conductor: A microservices orchestrator}.
\newblock
  \bibinfo{howpublished}{https://medium.com/netflix-techblog/netflix-conductor-a-microservices-orchestrator-2e8d4771bf40}.
\newblock


\bibitem[\protect\citeauthoryear{Barbier}{Barbier}{2006}]%
        {4216407}
\bibfield{author}{\bibinfo{person}{F. Barbier}.}
  \bibinfo{year}{2006}\natexlab{}.
\newblock \showarticletitle{MDE-based Design and Implementation of Autonomic
  Software Components}. In \bibinfo{booktitle}{\emph{2006 5th IEEE
  International Conference on Cognitive Informatics}},
  Vol.~\bibinfo{volume}{1}. \bibinfo{pages}{163--169}.
\newblock
\urldef\tempurl%
\url{https://doi.org/10.1109/COGINF.2006.365692}
\showDOI{\tempurl}


\bibitem[\protect\citeauthoryear{Bass, Weber, and Zhu}{Bass
  et~al\mbox{.}}{2015}]%
        {bass2015devops}
\bibfield{author}{\bibinfo{person}{L. Bass}, \bibinfo{person}{I. Weber}, {and}
  \bibinfo{person}{L. Zhu}.} \bibinfo{year}{2015}\natexlab{}.
\newblock \bibinfo{booktitle}{\emph{DevOps: A Software Architect's
  Perspective}}.
\newblock \bibinfo{publisher}{Pearson Education}.
\newblock
\showISBNx{9780134049878}
\urldef\tempurl%
\url{https://books.google.com.sa/books?id=fcwkCQAAQBAJ}
\showURL{%
\tempurl}


\bibitem[\protect\citeauthoryear{Behara}{Behara}{2018}]%
        {govern}
\bibfield{author}{\bibinfo{person}{Dr. Gopala~Krishna Behara}.}
  \bibinfo{year}{2018}\natexlab{}.
\newblock \bibinfo{title}{Microservices Governance: A Detailed Guide}.
\newblock
  \bibinfo{howpublished}{https://blog.leanix.net/en/microservices-governance}.
\newblock


\bibitem[\protect\citeauthoryear{Benatallah, Dumas, Sheng, and Ngu}{Benatallah
  et~al\mbox{.}}{2002}]%
        {p2p}
\bibfield{author}{\bibinfo{person}{B. Benatallah}, \bibinfo{person}{M. Dumas},
  \bibinfo{person}{Q.Z. Sheng}, {and} \bibinfo{person}{A.H.H. Ngu}.}
  \bibinfo{year}{2002}\natexlab{}.
\newblock \showarticletitle{Declarative composition and peer-to-peer
  provisioning of dynamic Web services}. In \bibinfo{booktitle}{\emph{Data
  Engineering, 2002. Proceedings. 18th International Conference on}}.
  \bibinfo{pages}{297--308}.
\newblock
\showISSN{1063-6382}
\urldef\tempurl%
\url{https://doi.org/10.1109/ICDE.2002.994738}
\showDOI{\tempurl}


\bibitem[\protect\citeauthoryear{Bencomo and Belaggoun}{Bencomo and
  Belaggoun}{2013}]%
        {ddn}
\bibfield{author}{\bibinfo{person}{Nelly Bencomo} {and} \bibinfo{person}{Amel
  Belaggoun}.} \bibinfo{year}{2013}\natexlab{}.
\newblock \showarticletitle{Supporting Decision-Making for Self-Adaptive
  Systems: From Goal Models to Dynamic Decision Networks}.
\newblock In \bibinfo{booktitle}{\emph{Requirements Engineering: Foundation for
  Software Quality}}, \bibfield{editor}{\bibinfo{person}{Joerg Doerr} {and}
  \bibinfo{person}{AndreasL. Opdahl}} (Eds.). \bibinfo{series}{Lecture Notes in
  Computer Science}, Vol.~\bibinfo{volume}{7830}. \bibinfo{publisher}{Springer
  Berlin Heidelberg}, \bibinfo{pages}{221--236}.
\newblock
\showISBNx{978-3-642-37421-0}
\urldef\tempurl%
\url{https://doi.org/10.1007/978-3-642-37422-7_16}
\showDOI{\tempurl}


\bibitem[\protect\citeauthoryear{Bencomo, Belaggoun, and Issarny}{Bencomo
  et~al\mbox{.}}{2013}]%
        {Bencomo:2013:DDN:2663546.2663565}
\bibfield{author}{\bibinfo{person}{Nelly Bencomo}, \bibinfo{person}{Amel
  Belaggoun}, {and} \bibinfo{person}{Valerie Issarny}.}
  \bibinfo{year}{2013}\natexlab{}.
\newblock \showarticletitle{Dynamic Decision Networks for Decision-making in
  Self-adaptive Systems: A Case Study}. In
  \bibinfo{booktitle}{\emph{Proceedings of the 8th International Symposium on
  Software Engineering for Adaptive and Self-Managing Systems}}
  \emph{(\bibinfo{series}{SEAMS '13})}. \bibinfo{publisher}{IEEE Press},
  \bibinfo{address}{Piscataway, NJ, USA}, \bibinfo{pages}{113--122}.
\newblock
\showISBNx{978-1-4673-4401-2}
\urldef\tempurl%
\url{http://dl.acm.org/citation.cfm?id=2663546.2663565}
\showURL{%
\tempurl}


\bibitem[\protect\citeauthoryear{Berger, Nguyen, and Benderius}{Berger
  et~al\mbox{.}}{2017}]%
        {7958428}
\bibfield{author}{\bibinfo{person}{C. Berger}, \bibinfo{person}{B. Nguyen},
  {and} \bibinfo{person}{O. Benderius}.} \bibinfo{year}{2017}\natexlab{}.
\newblock \showarticletitle{Containerized Development and Microservices for
  Self-Driving Vehicles: Experiences Best Practices}. In
  \bibinfo{booktitle}{\emph{2017 IEEE International Conference on Software
  Architecture Workshops (ICSAW)}}. \bibinfo{pages}{7--12}.
\newblock
\urldef\tempurl%
\url{https://doi.org/10.1109/ICSAW.2017.56}
\showDOI{\tempurl}


\bibitem[\protect\citeauthoryear{Betts}{Betts}{2017}]%
        {qna}
\bibfield{author}{\bibinfo{person}{Thomas Betts}.}
  \bibinfo{year}{2017}\natexlab{}.
\newblock \bibinfo{title}{Q\&A with Susan Fowler on Production-Ready
  Microservices}.
\newblock
  \bibinfo{howpublished}{https://www.infoq.com/news/2017/01/production-ready-microservices}.
\newblock


\bibitem[\protect\citeauthoryear{Bianco, Kotermanski, and Merson}{Bianco
  et~al\mbox{.}}{2007}]%
        {evalsoa}
\bibfield{author}{\bibinfo{person}{Philip Bianco}, \bibinfo{person}{Rick
  Kotermanski}, {and} \bibinfo{person}{Paulo~F. Merson}.}
  \bibinfo{year}{2007}\natexlab{}.
\newblock \bibinfo{booktitle}{\emph{Evaluating a Service-Oriented
  Architecture}}.
\newblock \bibinfo{type}{{T}echnical {R}eport} CMU/SEI-2007-TR-015.
  \bibinfo{institution}{Carnegie Mellon University}.
\newblock


\bibitem[\protect\citeauthoryear{Blair, Bencomo, and France}{Blair
  et~al\mbox{.}}{2009}]%
        {Blair:2009:MR:1638585.1638645}
\bibfield{author}{\bibinfo{person}{Gordon Blair}, \bibinfo{person}{Nelly
  Bencomo}, {and} \bibinfo{person}{Robert~B. France}.}
  \bibinfo{year}{2009}\natexlab{}.
\newblock \showarticletitle{Models@ Run.Time}.
\newblock \bibinfo{journal}{\emph{Computer}} \bibinfo{volume}{42},
  \bibinfo{number}{10} (\bibinfo{date}{Oct.} \bibinfo{year}{2009}),
  \bibinfo{pages}{22--27}.
\newblock
\showISSN{0018-9162}
\urldef\tempurl%
\url{https://doi.org/10.1109/MC.2009.326}
\showDOI{\tempurl}


\bibitem[\protect\citeauthoryear{Briand, Morasca, and Basili}{Briand
  et~al\mbox{.}}{1993}]%
        {Briand:1993:MAM:645542.658018}
\bibfield{author}{\bibinfo{person}{Lionel~C. Briand}, \bibinfo{person}{Sandro
  Morasca}, {and} \bibinfo{person}{Victor~R. Basili}.}
  \bibinfo{year}{1993}\natexlab{}.
\newblock \showarticletitle{Measuring and Assessing Maintainability at the End
  of High Level Design}. In \bibinfo{booktitle}{\emph{Proceedings of the
  Conference on Software Maintenance}} \emph{(\bibinfo{series}{ICSM '93})}.
  \bibinfo{publisher}{IEEE Computer Society}, \bibinfo{address}{Washington, DC,
  USA}, \bibinfo{pages}{88--97}.
\newblock
\showISBNx{0-8186-4600-4}
\urldef\tempurl%
\url{http://dl.acm.org/citation.cfm?id=645542.658018}
\showURL{%
\tempurl}


\bibitem[\protect\citeauthoryear{Brilhante, Costa, and Maritan}{Brilhante
  et~al\mbox{.}}{2017}]%
        {Brilhante:2017:AQB:3126858.3126873}
\bibfield{author}{\bibinfo{person}{Jonathan Brilhante},
  \bibinfo{person}{Rostand Costa}, {and} \bibinfo{person}{Tiago Maritan}.}
  \bibinfo{year}{2017}\natexlab{}.
\newblock \showarticletitle{Asynchronous Queue Based Approach for Building
  Reactive Microservices}. In \bibinfo{booktitle}{\emph{Proceedings of the 23rd
  Brazillian Symposium on Multimedia and the Web}}
  \emph{(\bibinfo{series}{WebMedia '17})}. \bibinfo{publisher}{ACM},
  \bibinfo{address}{New York, NY, USA}, \bibinfo{pages}{373--380}.
\newblock
\showISBNx{978-1-4503-5096-9}
\urldef\tempurl%
\url{https://doi.org/10.1145/3126858.3126873}
\showDOI{\tempurl}


\bibitem[\protect\citeauthoryear{Brogi, Canciani, Neri, Rinaldi, and
  Soldani}{Brogi et~al\mbox{.}}{2018}]%
        {10.1007/978-3-319-74781-1_16}
\bibfield{author}{\bibinfo{person}{Antonio Brogi}, \bibinfo{person}{Andrea
  Canciani}, \bibinfo{person}{Davide Neri}, \bibinfo{person}{Luca Rinaldi},
  {and} \bibinfo{person}{Jacopo Soldani}.} \bibinfo{year}{2018}\natexlab{}.
\newblock \showarticletitle{Towards a Reference Dataset of Microservice-Based
  Applications}. In \bibinfo{booktitle}{\emph{Software Engineering and Formal
  Methods}}, \bibfield{editor}{\bibinfo{person}{Antonio Cerone} {and}
  \bibinfo{person}{Marco Roveri}} (Eds.). \bibinfo{publisher}{Springer
  International Publishing}, \bibinfo{address}{Cham},
  \bibinfo{pages}{219--229}.
\newblock


\bibitem[\protect\citeauthoryear{Bryant}{Bryant}{[n. d.]}]%
        {econmic}
\bibfield{author}{\bibinfo{person}{Daniel Bryant}.} \bibinfo{year}{[n.
  d.]}\natexlab{}.
\newblock \bibinfo{title}{The Economics of Microservices: Phil Cal\c{c}ado
  Recommends Avoiding ``Microliths}.
\newblock
\newblock


\bibitem[\protect\citeauthoryear{Bryant}{Bryant}{2016}]%
        {deadly}
\bibfield{author}{\bibinfo{person}{Daniel Bryant}.}
  \bibinfo{year}{2016}\natexlab{}.
\newblock \bibinfo{title}{The Seven (More) Deadly Sins of Microservices}.
\newblock
  \bibinfo{howpublished}{https://www.infoq.com/presentations/microservices-antipatterns-2016}.
\newblock


\bibitem[\protect\citeauthoryear{Bryant}{Bryant}{2017a}]%
        {peopleimpact}
\bibfield{author}{\bibinfo{person}{Daniel Bryant}.}
  \bibinfo{year}{2017}\natexlab{a}.
\newblock \bibinfo{title}{Microservices: The Organisational and People Impact}.
\newblock \bibinfo{howpublished}{https://gotocph.com/2017/sessions/295}.
\newblock


\bibitem[\protect\citeauthoryear{Bryant}{Bryant}{2017b}]%
        {observe}
\bibfield{author}{\bibinfo{person}{Daniel Bryant}.}
  \bibinfo{year}{2017}\natexlab{b}.
\newblock \bibinfo{title}{Observability and Avoiding Alert Overload from
  Microservices at the Financial Times}.
\newblock
  \bibinfo{howpublished}{https://www.infoq.com/articles/observability-financial-times}.
\newblock


\bibitem[\protect\citeauthoryear{Bryant}{Bryant}{2017c}]%
        {shrinking}
\bibfield{author}{\bibinfo{person}{Daniel Bryant}.}
  \bibinfo{year}{2017}\natexlab{c}.
\newblock \bibinfo{title}{Shrinking Microservices to Functions: Adrian
  Cockcroft Discusses Serverless at microXchg}.
\newblock
  \bibinfo{howpublished}{https://www.infoq.com/news/2017/02/microxchg-microservice-functions}.
\newblock


\bibitem[\protect\citeauthoryear{Burns}{Burns}{2001}]%
        {cont1}
\bibfield{author}{\bibinfo{person}{R. Burns}.} \bibinfo{year}{2001}\natexlab{}.
\newblock \bibinfo{booktitle}{\emph{Advanced Control Engineering}}.
\newblock \bibinfo{publisher}{Butterworth-Heinemann}.
\newblock
\showISBNx{9780080498782}
\urldef\tempurl%
\url{https://books.google.co.uk/books?id=DovwVQu6ImsC}
\showURL{%
\tempurl}


\bibitem[\protect\citeauthoryear{Calcado}{Calcado}{[n. d.]}]%
        {freelunch}
\bibfield{author}{\bibinfo{person}{Phil Calcado}.} \bibinfo{year}{[n.
  d.]}\natexlab{}.
\newblock \bibinfo{title}{No Free Lunch, Indeed: Three Years of Micro-services
  at SoundCloud}.
\newblock
  \bibinfo{howpublished}{http://www.infoq.com/presentations/soundcloud-microservices}.
\newblock


\bibitem[\protect\citeauthoryear{Calinescu, Ghezzi, Kwiatkowska, and
  Mirandola}{Calinescu et~al\mbox{.}}{2012}]%
        {assur3}
\bibfield{author}{\bibinfo{person}{Radu Calinescu}, \bibinfo{person}{Carlo
  Ghezzi}, \bibinfo{person}{Marta Kwiatkowska}, {and} \bibinfo{person}{Raffaela
  Mirandola}.} \bibinfo{year}{2012}\natexlab{}.
\newblock \showarticletitle{Self-adaptive Software Needs Quantitative
  Verification at Runtime}.
\newblock \bibinfo{journal}{\emph{Commun. ACM}} \bibinfo{volume}{55},
  \bibinfo{number}{9} (\bibinfo{date}{Sept.} \bibinfo{year}{2012}),
  \bibinfo{pages}{69--77}.
\newblock
\showISSN{0001-0782}
\urldef\tempurl%
\url{https://doi.org/10.1145/2330667.2330686}
\showDOI{\tempurl}


\bibitem[\protect\citeauthoryear{Calinescu, Rafiq, Johnson, and
  Bakir}{Calinescu et~al\mbox{.}}{2014}]%
        {assur2}
\bibfield{author}{\bibinfo{person}{Radu Calinescu}, \bibinfo{person}{Yasmin
  Rafiq}, \bibinfo{person}{Kenneth Johnson}, {and} \bibinfo{person}{Mehmet~Emin
  Bakir}.} \bibinfo{year}{2014}\natexlab{}.
\newblock \showarticletitle{Adaptive Model Learning for Continual Verification
  of Non-functional Properties}. In \bibinfo{booktitle}{\emph{Proceedings of
  the 5th ACM/SPEC International Conference on Performance Engineering}}
  \emph{(\bibinfo{series}{ICPE '14})}. \bibinfo{publisher}{ACM},
  \bibinfo{address}{New York, NY, USA}, \bibinfo{pages}{87--98}.
\newblock
\showISBNx{978-1-4503-2733-6}
\urldef\tempurl%
\url{https://doi.org/10.1145/2568088.2568094}
\showDOI{\tempurl}


\bibitem[\protect\citeauthoryear{Cardoso, Sheth, Miller, Arnold, and
  Kochut}{Cardoso et~al\mbox{.}}{2004}]%
        {workflow}
\bibfield{author}{\bibinfo{person}{Jorge Cardoso}, \bibinfo{person}{Amit
  Sheth}, \bibinfo{person}{John Miller}, \bibinfo{person}{Jonathan Arnold},
  {and} \bibinfo{person}{Krys Kochut}.} \bibinfo{year}{2004}\natexlab{}.
\newblock \showarticletitle{Quality of service for workflows and web service
  processes}.
\newblock \bibinfo{journal}{\emph{Journal of Web Semantics}}
  \bibinfo{volume}{1}, \bibinfo{number}{3} (\bibinfo{year}{2004}),
  \bibinfo{pages}{281 -- 308}.
\newblock
\showISSN{1570-8268}
\urldef\tempurl%
\url{https://doi.org/10.1016/j.websem.2004.03.001}
\showDOI{\tempurl}


\bibitem[\protect\citeauthoryear{Carlson}{Carlson}{2017}]%
        {whataremicro}
\bibfield{author}{\bibinfo{person}{Lucas Carlson}.}
  \bibinfo{year}{2017}\natexlab{}.
\newblock \bibinfo{title}{What are microservices? Lightweight software
  development explained}.
\newblock
  \bibinfo{howpublished}{https://www.infoworld.com/article/3237697/application-development/what-are-microservices-lightweight-software-development-explained.html}.
\newblock


\bibitem[\protect\citeauthoryear{Casati, Ilnicki, Jin, Krishnamoorthy, and
  Shan}{Casati et~al\mbox{.}}{2000}]%
        {Casati:2000:ADS:646088.679914}
\bibfield{author}{\bibinfo{person}{Fabio Casati}, \bibinfo{person}{Ski
  Ilnicki}, \bibinfo{person}{Li-jie Jin}, \bibinfo{person}{Vasudev
  Krishnamoorthy}, {and} \bibinfo{person}{Ming-Chien Shan}.}
  \bibinfo{year}{2000}\natexlab{}.
\newblock \showarticletitle{Adaptive and Dynamic Service Composition in eFlow}.
  \bibinfo{howpublished}{http://dl.acm.org/citation.cfm?id=646088.679914}. In
  \bibinfo{booktitle}{\emph{Proceedings of the 12th International Conference on
  Advanced Information Systems Engineering}} \emph{(\bibinfo{series}{CAiSE
  '00})}. \bibinfo{publisher}{Springer-Verlag}, \bibinfo{address}{London, UK,
  UK}, \bibinfo{pages}{13--31}.
\newblock
\showISBNx{3-540-67630-9}


\bibitem[\protect\citeauthoryear{Cerny}{Cerny}{2018}]%
        {doi:10.1080/17517575.2018.1462406}
\bibfield{author}{\bibinfo{person}{Tomas Cerny}.}
  \bibinfo{year}{2018}\natexlab{}.
\newblock \showarticletitle{Aspect-oriented challenges in system integration
  with microservices, SOA and IoT}.
\newblock \bibinfo{journal}{\emph{Enterprise Information Systems}}
  \bibinfo{volume}{0}, \bibinfo{number}{0} (\bibinfo{year}{2018}),
  \bibinfo{pages}{1--23}.
\newblock
\urldef\tempurl%
\url{https://doi.org/10.1080/17517575.2018.1462406}
\showDOI{\tempurl}


\bibitem[\protect\citeauthoryear{Cerny, Donahoo, and Pechanec}{Cerny
  et~al\mbox{.}}{2017}]%
        {Cerny:2017:DCS:3129676.3129682}
\bibfield{author}{\bibinfo{person}{Tomas Cerny}, \bibinfo{person}{Michael~J.
  Donahoo}, {and} \bibinfo{person}{Jiri Pechanec}.}
  \bibinfo{year}{2017}\natexlab{}.
\newblock \showarticletitle{Disambiguation and Comparison of SOA, Microservices
  and Self-Contained Systems}. In \bibinfo{booktitle}{\emph{Proceedings of the
  International Conference on Research in Adaptive and Convergent Systems}}
  \emph{(\bibinfo{series}{RACS '17})}. \bibinfo{publisher}{ACM},
  \bibinfo{address}{New York, NY, USA}, \bibinfo{pages}{228--235}.
\newblock
\showISBNx{978-1-4503-5027-3}
\urldef\tempurl%
\url{https://doi.org/10.1145/3129676.3129682}
\showDOI{\tempurl}


\bibitem[\protect\citeauthoryear{Cerny, Donahoo, and Trnka}{Cerny
  et~al\mbox{.}}{2018}]%
        {Cerny:2018:CUM:3183628.3183631}
\bibfield{author}{\bibinfo{person}{Tomas Cerny}, \bibinfo{person}{Michael~J.
  Donahoo}, {and} \bibinfo{person}{Michal Trnka}.}
  \bibinfo{year}{2018}\natexlab{}.
\newblock \showarticletitle{Contextual Understanding of Microservice
  Architecture: Current and Future Directions}.
\newblock \bibinfo{journal}{\emph{SIGAPP Appl. Comput. Rev.}}
  \bibinfo{volume}{17}, \bibinfo{number}{4} (\bibinfo{date}{Jan.}
  \bibinfo{year}{2018}), \bibinfo{pages}{29--45}.
\newblock
\showISSN{1559-6915}
\urldef\tempurl%
\url{https://doi.org/10.1145/3183628.3183631}
\showDOI{\tempurl}


\bibitem[\protect\citeauthoryear{Channabasavaiah, Holley, and
  Tuggle}{Channabasavaiah et~al\mbox{.}}{2003}]%
        {channabasavaiah2003migrating}
\bibfield{author}{\bibinfo{person}{Kishore Channabasavaiah},
  \bibinfo{person}{Kerrie Holley}, {and} \bibinfo{person}{Edward Tuggle}.}
  \bibinfo{year}{2003}\natexlab{}.
\newblock \showarticletitle{Migrating to a service-oriented architecture}.
\newblock \bibinfo{journal}{\emph{IBM DeveloperWorks}}  \bibinfo{volume}{16}
  (\bibinfo{year}{2003}), \bibinfo{pages}{727--728}.
\newblock


\bibitem[\protect\citeauthoryear{Chen}{Chen}{2018}]%
        {contdevops}
\bibfield{author}{\bibinfo{person}{Lianping Chen}.}
  \bibinfo{year}{2018}\natexlab{}.
\newblock \showarticletitle{Microservices: Architecting for Continuous Delivery
  and DevOps}. In \bibinfo{booktitle}{\emph{IEEE International Conference on
  Software Architecture (ICSA 2018)}}.
\newblock


\bibitem[\protect\citeauthoryear{Cheng, de~Lemos, Giese, Inverardi, Magee,
  Andersson, Becker, Bencomo, Brun, Cukic, Di~Marzo~Serugendo, Dustdar,
  Finkelstein, Gacek, Geihs, Grassi, Karsai, Kienle, Kramer, Litoiu, Malek,
  Mirandola, Müller, Park, Shaw, Tichy, Tivoli, Weyns, and Whittle}{Cheng
  et~al\mbox{.}}{2009}]%
        {roadmap}
\bibfield{author}{\bibinfo{person}{BettyH.C. Cheng}, \bibinfo{person}{Rogerio
  de Lemos}, \bibinfo{person}{Holger Giese}, \bibinfo{person}{Paola Inverardi},
  \bibinfo{person}{Jeff Magee}, \bibinfo{person}{Jesper Andersson},
  \bibinfo{person}{Basil Becker}, \bibinfo{person}{Nelly Bencomo},
  \bibinfo{person}{Yuriy Brun}, \bibinfo{person}{Bojan Cukic},
  \bibinfo{person}{Giovanna Di~Marzo~Serugendo}, \bibinfo{person}{Schahram
  Dustdar}, \bibinfo{person}{Anthony Finkelstein}, \bibinfo{person}{Cristina
  Gacek}, \bibinfo{person}{Kurt Geihs}, \bibinfo{person}{Vincenzo Grassi},
  \bibinfo{person}{Gabor Karsai}, \bibinfo{person}{HolgerM. Kienle},
  \bibinfo{person}{Jeff Kramer}, \bibinfo{person}{Marin Litoiu},
  \bibinfo{person}{Sam Malek}, \bibinfo{person}{Raffaela Mirandola},
  \bibinfo{person}{HausiA. Müller}, \bibinfo{person}{Sooyong Park},
  \bibinfo{person}{Mary Shaw}, \bibinfo{person}{Matthias Tichy},
  \bibinfo{person}{Massimo Tivoli}, \bibinfo{person}{Danny Weyns}, {and}
  \bibinfo{person}{Jon Whittle}.} \bibinfo{year}{2009}\natexlab{}.
\newblock \showarticletitle{Software Engineering for Self-Adaptive Systems: A
  Research Roadmap}.
\newblock In \bibinfo{booktitle}{\emph{Software Engineering for Self-Adaptive
  Systems}}, \bibfield{editor}{\bibinfo{person}{Betty~H.C. Cheng},
  \bibinfo{person}{Rogerio de~Lemos}, \bibinfo{person}{Holger Giese},
  \bibinfo{person}{Paola Inverardi}, {and} \bibinfo{person}{Jeff Magee}}
  (Eds.). \bibinfo{series}{Lecture Notes in Computer Science},
  Vol.~\bibinfo{volume}{5525}. \bibinfo{publisher}{Springer Berlin Heidelberg},
  \bibinfo{pages}{1--26}.
\newblock
\showISBNx{978-3-642-02160-2}
\urldef\tempurl%
\url{https://doi.org/10.1007/978-3-642-02161-9_1}
\showDOI{\tempurl}


\bibitem[\protect\citeauthoryear{Cheng, Garlan, and Schmerl}{Cheng
  et~al\mbox{.}}{2005}]%
        {explicit}
\bibfield{author}{\bibinfo{person}{Shang-Wen Cheng}, \bibinfo{person}{David
  Garlan}, {and} \bibinfo{person}{Bradley Schmerl}.}
  \bibinfo{year}{2005}\natexlab{}.
\newblock \showarticletitle{Self-star Properties in Complex Information
  Systems}.
\newblock In \bibinfo{booktitle}{\emph{Making Self-adaptation an Engineering
  Reality}}, \bibfield{editor}{\bibinfo{person}{Ozalp Babaoglu},
  \bibinfo{person}{M\'{a}rk Jelasity}, \bibinfo{person}{Alberto Montresor},
  \bibinfo{person}{Christof Fetzer}, {and} \bibinfo{person}{Stefano Leonardi}}
  (Eds.). \bibinfo{publisher}{Springer-Verlag}, \bibinfo{address}{Berlin,
  Heidelberg}, \bibinfo{pages}{158--173}.
\newblock
\showISBNx{3-540-26009-9, 978-3-540-26009-7}
\urldef\tempurl%
\url{http://dl.acm.org/citation.cfm?id=2167575.2167589}
\showURL{%
\tempurl}


\bibitem[\protect\citeauthoryear{Chris~Scaffidi}{Chris~Scaffidi}{2005}]%
        {shaw}
\bibfield{author}{\bibinfo{person}{Shawn Butler Mary~Shaw Chris~Scaffidi,
  Ashish~Arora}.} \bibinfo{year}{2005}\natexlab{}.
\newblock \showarticletitle{A Value-Based Approach to Predicting System
  Properties from Design}. In \bibinfo{booktitle}{\emph{5th EDSER}}.
\newblock


\bibitem[\protect\citeauthoryear{Ciuffoletti}{Ciuffoletti}{2015}]%
        {CIUFFOLETTI2015163}
\bibfield{author}{\bibinfo{person}{Augusto Ciuffoletti}.}
  \bibinfo{year}{2015}\natexlab{}.
\newblock \showarticletitle{Automated Deployment of a Microservice-based
  Monitoring Infrastructure}.
\newblock \bibinfo{journal}{\emph{Procedia Computer Science}}
  \bibinfo{volume}{68} (\bibinfo{year}{2015}), \bibinfo{pages}{163 -- 172}.
\newblock
\showISSN{1877-0509}
\urldef\tempurl%
\url{https://doi.org/10.1016/j.procs.2015.09.232}
\showDOI{\tempurl}
\newblock
\shownote{1st International Conference on Cloud Forward: From Distributed to
  Complete Computing.}


\bibitem[\protect\citeauthoryear{Cámara, Garlan, Kang, Peng, and
  Schmerl}{Cámara et~al\mbox{.}}{2017}]%
        {mapek}
\bibfield{author}{\bibinfo{person}{Javier Cámara}, \bibinfo{person}{David
  Garlan}, \bibinfo{person}{Won~Gu Kang}, \bibinfo{person}{Wenxin Peng}, {and}
  \bibinfo{person}{Bradley Schmerl}.} \bibinfo{year}{2017}\natexlab{}.
\newblock \bibinfo{booktitle}{\emph{Uncertainty in Self-Adaptive
  SystemsCategories, Management, and Perspectives}}.
\newblock \bibinfo{type}{{T}echnical {R}eport} CMU-ISR-17-110.
  \bibinfo{institution}{Carnegie Mellon University}.
\newblock


\bibitem[\protect\citeauthoryear{Cockroft}{Cockroft}{2015a}]%
        {gluecon}
\bibfield{author}{\bibinfo{person}{Adrian Cockroft}.}
  \bibinfo{year}{2015}\natexlab{a}.
\newblock \bibinfo{title}{Gluecon Monitoring Microservices and Containers: A
  Challenge}.
\newblock
  \bibinfo{howpublished}{http://www.slideshare.net/adriancockcroft/gluecon-monitoring-microservices-and-containers-a-challenge}.
\newblock


\bibitem[\protect\citeauthoryear{Cockroft}{Cockroft}{2015b}]%
        {adrianstate}
\bibfield{author}{\bibinfo{person}{Adrian Cockroft}.}
  \bibinfo{year}{2015}\natexlab{b}.
\newblock \bibinfo{title}{State of the Art in Microservices}.
\newblock
  \bibinfo{howpublished}{https://www.infoq.com/presentations/microservices-comparison-evolution}.
\newblock


\bibitem[\protect\citeauthoryear{Conway}{Conway}{1968}]%
        {conway1968committees}
\bibfield{author}{\bibinfo{person}{Melvin~E Conway}.}
  \bibinfo{year}{1968}\natexlab{}.
\newblock \showarticletitle{How do committees invent}.
\newblock  (\bibinfo{year}{1968}).
\newblock


\bibitem[\protect\citeauthoryear{Curlett}{Curlett}{2016}]%
        {tame}
\bibfield{author}{\bibinfo{person}{Conor Curlett}.}
  \bibinfo{year}{2016}\natexlab{}.
\newblock \bibinfo{title}{Tame microservices complexity with APIs}.
\newblock
  \bibinfo{howpublished}{https://www.infoworld.com/article/3111349/application-development/tame-microservices-complexity-with-apis.html}.
\newblock


\bibitem[\protect\citeauthoryear{Danilovic and Browning}{Danilovic and
  Browning}{2007}]%
        {dsm}
\bibfield{author}{\bibinfo{person}{Mike Danilovic} {and}
  \bibinfo{person}{Tyson~R. Browning}.} \bibinfo{year}{2007}\natexlab{}.
\newblock \showarticletitle{Managing complex product development projects with
  design structure matrices and domain mapping matrices}.
\newblock \bibinfo{journal}{\emph{International Journal of Project Management}}
  \bibinfo{volume}{25}, \bibinfo{number}{3} (\bibinfo{year}{2007}),
  \bibinfo{pages}{300 -- 314}.
\newblock
\showISSN{0263-7863}
\urldef\tempurl%
\url{https://doi.org/10.1016/j.ijproman.2006.11.003}
\showDOI{\tempurl}


\bibitem[\protect\citeauthoryear{Davies}{Davies}{2004}]%
        {Davies01102004}
\bibfield{author}{\bibinfo{person}{Andrew Davies}.}
  \bibinfo{year}{2004}\natexlab{}.
\newblock \showarticletitle{Moving base into high-value integrated solutions: a
  value stream approach}.
\newblock \bibinfo{journal}{\emph{Industrial and Corporate Change}}
  \bibinfo{volume}{13}, \bibinfo{number}{5} (\bibinfo{year}{2004}),
  \bibinfo{pages}{727--756}.
\newblock
\urldef\tempurl%
\url{https://doi.org/10.1093/icc/dth029}
\showDOI{\tempurl}
\showeprint{http://icc.oxfordjournals.org/content/13/5/727.full.pdf+html}


\bibitem[\protect\citeauthoryear{Daya, Van~Duy, Eati, Ferreira, Glozic, Gucer,
  Gupta, Joshi, Lampkin, Martins, et~al\mbox{.}}{Daya et~al\mbox{.}}{2015}]%
        {bluemix}
\bibfield{author}{\bibinfo{person}{S. Daya}, \bibinfo{person}{N. Van~Duy},
  \bibinfo{person}{K. Eati}, \bibinfo{person}{C.M. Ferreira},
  \bibinfo{person}{D. Glozic}, \bibinfo{person}{V. Gucer}, \bibinfo{person}{M.
  Gupta}, \bibinfo{person}{S. Joshi}, \bibinfo{person}{V. Lampkin},
  \bibinfo{person}{M. Martins}, {et~al\mbox{.}}}
  \bibinfo{year}{2015}\natexlab{}.
\newblock \bibinfo{booktitle}{\emph{Microservices from Theory to Practice:
  Creating Applications in IBM Bluemix Using the Microservices Approach}}.
\newblock \bibinfo{publisher}{IBM Redbooks}.
\newblock
\showISBNx{9780738440811}
\urldef\tempurl%
\url{https://books.google.co.uk/books?id=eOZyCgAAQBAJ}
\showURL{%
\tempurl}


\bibitem[\protect\citeauthoryear{de~Lemos, Giese, Müller, Shaw, Andersson,
  Litoiu, Schmerl, Tamura, Villegas, Vogel, Weyns, Baresi, Becker, Bencomo,
  Brun, Cukic, Desmarais, Dustdar, Engels, Geihs, Göschka, Gorla, Grassi,
  Inverardi, Karsai, Kramer, Lopes, Magee, Malek, Mankovskii, Mirandola,
  Mylopoulos, Nierstrasz, Pezzè, Prehofer, Schäfer, Schlichting, Smith,
  Sousa, Tahvildari, Wong, and Wuttke}{de~Lemos et~al\mbox{.}}{2013}]%
        {secroadmap}
\bibfield{author}{\bibinfo{person}{Rogerio de Lemos}, \bibinfo{person}{Holger
  Giese}, \bibinfo{person}{HausiA. Müller}, \bibinfo{person}{Mary Shaw},
  \bibinfo{person}{Jesper Andersson}, \bibinfo{person}{Marin Litoiu},
  \bibinfo{person}{Bradley Schmerl}, \bibinfo{person}{Gabriel Tamura},
  \bibinfo{person}{NorhaM. Villegas}, \bibinfo{person}{Thomas Vogel},
  \bibinfo{person}{Danny Weyns}, \bibinfo{person}{Luciano Baresi},
  \bibinfo{person}{Basil Becker}, \bibinfo{person}{Nelly Bencomo},
  \bibinfo{person}{Yuriy Brun}, \bibinfo{person}{Bojan Cukic},
  \bibinfo{person}{Ron Desmarais}, \bibinfo{person}{Schahram Dustdar},
  \bibinfo{person}{Gregor Engels}, \bibinfo{person}{Kurt Geihs},
  \bibinfo{person}{KarlM. Göschka}, \bibinfo{person}{Alessandra Gorla},
  \bibinfo{person}{Vincenzo Grassi}, \bibinfo{person}{Paola Inverardi},
  \bibinfo{person}{Gabor Karsai}, \bibinfo{person}{Jeff Kramer},
  \bibinfo{person}{Antónia Lopes}, \bibinfo{person}{Jeff Magee},
  \bibinfo{person}{Sam Malek}, \bibinfo{person}{Serge Mankovskii},
  \bibinfo{person}{Raffaela Mirandola}, \bibinfo{person}{John Mylopoulos},
  \bibinfo{person}{Oscar Nierstrasz}, \bibinfo{person}{Mauro Pezzè},
  \bibinfo{person}{Christian Prehofer}, \bibinfo{person}{Wilhelm Schäfer},
  \bibinfo{person}{Rick Schlichting}, \bibinfo{person}{DennisB. Smith},
  \bibinfo{person}{JoãoPedro Sousa}, \bibinfo{person}{Ladan Tahvildari},
  \bibinfo{person}{Kenny Wong}, {and} \bibinfo{person}{Jochen Wuttke}.}
  \bibinfo{year}{2013}\natexlab{}.
\newblock \showarticletitle{Software Engineering for Self-Adaptive Systems: A
  Second Research Roadmap}.
\newblock In \bibinfo{booktitle}{\emph{Software Engineering for Self-Adaptive
  Systems II}}, \bibfield{editor}{\bibinfo{person}{Rogerio de~Lemos},
  \bibinfo{person}{Holger Giese}, \bibinfo{person}{HausiA. Muller}, {and}
  \bibinfo{person}{Mary Shaw}} (Eds.). \bibinfo{series}{Lecture Notes in
  Computer Science}, Vol.~\bibinfo{volume}{7475}. \bibinfo{publisher}{Springer
  Berlin Heidelberg}, \bibinfo{pages}{1--32}.
\newblock
\showISBNx{978-3-642-35812-8}
\urldef\tempurl%
\url{https://doi.org/10.1007/978-3-642-35813-5_1}
\showDOI{\tempurl}


\bibitem[\protect\citeauthoryear{Dehghani}{Dehghani}{2015}]%
        {front}
\bibfield{author}{\bibinfo{person}{Zhamak Dehghani}.}
  \bibinfo{year}{2015}\natexlab{}.
\newblock \bibinfo{title}{Zhamak Dehghani Real World Microservices: Lessons
  from the Frontline}.
\newblock \bibinfo{howpublished}{https://youtu.be/hsoovFbpAoE}.
\newblock


\bibitem[\protect\citeauthoryear{Derakhshanmanesh and Grieger}{Derakhshanmanesh
  and Grieger}{[n. d.]}]%
        {derakhshanmanesh2016model}
\bibfield{author}{\bibinfo{person}{Mahdi Derakhshanmanesh} {and}
  \bibinfo{person}{Marvin Grieger}.} \bibinfo{year}{[n. d.]}\natexlab{}.
\newblock \showarticletitle{Model-Integrating Microservices: A Vision Paper.}.
  In \bibinfo{booktitle}{\emph{Software Engineering Workshops}}.
\newblock


\bibitem[\protect\citeauthoryear{Dietrich, McCartin, Tempero, and
  Shah}{Dietrich et~al\mbox{.}}{2010}]%
        {barrier}
\bibfield{author}{\bibinfo{person}{Jens Dietrich}, \bibinfo{person}{Catherine
  McCartin}, \bibinfo{person}{Ewan Tempero}, {and} \bibinfo{person}{SyedM.Ali
  Shah}.} \bibinfo{year}{2010}\natexlab{}.
\newblock \showarticletitle{Barriers to Modularity - An Empirical Study to
  Assess the Potential for Modularisation of Java Programs}.
\newblock In \bibinfo{booktitle}{\emph{Research into Practice – Reality and
  Gaps}}, \bibfield{editor}{\bibinfo{person}{GeorgeT. Heineman},
  \bibinfo{person}{Jan Kofron}, {and} \bibinfo{person}{Frantisek Plasil}}
  (Eds.). \bibinfo{series}{Lecture Notes in Computer Science},
  Vol.~\bibinfo{volume}{6093}. \bibinfo{publisher}{Springer Berlin Heidelberg},
  \bibinfo{pages}{135--150}.
\newblock
\showISBNx{978-3-642-13820-1}
\urldef\tempurl%
\url{https://doi.org/10.1007/978-3-642-13821-8_11}
\showDOI{\tempurl}


\bibitem[\protect\citeauthoryear{Dobrica and Niemela}{Dobrica and
  Niemela}{2002}]%
        {1019479}
\bibfield{author}{\bibinfo{person}{L. Dobrica} {and} \bibinfo{person}{E.
  Niemela}.} \bibinfo{year}{2002}\natexlab{}.
\newblock \showarticletitle{A survey on software architecture analysis
  methods}.
\newblock \bibinfo{journal}{\emph{Software Engineering, IEEE Transactions on}}
  \bibinfo{volume}{28}, \bibinfo{number}{7} (\bibinfo{date}{Jul}
  \bibinfo{year}{2002}), \bibinfo{pages}{638--653}.
\newblock
\showISSN{0098-5589}
\urldef\tempurl%
\url{https://doi.org/10.1109/TSE.2002.1019479}
\showDOI{\tempurl}


\bibitem[\protect\citeauthoryear{Dobson, Denazis, Fern\'{a}ndez, Ga\"{\i}ti,
  Gelenbe, Massacci, Nixon, Saffre, Schmidt, and Zambonelli}{Dobson
  et~al\mbox{.}}{2006}]%
        {loop}
\bibfield{author}{\bibinfo{person}{Simon Dobson}, \bibinfo{person}{Spyros
  Denazis}, \bibinfo{person}{Antonio Fern\'{a}ndez}, \bibinfo{person}{Dominique
  Ga\"{\i}ti}, \bibinfo{person}{Erol Gelenbe}, \bibinfo{person}{Fabio
  Massacci}, \bibinfo{person}{Paddy Nixon}, \bibinfo{person}{Fabrice Saffre},
  \bibinfo{person}{Nikita Schmidt}, {and} \bibinfo{person}{Franco Zambonelli}.}
  \bibinfo{year}{2006}\natexlab{}.
\newblock \showarticletitle{A Survey of Autonomic Communications}.
\newblock \bibinfo{journal}{\emph{ACM Trans. Auton. Adapt. Syst.}}
  \bibinfo{volume}{1}, \bibinfo{number}{2} (\bibinfo{date}{Dec.}
  \bibinfo{year}{2006}), \bibinfo{pages}{223--259}.
\newblock
\showISSN{1556-4665}
\urldef\tempurl%
\url{https://doi.org/10.1145/1186778.1186782}
\showDOI{\tempurl}


\bibitem[\protect\citeauthoryear{Dragoni, Dustdar, Larsen, and Mazzara}{Dragoni
  et~al\mbox{.}}{2017a}]%
        {DBLP:journals/corr/DragoniDLM17}
\bibfield{author}{\bibinfo{person}{Nicola Dragoni}, \bibinfo{person}{Schahram
  Dustdar}, \bibinfo{person}{Stephan~Thordal Larsen}, {and}
  \bibinfo{person}{Manuel Mazzara}.} \bibinfo{year}{2017}\natexlab{a}.
\newblock \showarticletitle{Microservices: Migration of a Mission Critical
  System}.
\newblock \bibinfo{journal}{\emph{CoRR}}  \bibinfo{volume}{abs/1704.04173}
  (\bibinfo{year}{2017}).
\newblock
\showeprint[arxiv]{1704.04173}
\urldef\tempurl%
\url{http://arxiv.org/abs/1704.04173}
\showURL{%
\tempurl}


\bibitem[\protect\citeauthoryear{Dragoni, Giallorenzo, Lluch{-}Lafuente,
  Mazzara, Montesi, Mustafin, and Safina}{Dragoni et~al\mbox{.}}{2016}]%
        {DBLP:journals/corr/DragoniGLMMMS16}
\bibfield{author}{\bibinfo{person}{Nicola Dragoni}, \bibinfo{person}{Saverio
  Giallorenzo}, \bibinfo{person}{Alberto Lluch{-}Lafuente},
  \bibinfo{person}{Manuel Mazzara}, \bibinfo{person}{Fabrizio Montesi},
  \bibinfo{person}{Ruslan Mustafin}, {and} \bibinfo{person}{Larisa Safina}.}
  \bibinfo{year}{2016}\natexlab{}.
\newblock \showarticletitle{Microservices: yesterday, today, and tomorrow}.
\newblock \bibinfo{journal}{\emph{CoRR}}  \bibinfo{volume}{abs/1606.04036}
  (\bibinfo{year}{2016}).
\newblock
\showeprint[arxiv]{1606.04036}
\urldef\tempurl%
\url{http://arxiv.org/abs/1606.04036}
\showURL{%
\tempurl}


\bibitem[\protect\citeauthoryear{Dragoni, Lanese, Larsen, Mazzara, Mustafin,
  and Safina}{Dragoni et~al\mbox{.}}{2017b}]%
        {DBLP:journals/corr/DragoniLLMMS17}
\bibfield{author}{\bibinfo{person}{Nicola Dragoni}, \bibinfo{person}{Ivan
  Lanese}, \bibinfo{person}{Stephan~Thordal Larsen}, \bibinfo{person}{Manuel
  Mazzara}, \bibinfo{person}{Ruslan Mustafin}, {and} \bibinfo{person}{Larisa
  Safina}.} \bibinfo{year}{2017}\natexlab{b}.
\newblock \showarticletitle{Microservices: How To Make Your Application Scale}.
\newblock \bibinfo{journal}{\emph{CoRR}}  \bibinfo{volume}{abs/1702.07149}
  (\bibinfo{year}{2017}).
\newblock
\showeprint[arxiv]{1702.07149}
\urldef\tempurl%
\url{http://arxiv.org/abs/1702.07149}
\showURL{%
\tempurl}


\bibitem[\protect\citeauthoryear{ElectricCloud}{ElectricCloud}{[n. d.]}]%
        {deploy}
\bibfield{author}{\bibinfo{person}{ElectricCloud}.} \bibinfo{year}{[n.
  d.]}\natexlab{}.
\newblock \bibinfo{title}{Deployment Automation}.
\newblock
  \bibinfo{howpublished}{http://electric-cloud.com/wiki/display/releasemanagement/Deployment+Automation\#DeploymentAutomation-DeploymentAutomationOverview}.
\newblock


\bibitem[\protect\citeauthoryear{Erl}{Erl}{2005}]%
        {erl2005service}
\bibfield{author}{\bibinfo{person}{T. Erl}.} \bibinfo{year}{2005}\natexlab{}.
\newblock \bibinfo{booktitle}{\emph{Service-Oriented Architecture: Concepts,
  Technology, and Design}}.
\newblock \bibinfo{publisher}{Pearson Education}.
\newblock
\showISBNx{9780132715829}
\urldef\tempurl%
\url{https://books.google.co.uk/books?id=y2MALc9HOF8C}
\showURL{%
\tempurl}


\bibitem[\protect\citeauthoryear{Erl}{Erl}{2007}]%
        {erl2007soa}
\bibfield{author}{\bibinfo{person}{T. Erl}.} \bibinfo{year}{2007}\natexlab{}.
\newblock \bibinfo{booktitle}{\emph{SOA Principles of Service Design}}.
\newblock \bibinfo{publisher}{Pearson Education}.
\newblock
\showISBNx{9780132715836}
\urldef\tempurl%
\url{https://books.google.co.uk/books?id=mkQJvjR2sX0C}
\showURL{%
\tempurl}


\bibitem[\protect\citeauthoryear{Esposte, Kon, Costa, and Lago}{Esposte
  et~al\mbox{.}}{2017}]%
        {Esposte2017InterSCityAS}
\bibfield{author}{\bibinfo{person}{Arthur M.~Del Esposte},
  \bibinfo{person}{Fabio Kon}, \bibinfo{person}{F{\'a}bio~M. Costa}, {and}
  \bibinfo{person}{Nelson Lago}.} \bibinfo{year}{2017}\natexlab{}.
\newblock \showarticletitle{InterSCity: A Scalable Microservice-based Open
  Source Platform for Smart Cities}. In \bibinfo{booktitle}{\emph{The 6th
  International Conference on Smart Cities and Green ICT}}.
\newblock


\bibitem[\protect\citeauthoryear{Evans}{Evans}{[n. d.]}]%
        {domdriv}
\bibfield{author}{\bibinfo{person}{Eric~J. Evans}.} \bibinfo{year}{[n.
  d.]}\natexlab{}.
\newblock \bibinfo{booktitle}{\emph{Domain-Driven Design: Tackling Complexity
  in the Heart of Software}}.
\newblock


\bibitem[\protect\citeauthoryear{Fabrizio~Montesi}{Fabrizio~Montesi}{2017}]%
        {workinprog}
\bibfield{author}{\bibinfo{person}{Dan Sebastian~Thrane Fabrizio~Montesi}.}
  \bibinfo{year}{2017}\natexlab{}.
\newblock \showarticletitle{Packaging Microservices (Work in Progress)}. In
  \bibinfo{booktitle}{\emph{IFIP International Federation for Information
  Processing 2017}}. \bibinfo{publisher}{Springer}, \bibinfo{pages}{131--137}.
\newblock


\bibitem[\protect\citeauthoryear{Fairbanks}{Fairbanks}{2010}]%
        {fairbanks2010just}
\bibfield{author}{\bibinfo{person}{G. Fairbanks}.}
  \bibinfo{year}{2010}\natexlab{}.
\newblock \bibinfo{booktitle}{\emph{Just Enough Software Architecture: A
  Risk-Driven Approach}}.
\newblock \bibinfo{publisher}{Marshall \& Brainerd}.
\newblock
\showISBNx{9780984618101}
\urldef\tempurl%
\url{https://books.google.co.uk/books?id=5UZ-AQAAQBAJ}
\showURL{%
\tempurl}


\bibitem[\protect\citeauthoryear{Farcic}{Farcic}{2018}]%
        {devopstoolkit}
\bibfield{author}{\bibinfo{person}{Victor Farcic}.}
  \bibinfo{year}{2018}\natexlab{}.
\newblock \bibinfo{booktitle}{\emph{The DevOps 2.0 Toolkit Automating the
  Continuous Deployment Pipeline with Containerized Microservices}}.
\newblock \bibinfo{publisher}{Leanpub}.
\newblock


\bibitem[\protect\citeauthoryear{Filieri, Ghezzi, and Tamburrelli}{Filieri
  et~al\mbox{.}}{2011}]%
        {assur1}
\bibfield{author}{\bibinfo{person}{Antonio Filieri}, \bibinfo{person}{Carlo
  Ghezzi}, {and} \bibinfo{person}{Giordano Tamburrelli}.}
  \bibinfo{year}{2011}\natexlab{}.
\newblock \showarticletitle{A formal approach to adaptive software: continuous
  assurance of non-functional requirements}.
\newblock \bibinfo{journal}{\emph{Formal Aspects of Computing}}
  \bibinfo{volume}{24}, \bibinfo{number}{2} (\bibinfo{year}{2011}),
  \bibinfo{pages}{163--186}.
\newblock
\showISSN{1433-299X}
\urldef\tempurl%
\url{https://doi.org/10.1007/s00165-011-0207-2}
\showDOI{\tempurl}


\bibitem[\protect\citeauthoryear{Fisher}{Fisher}{2014}]%
        {enterpriselandscape}
\bibfield{author}{\bibinfo{person}{Tim Fisher}.}
  \bibinfo{year}{2014}\natexlab{}.
\newblock \bibinfo{title}{Digital Transformation is underpinned by the
  Enterprise IT Landscape}.
\newblock
  \bibinfo{howpublished}{https://www.capgemini.com/resources/digital-transformation-is-underpinned-by-the-enterprise-it-landscape/}.
\newblock


\bibitem[\protect\citeauthoryear{Fleurey, Dehlen, Bencomo, Morin, and
  J{\'e}z{\'e}quel}{Fleurey et~al\mbox{.}}{2009}]%
        {Fleurey:2009:MVD:1537875.1537890}
\bibfield{author}{\bibinfo{person}{Franck Fleurey}, \bibinfo{person}{Vegard
  Dehlen}, \bibinfo{person}{Nelly Bencomo}, \bibinfo{person}{Brice Morin},
  {and} \bibinfo{person}{Jean-Marc J{\'e}z{\'e}quel}.}
  \bibinfo{year}{2009}\natexlab{}.
\newblock \showarticletitle{Models in Software Engineering}.
\newblock \bibinfo{publisher}{Springer-Verlag}, \bibinfo{address}{Berlin,
  Heidelberg}, Chapter Modeling and Validating Dynamic Adaptation,
  \bibinfo{pages}{97--108}.
\newblock
\showISBNx{978-3-642-01647-9}
\urldef\tempurl%
\url{https://doi.org/10.1007/978-3-642-01648-6_11}
\showDOI{\tempurl}


\bibitem[\protect\citeauthoryear{Floch, Hallsteinsen, Stav, Eliassen, Lund, and
  Gjorven}{Floch et~al\mbox{.}}{2006}]%
        {Floch:2006:UAM:1128592.1128711}
\bibfield{author}{\bibinfo{person}{Jacqueline Floch}, \bibinfo{person}{Svein
  Hallsteinsen}, \bibinfo{person}{Erlend Stav}, \bibinfo{person}{Frank
  Eliassen}, \bibinfo{person}{Ketil Lund}, {and} \bibinfo{person}{Eli
  Gjorven}.} \bibinfo{year}{2006}\natexlab{}.
\newblock \showarticletitle{Using Architecture Models for Runtime
  Adaptability}.
\newblock \bibinfo{journal}{\emph{IEEE Softw.}} \bibinfo{volume}{23},
  \bibinfo{number}{2} (\bibinfo{date}{March} \bibinfo{year}{2006}),
  \bibinfo{pages}{62--70}.
\newblock
\showISSN{0740-7459}
\urldef\tempurl%
\url{https://doi.org/10.1109/MS.2006.61}
\showDOI{\tempurl}


\bibitem[\protect\citeauthoryear{Florio and Nitto}{Florio and Nitto}{2016}]%
        {7573165}
\bibfield{author}{\bibinfo{person}{L. Florio} {and} \bibinfo{person}{E.~D.
  Nitto}.} \bibinfo{year}{2016}\natexlab{}.
\newblock \showarticletitle{Gru: An Approach to Introduce Decentralized
  Autonomic Behavior in Microservices Architectures}. In
  \bibinfo{booktitle}{\emph{2016 IEEE International Conference on Autonomic
  Computing (ICAC)}}. \bibinfo{pages}{357--362}.
\newblock
\urldef\tempurl%
\url{https://doi.org/10.1109/ICAC.2016.25}
\showDOI{\tempurl}


\bibitem[\protect\citeauthoryear{Flygare and Holmqvist}{Flygare and
  Holmqvist}{2017}]%
        {Flygare1119785}
\bibfield{author}{\bibinfo{person}{Robin Flygare} {and} \bibinfo{person}{Anthon
  Holmqvist}.} \bibinfo{year}{2017}\natexlab{}.
\newblock \emph{\bibinfo{title}{Performance characteristics between monolithic
  and microservice-based systems}}.
\newblock \bibinfo{thesistype}{Master's\ thesis}. \bibinfo{school}{, Department
  of Software Engineering}.
\newblock


\bibitem[\protect\citeauthoryear{Ford}{Ford}{2018}]%
        {nealford}
\bibfield{author}{\bibinfo{person}{Neal Ford}.}
  \bibinfo{year}{2018}\natexlab{}.
\newblock \bibinfo{title}{Building Microservice Architectures}.
\newblock
\newblock
\urldef\tempurl%
\url{http://nealford.com/downloads/Building_Microservice_Architectures_Neal_Ford.pdf}
\showURL{%
\tempurl}


\bibitem[\protect\citeauthoryear{Fowler}{Fowler}{2004a}]%
        {intercept}
\bibfield{author}{\bibinfo{person}{Martin Fowler}.}
  \bibinfo{year}{2004}\natexlab{a}.
\newblock \bibinfo{title}{Event Interception}.
\newblock
  \bibinfo{howpublished}{http://www.martinfowler.com/bliki/EventInterception.html}.
\newblock


\bibitem[\protect\citeauthoryear{Fowler}{Fowler}{2004b}]%
        {strang}
\bibfield{author}{\bibinfo{person}{Martin Fowler}.}
  \bibinfo{year}{2004}\natexlab{b}.
\newblock \bibinfo{title}{Strangler Application}.
\newblock
  \bibinfo{howpublished}{https://www.martinfowler.com/bliki/StranglerApplication.html}.
\newblock


\bibitem[\protect\citeauthoryear{Fowler}{Fowler}{2006}]%
        {eventcol}
\bibfield{author}{\bibinfo{person}{Martin Fowler}.}
  \bibinfo{year}{2006}\natexlab{}.
\newblock \bibinfo{title}{Event Collaboration}.
\newblock
  \bibinfo{howpublished}{https://martinfowler.com/eaaDev/EventCollaboration.html}.
\newblock


\bibitem[\protect\citeauthoryear{Fowler}{Fowler}{2014}]%
        {bounded}
\bibfield{author}{\bibinfo{person}{Martin Fowler}.}
  \bibinfo{year}{2014}\natexlab{}.
\newblock \bibinfo{title}{Bounded Context}.
\newblock
  \bibinfo{howpublished}{https://martinfowler.com/bliki/BoundedContext.html}.
\newblock


\bibitem[\protect\citeauthoryear{Fowler}{Fowler}{2015}]%
        {martin}
\bibfield{author}{\bibinfo{person}{Martin Fowler}.}
  \bibinfo{year}{2015}\natexlab{}.
\newblock \bibinfo{title}{Microservices --- Martin Fowler}.
\newblock \bibinfo{howpublished}{https://www.youtube.com/watch?v=wgdBVIX9ifA}.
\newblock


\bibitem[\protect\citeauthoryear{Francesco, Malavolta, and Lago}{Francesco
  et~al\mbox{.}}{2017}]%
        {7930195}
\bibfield{author}{\bibinfo{person}{P.~D. Francesco}, \bibinfo{person}{I.
  Malavolta}, {and} \bibinfo{person}{P. Lago}.}
  \bibinfo{year}{2017}\natexlab{}.
\newblock \showarticletitle{Research on Architecting Microservices: Trends,
  Focus, and Potential for Industrial Adoption}. In
  \bibinfo{booktitle}{\emph{2017 IEEE International Conference on Software
  Architecture (ICSA)}}. \bibinfo{pages}{21--30}.
\newblock
\urldef\tempurl%
\url{https://doi.org/10.1109/ICSA.2017.24}
\showDOI{\tempurl}


\bibitem[\protect\citeauthoryear{Gamma, Helm, Johnson, and Vlissides}{Gamma
  et~al\mbox{.}}{1994}]%
        {gamma}
\bibfield{author}{\bibinfo{person}{E. Gamma}, \bibinfo{person}{R. Helm},
  \bibinfo{person}{R. Johnson}, {and} \bibinfo{person}{J. Vlissides}.}
  \bibinfo{year}{1994}\natexlab{}.
\newblock \bibinfo{booktitle}{\emph{Design Patterns: Elements of Reusable
  Object-Oriented Software}}.
\newblock \bibinfo{publisher}{Pearson Education}.
\newblock
\showISBNx{9780321700698}
\urldef\tempurl%
\url{https://books.google.co.uk/books?id=6oHuKQe3TjQC}
\showURL{%
\tempurl}


\bibitem[\protect\citeauthoryear{Garlan, Cheng, Huang, Schmerl, and
  Steenkiste}{Garlan et~al\mbox{.}}{2004}]%
        {rainbow}
\bibfield{author}{\bibinfo{person}{David Garlan}, \bibinfo{person}{Shang-Wen
  Cheng}, \bibinfo{person}{An-Cheng Huang}, \bibinfo{person}{Bradley Schmerl},
  {and} \bibinfo{person}{Peter Steenkiste}.} \bibinfo{year}{2004}\natexlab{}.
\newblock \showarticletitle{Rainbow: Architecture-Based Self-Adaptation with
  Reusable Infrastructure}.
\newblock \bibinfo{journal}{\emph{Computer}} \bibinfo{volume}{37},
  \bibinfo{number}{10} (\bibinfo{date}{Oct.} \bibinfo{year}{2004}),
  \bibinfo{pages}{46--54}.
\newblock
\showISSN{0018-9162}
\urldef\tempurl%
\url{https://doi.org/10.1109/MC.2004.175}
\showDOI{\tempurl}


\bibitem[\protect\citeauthoryear{Garriga}{Garriga}{2018}]%
        {10.1007/978-3-319-74781-1_15}
\bibfield{author}{\bibinfo{person}{Martin Garriga}.}
  \bibinfo{year}{2018}\natexlab{}.
\newblock \showarticletitle{Towards a Taxonomy of Microservices Architectures}.
  In \bibinfo{booktitle}{\emph{Software Engineering and Formal Methods}},
  \bibfield{editor}{\bibinfo{person}{Antonio Cerone} {and}
  \bibinfo{person}{Marco Roveri}} (Eds.). \bibinfo{publisher}{Springer
  International Publishing}, \bibinfo{address}{Cham},
  \bibinfo{pages}{203--218}.
\newblock
\showISBNx{978-3-319-74781-1}


\bibitem[\protect\citeauthoryear{Geerinck}{Geerinck}{[n. d.]}]%
        {pop00133}
\bibfield{author}{\bibinfo{person}{X Geerinck}.} \bibinfo{year}{[n.
  d.]}\natexlab{}.
\newblock \emph{\bibinfo{title}{An Architecture for Resource Analysis,
  Prediction and Visualization in Microservice Deployments}}.
\newblock \bibinfo{thesistype}{Master's\ thesis}. \bibinfo{school}{University
  of Ghent}.
\newblock
\urldef\tempurl%
\url{http://lib.ugent.be/fulltxt/RUG01/002/367/136/RUG01-002367136_2017_0001_AC.pdf}
\showURL{%
\tempurl}


\bibitem[\protect\citeauthoryear{George}{George}{2015}]%
        {chal}
\bibfield{author}{\bibinfo{person}{Fred George}.}
  \bibinfo{year}{2015}\natexlab{}.
\newblock \bibinfo{title}{Challenges in Implementing MicroServices by Fred
  George}.
\newblock
\newblock


\bibitem[\protect\citeauthoryear{Georgiadis, Magee, and Kramer}{Georgiadis
  et~al\mbox{.}}{2002}]%
        {dec2}
\bibfield{author}{\bibinfo{person}{Ioannis Georgiadis}, \bibinfo{person}{Jeff
  Magee}, {and} \bibinfo{person}{Jeff Kramer}.}
  \bibinfo{year}{2002}\natexlab{}.
\newblock \showarticletitle{Self-organising Software Architectures for
  Distributed Systems}. In \bibinfo{booktitle}{\emph{Proceedings of the First
  Workshop on Self-healing Systems}} \emph{(\bibinfo{series}{WOSS '02})}.
  \bibinfo{publisher}{ACM}, \bibinfo{address}{New York, NY, USA},
  \bibinfo{pages}{33--38}.
\newblock
\showISBNx{1-58113-609-9}
\urldef\tempurl%
\url{https://doi.org/10.1145/582128.582135}
\showDOI{\tempurl}


\bibitem[\protect\citeauthoryear{Giallorenzo, Lanese, Mauro, and
  Gabbrielli}{Giallorenzo et~al\mbox{.}}{2017}]%
        {giallorenzo2017programming}
\bibfield{author}{\bibinfo{person}{Saverio Giallorenzo}, \bibinfo{person}{Ivan
  Lanese}, \bibinfo{person}{Jacopo Mauro}, {and} \bibinfo{person}{Maurizio
  Gabbrielli}.} \bibinfo{year}{2017}\natexlab{}.
\newblock \showarticletitle{Programming Adaptive Microservice Applications: An
  AIOCJ Tutorial}.
\newblock \bibinfo{journal}{\emph{Behavioural Types: from Theory to Tools}}
  (\bibinfo{year}{2017}), \bibinfo{pages}{147}.
\newblock


\bibitem[\protect\citeauthoryear{Godwin}{Godwin}{2016}]%
        {bbc}
\bibfield{author}{\bibinfo{person}{Stephen Godwin}.}
  \bibinfo{year}{2016}\natexlab{}.
\newblock \bibinfo{title}{Cloud-based Microservices powering BBC iPlayer}.
\newblock
\newblock
\urldef\tempurl%
\url{https://www.infoq.com/presentations/bbc-microservices-aws?utm_campaign=infoq_content\&utm_source=infoq\&utm_medium=feed\&utm_term=Microservices}
\showURL{%
\tempurl}


\bibitem[\protect\citeauthoryear{Haselb{\"o}ck, Weinreich, and
  Buchgeher}{Haselb{\"o}ck et~al\mbox{.}}{2017}]%
        {10.1007/978-3-319-65831-5_11}
\bibfield{author}{\bibinfo{person}{Stefan Haselb{\"o}ck},
  \bibinfo{person}{Rainer Weinreich}, {and} \bibinfo{person}{Georg Buchgeher}.}
  \bibinfo{year}{2017}\natexlab{}.
\newblock \showarticletitle{Decision Models for Microservices: Design Areas,
  Stakeholders, Use Cases, and Requirements}. In
  \bibinfo{booktitle}{\emph{Software Architecture}},
  \bibfield{editor}{\bibinfo{person}{Ant{\'o}nia Lopes} {and}
  \bibinfo{person}{Rog{\'e}rio de~Lemos}} (Eds.). \bibinfo{publisher}{Springer
  International Publishing}, \bibinfo{address}{Cham},
  \bibinfo{pages}{155--170}.
\newblock
\showISBNx{978-3-319-65831-5}


\bibitem[\protect\citeauthoryear{Hassan, Ali, and Bahsoon}{Hassan
  et~al\mbox{.}}{2017}]%
        {icsa2017}
\bibfield{author}{\bibinfo{person}{S. Hassan}, \bibinfo{person}{N. Ali}, {and}
  \bibinfo{person}{R. Bahsoon}.} \bibinfo{year}{2017}\natexlab{}.
\newblock \showarticletitle{Microservice Ambients: An Architectural
  Meta-Modelling Approach for Microservice Granularity}. In
  \bibinfo{booktitle}{\emph{2017 IEEE International Conference on Software
  Architecture (ICSA)}}. \bibinfo{pages}{1--10}.
\newblock
\urldef\tempurl%
\url{https://doi.org/10.1109/ICSA.2017.32}
\showDOI{\tempurl}


\bibitem[\protect\citeauthoryear{Hassan and Bahsoon}{Hassan and
  Bahsoon}{2016}]%
        {scc}
\bibfield{author}{\bibinfo{person}{S. Hassan} {and} \bibinfo{person}{R.
  Bahsoon}.} \bibinfo{year}{2016}\natexlab{}.
\newblock \showarticletitle{Microservices and Their Design Trade-Offs: A
  Self-Adaptive Roadmap}. In \bibinfo{booktitle}{\emph{2016 IEEE International
  Conference on Services Computing (SCC)}}. \bibinfo{pages}{813--818}.
\newblock
\urldef\tempurl%
\url{https://doi.org/10.1109/SCC.2016.113}
\showDOI{\tempurl}


\bibitem[\protect\citeauthoryear{Hasselbring}{Hasselbring}{2016}]%
        {Hasselbring:2016:MSK:2851553.2858659}
\bibfield{author}{\bibinfo{person}{Wilhelm Hasselbring}.}
  \bibinfo{year}{2016}\natexlab{}.
\newblock \showarticletitle{Microservices for Scalability: Keynote Talk
  Abstract}. In \bibinfo{booktitle}{\emph{Proceedings of the 7th ACM/SPEC on
  International Conference on Performance Engineering}}
  \emph{(\bibinfo{series}{ICPE '16})}. \bibinfo{publisher}{ACM},
  \bibinfo{address}{New York, NY, USA}, \bibinfo{pages}{133--134}.
\newblock
\showISBNx{978-1-4503-4080-9}
\urldef\tempurl%
\url{https://doi.org/10.1145/2851553.2858659}
\showDOI{\tempurl}


\bibitem[\protect\citeauthoryear{Haupt, Leymann, Scherer, and
  Vukojevic-Haupt}{Haupt et~al\mbox{.}}{2017}]%
        {7930199}
\bibfield{author}{\bibinfo{person}{F. Haupt}, \bibinfo{person}{F. Leymann},
  \bibinfo{person}{A. Scherer}, {and} \bibinfo{person}{K. Vukojevic-Haupt}.}
  \bibinfo{year}{2017}\natexlab{}.
\newblock \showarticletitle{A Framework for the Structural Analysis of REST
  APIs}. In \bibinfo{booktitle}{\emph{2017 IEEE International Conference on
  Software Architecture (ICSA)}}. \bibinfo{pages}{55--58}.
\newblock
\urldef\tempurl%
\url{https://doi.org/10.1109/ICSA.2017.40}
\showDOI{\tempurl}


\bibitem[\protect\citeauthoryear{Haywood}{Haywood}{2017}]%
        {infoqmag}
\bibfield{author}{\bibinfo{person}{Dan Haywood}.}
  \bibinfo{year}{2017}\natexlab{}.
\newblock \bibinfo{title}{In Defence of the Monolith, Part 1}.
\newblock
  \bibinfo{howpublished}{https://www.infoq.com/articles/monolith-defense-part-1}.
\newblock


\bibitem[\protect\citeauthoryear{Heinrich, Zirkelbach, and Jung}{Heinrich
  et~al\mbox{.}}{2017}]%
        {7958486}
\bibfield{author}{\bibinfo{person}{R. Heinrich}, \bibinfo{person}{C.
  Zirkelbach}, {and} \bibinfo{person}{R. Jung}.}
  \bibinfo{year}{2017}\natexlab{}.
\newblock \showarticletitle{Architectural Runtime Modeling and Visualization
  for Quality-Aware DevOps in Cloud Applications}. In
  \bibinfo{booktitle}{\emph{2017 IEEE International Conference on Software
  Architecture Workshops (ICSAW)}}. \bibinfo{pages}{199--201}.
\newblock
\urldef\tempurl%
\url{https://doi.org/10.1109/ICSAW.2017.33}
\showDOI{\tempurl}


\bibitem[\protect\citeauthoryear{Herzberg, Hochreiner, Kopp, and
  Lenhard}{Herzberg et~al\mbox{.}}{2018}]%
        {stateofpractice}
\bibfield{author}{\bibinfo{person}{N. Herzberg}, \bibinfo{person}{C.
  Hochreiner}, \bibinfo{person}{O. Kopp}, {and} \bibinfo{person}{J. Lenhard}.}
  \bibinfo{year}{2018}\natexlab{}.
\newblock \showarticletitle{Challenges of Microservices Architecture: A Survey
  on the State of the Practice}. In \bibinfo{booktitle}{\emph{10th ZEUS
  Workshop, ZEUS 2018}}.
\newblock


\bibitem[\protect\citeauthoryear{Humble}{Humble}{2018}]%
        {buzzfeed}
\bibfield{author}{\bibinfo{person}{Charles Humble}.}
  \bibinfo{year}{2018}\natexlab{}.
\newblock \bibinfo{title}{How BuzzFeed Migrated from a Perl Monolith to Go and
  Python Microservices}.
\newblock
  \bibinfo{howpublished}{https://www.infoq.com/articles/buzzfeed-microservices-migration}.
\newblock


\bibitem[\protect\citeauthoryear{Huston}{Huston}{2018}]%
        {smartbear}
\bibfield{author}{\bibinfo{person}{Tom Huston}.}
  \bibinfo{year}{2018}\natexlab{}.
\newblock \bibinfo{title}{What is Microservice Architecture?}
\newblock
  \bibinfo{howpublished}{https://smartbear.com/learn/api-design/what-are-microservices/}.
\newblock


\bibitem[\protect\citeauthoryear{Huttunen}{Huttunen}{2017}]%
        {pop00049}
\bibfield{author}{\bibinfo{person}{J Huttunen}.}
  \bibinfo{year}{2017}\natexlab{}.
\newblock \emph{\bibinfo{title}{Micro service Testing Practices in Public
  Sector Software Projects}}.
\newblock \bibinfo{thesistype}{Master's\ thesis}. \bibinfo{school}{Aalto
  University}.
\newblock
\urldef\tempurl%
\url{https://aaltodoc.aalto.fi/handle/123456789/26673}
\showURL{%
\tempurl}


\bibitem[\protect\citeauthoryear{IBM}{IBM}{2006}]%
        {ibm}
\bibfield{author}{\bibinfo{person}{IBM}.} \bibinfo{year}{2006}\natexlab{}.
\newblock \bibinfo{booktitle}{\emph{An architectural blueprint for autonomic
  computing}}.
\newblock \bibinfo{type}{{T}echnical {R}eport}. \bibinfo{institution}{IBM}.
\newblock


\bibitem[\protect\citeauthoryear{Iffland}{Iffland}{2016}]%
        {balazs}
\bibfield{author}{\bibinfo{person}{David Iffland}.}
  \bibinfo{year}{2016}\natexlab{}.
\newblock \bibinfo{title}{Q\&A with Intuit's Alex Balazs}.
\newblock
  \bibinfo{howpublished}{https://www.infoq.com/articles/intuit-alex-balazs-node-services}.
\newblock


\bibitem[\protect\citeauthoryear{Ismail, Rolnick, Fabijan, and Wallgren}{Ismail
  et~al\mbox{.}}{2016}]%
        {lessonslearneddeploy}
\bibfield{author}{\bibinfo{person}{Usman Ismail}, \bibinfo{person}{Daniel
  Rolnick}, \bibinfo{person}{Darko Fabijan}, {and} \bibinfo{person}{Anders
  Wallgren}.} \bibinfo{year}{2016}\natexlab{}.
\newblock \bibinfo{title}{Challenges of micro-service deployments}.
\newblock \bibinfo{howpublished}{http://techtraits.com/microservice.html}.
\newblock


\bibitem[\protect\citeauthoryear{Ivkovic and Kontogiannis}{Ivkovic and
  Kontogiannis}{2006}]%
        {1602365}
\bibfield{author}{\bibinfo{person}{I. Ivkovic} {and} \bibinfo{person}{K.
  Kontogiannis}.} \bibinfo{year}{2006}\natexlab{}.
\newblock \showarticletitle{A framework for software architecture refactoring
  using model transformations and semantic annotations}. In
  \bibinfo{booktitle}{\emph{Conference on Software Maintenance and
  Reengineering (CSMR'06)}}. \bibinfo{pages}{10 pp.--144}.
\newblock
\showISSN{1534-5351}
\urldef\tempurl%
\url{https://doi.org/10.1109/CSMR.2006.3}
\showDOI{\tempurl}


\bibitem[\protect\citeauthoryear{Jamshidi, Pahl, Chinenyeze, and Liu}{Jamshidi
  et~al\mbox{.}}{2015}]%
        {10.1007/978-3-319-22885-3_2}
\bibfield{author}{\bibinfo{person}{Pooyan Jamshidi}, \bibinfo{person}{Claus
  Pahl}, \bibinfo{person}{Samuel Chinenyeze}, {and} \bibinfo{person}{Xiaodong
  Liu}.} \bibinfo{year}{2015}\natexlab{}.
\newblock \showarticletitle{Cloud Migration Patterns: A Multi-cloud Service
  Architecture Perspective}. In \bibinfo{booktitle}{\emph{Service-Oriented
  Computing - ICSOC 2014 Workshops}}, \bibfield{editor}{\bibinfo{person}{Farouk
  Toumani}, \bibinfo{person}{Barbara Pernici}, \bibinfo{person}{Daniela
  Grigori}, \bibinfo{person}{Djamal Benslimane}, \bibinfo{person}{Jan
  Mendling}, \bibinfo{person}{Nejib Ben Hadj-Alouane}, \bibinfo{person}{Brian
  Blake}, \bibinfo{person}{Olivier Perrin}, \bibinfo{person}{Iman
  Saleh~Moustafa}, {and} \bibinfo{person}{Sami Bhiri}} (Eds.).
  \bibinfo{publisher}{Springer International Publishing},
  \bibinfo{address}{Cham}, \bibinfo{pages}{6--19}.
\newblock
\showISBNx{978-3-319-22885-3}


\bibitem[\protect\citeauthoryear{Ji and Liu}{Ji and Liu}{2016}]%
        {sla}
\bibfield{author}{\bibinfo{person}{ZL Ji} {and} \bibinfo{person}{Y Liu}.}
  \bibinfo{year}{2016}\natexlab{}.
\newblock \showarticletitle{A dynamic deployment method of micro service
  oriented to SLA}.
\newblock \bibinfo{journal}{\emph{International Journal of Computer Science
  Issues}} (\bibinfo{year}{2016}).
\newblock


\bibitem[\protect\citeauthoryear{Joselyne, Kanagwa, and Balikuddembe}{Joselyne
  et~al\mbox{.}}{2017}]%
        {joselyneframework}
\bibfield{author}{\bibinfo{person}{Munezero~Immaculee Joselyne},
  \bibinfo{person}{Benjamin Kanagwa}, {and} \bibinfo{person}{Joseph
  Balikuddembe}.} \bibinfo{year}{2017}\natexlab{}.
\newblock \showarticletitle{A framework to Modernize SME Application in
  Emerging Economies: Microservice Architecture Pattern Approach}.
\newblock  (\bibinfo{year}{2017}).
\newblock


\bibitem[\protect\citeauthoryear{Joshi, Hiltunen, Sanders, and
  Schlichting}{Joshi et~al\mbox{.}}{2005}]%
        {1541182}
\bibfield{author}{\bibinfo{person}{K.~R. Joshi}, \bibinfo{person}{M.~A.
  Hiltunen}, \bibinfo{person}{W.~H. Sanders}, {and} \bibinfo{person}{R.~D.
  Schlichting}.} \bibinfo{year}{2005}\natexlab{}.
\newblock \showarticletitle{Automatic model-driven recovery in distributed
  systems}. In \bibinfo{booktitle}{\emph{24th IEEE Symposium on Reliable
  Distributed Systems (SRDS'05)}}. \bibinfo{pages}{25--36}.
\newblock
\showISSN{1060-9857}
\urldef\tempurl%
\url{https://doi.org/10.1109/RELDIS.2005.11}
\showDOI{\tempurl}


\bibitem[\protect\citeauthoryear{Kaiser, Podjarny, Levine, Burgess, and
  Stenberg}{Kaiser et~al\mbox{.}}{2018}]%
        {future}
\bibfield{author}{\bibinfo{person}{Susanne Kaiser}, \bibinfo{person}{Guy
  Podjarny}, \bibinfo{person}{Idit Levine}, \bibinfo{person}{Mark Burgess},
  {and} \bibinfo{person}{Jan Stenberg}.} \bibinfo{year}{2018}\natexlab{}.
\newblock \bibinfo{title}{The Future of Microservices and Distributed Systems:
  QCon London Microservices Panel Discussion}.
\newblock
  \bibinfo{howpublished}{https://www.infoq.com/news/2018/03/microservices-future}.
\newblock


\bibitem[\protect\citeauthoryear{Kalske, M{\"a}kitalo, and Mikkonen}{Kalske
  et~al\mbox{.}}{2018}]%
        {10.1007/978-3-319-74433-9_3}
\bibfield{author}{\bibinfo{person}{Miika Kalske}, \bibinfo{person}{Niko
  M{\"a}kitalo}, {and} \bibinfo{person}{Tommi Mikkonen}.}
  \bibinfo{year}{2018}\natexlab{}.
\newblock \showarticletitle{Challenges When Moving from Monolith to
  Microservice Architecture}. In \bibinfo{booktitle}{\emph{Current Trends in
  Web Engineering}}, \bibfield{editor}{\bibinfo{person}{Irene Garrig{\'o}s}
  {and} \bibinfo{person}{Manuel Wimmer}} (Eds.). \bibinfo{publisher}{Springer
  International Publishing}, \bibinfo{address}{Cham}, \bibinfo{pages}{32--47}.
\newblock


\bibitem[\protect\citeauthoryear{Kecskemeti, Marosi, and Kertesz}{Kecskemeti
  et~al\mbox{.}}{2016}]%
        {7568389}
\bibfield{author}{\bibinfo{person}{G. Kecskemeti}, \bibinfo{person}{A.~C.
  Marosi}, {and} \bibinfo{person}{A. Kertesz}.}
  \bibinfo{year}{2016}\natexlab{}.
\newblock \showarticletitle{The ENTICE approach to decompose monolithic
  services into microservices}. In \bibinfo{booktitle}{\emph{2016 International
  Conference on High Performance Computing Simulation (HPCS)}}.
  \bibinfo{pages}{591--596}.
\newblock
\urldef\tempurl%
\url{https://doi.org/10.1109/HPCSim.2016.7568389}
\showDOI{\tempurl}


\bibitem[\protect\citeauthoryear{Kessler}{Kessler}{2014}]%
        {astro}
\bibfield{author}{\bibinfo{person}{Fondazione~Brundo Kessler}.}
  \bibinfo{year}{2014}\natexlab{}.
\newblock \bibinfo{title}{ASTRO-CAptEvo}.
\newblock \bibinfo{howpublished}{Online}.
\newblock
\urldef\tempurl%
\url{http://das.fbk.eu/astro-captevo}
\showURL{%
Retrieved 5 August 2017 from \tempurl}


\bibitem[\protect\citeauthoryear{Khazaei, Barna, Beigi-Mohammadi, and
  Litoiu}{Khazaei et~al\mbox{.}}{2016}]%
        {7830692}
\bibfield{author}{\bibinfo{person}{H. Khazaei}, \bibinfo{person}{C. Barna},
  \bibinfo{person}{N. Beigi-Mohammadi}, {and} \bibinfo{person}{M. Litoiu}.}
  \bibinfo{year}{2016}\natexlab{}.
\newblock \showarticletitle{Efficiency Analysis of Provisioning Microservices}.
  In \bibinfo{booktitle}{\emph{2016 IEEE International Conference on Cloud
  Computing Technology and Science (CloudCom)}}. \bibinfo{pages}{261--268}.
\newblock
\urldef\tempurl%
\url{https://doi.org/10.1109/CloudCom.2016.0051}
\showDOI{\tempurl}


\bibitem[\protect\citeauthoryear{Killalea}{Killalea}{2016}]%
        {Killalea:2016:HDM:2975594.2948985}
\bibfield{author}{\bibinfo{person}{Tom Killalea}.}
  \bibinfo{year}{2016}\natexlab{}.
\newblock \showarticletitle{The Hidden Dividends of Microservices}.
\newblock \bibinfo{journal}{\emph{Commun. ACM}} \bibinfo{volume}{59},
  \bibinfo{number}{8} (\bibinfo{date}{July} \bibinfo{year}{2016}),
  \bibinfo{pages}{42--45}.
\newblock
\showISSN{0001-0782}
\urldef\tempurl%
\url{https://doi.org/10.1145/2948985}
\showDOI{\tempurl}


\bibitem[\protect\citeauthoryear{Kitchenham}{Kitchenham}{2007}]%
        {slr}
\bibfield{author}{\bibinfo{person}{Barbara Kitchenham}.}
  \bibinfo{year}{2007}\natexlab{}.
\newblock \bibinfo{booktitle}{\emph{Guidelines for performing Systematic
  Literature Reviews in Software Engineering}}.
\newblock \bibinfo{type}{{T}echnical {R}eport}. \bibinfo{institution}{Keele
  University and University of Durham}.
\newblock


\bibitem[\protect\citeauthoryear{Kitchenham, Dyba, and Jorgensen}{Kitchenham
  et~al\mbox{.}}{2004}]%
        {Kitchenham:2004:ESE:998675.999432}
\bibfield{author}{\bibinfo{person}{Barbara~A. Kitchenham},
  \bibinfo{person}{Tore Dyba}, {and} \bibinfo{person}{Magne Jorgensen}.}
  \bibinfo{year}{2004}\natexlab{}.
\newblock \showarticletitle{Evidence-Based Software Engineering}. In
  \bibinfo{booktitle}{\emph{Proceedings of the 26th International Conference on
  Software Engineering}} \emph{(\bibinfo{series}{ICSE '04})}.
  \bibinfo{publisher}{IEEE Computer Society}, \bibinfo{address}{Washington, DC,
  USA}, \bibinfo{pages}{273--281}.
\newblock
\showISBNx{0-7695-2163-0}
\urldef\tempurl%
\url{http://dl.acm.org/citation.cfm?id=998675.999432}
\showURL{%
\tempurl}


\bibitem[\protect\citeauthoryear{Kleindienst}{Kleindienst}{2017}]%
        {Kleindienst2017}
\bibfield{author}{\bibinfo{person}{Patrick Kleindienst}.}
  \bibinfo{year}{2017}\natexlab{}.
\newblock \emph{\bibinfo{title}{Implementation and evaluation of a hybrid
  microservice infrastructure}}.
\newblock \bibinfo{thesistype}{Master's\ thesis}. \bibinfo{school}{Hochschule
  der Medien, Department 1: Printing and Media}.
\newblock


\bibitem[\protect\citeauthoryear{Klinaku, Frank, and Becker}{Klinaku
  et~al\mbox{.}}{2018}]%
        {Klinaku:2018:CEC:3185768.3186296}
\bibfield{author}{\bibinfo{person}{Floriment Klinaku}, \bibinfo{person}{Markus
  Frank}, {and} \bibinfo{person}{Steffen Becker}.}
  \bibinfo{year}{2018}\natexlab{}.
\newblock \showarticletitle{CAUS: An Elasticity Controller for a Containerized
  Microservice}. In \bibinfo{booktitle}{\emph{Companion of the 2018 ACM/SPEC
  International Conference on Performance Engineering}}
  \emph{(\bibinfo{series}{ICPE '18})}. \bibinfo{publisher}{ACM},
  \bibinfo{address}{New York, NY, USA}, \bibinfo{pages}{93--98}.
\newblock
\showISBNx{978-1-4503-5629-9}
\urldef\tempurl%
\url{https://doi.org/10.1145/3185768.3186296}
\showDOI{\tempurl}


\bibitem[\protect\citeauthoryear{Klock, Werf, Guelen, and Jansen}{Klock
  et~al\mbox{.}}{2017}]%
        {7930194}
\bibfield{author}{\bibinfo{person}{S. Klock}, \bibinfo{person}{J.~M. E. M.
  V.~D. Werf}, \bibinfo{person}{J.~P. Guelen}, {and} \bibinfo{person}{S.
  Jansen}.} \bibinfo{year}{2017}\natexlab{}.
\newblock \showarticletitle{Workload-Based Clustering of Coherent Feature Sets
  in Microservice Architectures}. In \bibinfo{booktitle}{\emph{2017 IEEE
  International Conference on Software Architecture (ICSA)}}.
  \bibinfo{pages}{11--20}.
\newblock
\urldef\tempurl%
\url{https://doi.org/10.1109/ICSA.2017.38}
\showDOI{\tempurl}


\bibitem[\protect\citeauthoryear{Koutsouras, Kougioumoutzakis, and
  Kantzavelou}{Koutsouras et~al\mbox{.}}{2015}]%
        {Koutsouras:2015:ASI:2801948.2802030}
\bibfield{author}{\bibinfo{person}{Aristeidis Koutsouras},
  \bibinfo{person}{Dimosthenis Kougioumoutzakis}, {and} \bibinfo{person}{Ioanna
  Kantzavelou}.} \bibinfo{year}{2015}\natexlab{}.
\newblock \showarticletitle{Assessment and Security Issues in Cloud Computing
  Services}. In \bibinfo{booktitle}{\emph{Proceedings of the 19th Panhellenic
  Conference on Informatics}} \emph{(\bibinfo{series}{PCI '15})}.
  \bibinfo{publisher}{ACM}, \bibinfo{address}{New York, NY, USA},
  \bibinfo{pages}{165--166}.
\newblock
\showISBNx{978-1-4503-3551-5}
\urldef\tempurl%
\url{https://doi.org/10.1145/2801948.2802030}
\showDOI{\tempurl}


\bibitem[\protect\citeauthoryear{Krafzig, Banke, and Slama}{Krafzig
  et~al\mbox{.}}{2005}]%
        {krafzig2005enterprise}
\bibfield{author}{\bibinfo{person}{D. Krafzig}, \bibinfo{person}{K. Banke},
  {and} \bibinfo{person}{D. Slama}.} \bibinfo{year}{2005}\natexlab{}.
\newblock \bibinfo{booktitle}{\emph{Enterprise SOA: Service-oriented
  Architecture Best Practices}}.
\newblock \bibinfo{publisher}{Prentice Hall Professional Technical Reference}.
\newblock
\showISBNx{9780131465756}
\showLCCN{2004110321}
\urldef\tempurl%
\url{https://books.google.co.uk/books?id=R7oGhITYUuUC}
\showURL{%
\tempurl}


\bibitem[\protect\citeauthoryear{Krajicek}{Krajicek}{2015}]%
        {solid}
\bibfield{author}{\bibinfo{person}{Ondrej Krajicek}.}
  \bibinfo{year}{2015}\natexlab{}.
\newblock \bibinfo{title}{Micro-services or Solid Services}.
\newblock
  \bibinfo{howpublished}{http://www.infoq.com/presentations/microservices-solid}.
\newblock


\bibitem[\protect\citeauthoryear{Krause}{Krause}{2015}]%
        {krause2015microservices}
\bibfield{author}{\bibinfo{person}{L. Krause}.}
  \bibinfo{year}{2015}\natexlab{}.
\newblock \bibinfo{booktitle}{\emph{Microservices: Patterns and Applications:
  Designing Fine-Grained Services by Applying Patterns}}.
\newblock \bibinfo{publisher}{Lucas Krause}.
\newblock
\showISBNx{9780692424278}
\urldef\tempurl%
\url{https://books.google.co.uk/books?id=dd5-rgEACAAJ}
\showURL{%
\tempurl}


\bibitem[\protect\citeauthoryear{Krylovskiy, Jahn, and Patti}{Krylovskiy
  et~al\mbox{.}}{2015}]%
        {7300793}
\bibfield{author}{\bibinfo{person}{A. Krylovskiy}, \bibinfo{person}{M. Jahn},
  {and} \bibinfo{person}{E. Patti}.} \bibinfo{year}{2015}\natexlab{}.
\newblock \showarticletitle{Designing a Smart City Internet of Things Platform
  with Microservice Architecture}. In \bibinfo{booktitle}{\emph{2015 3rd
  International Conference on Future Internet of Things and Cloud}}.
  \bibinfo{pages}{25--30}.
\newblock
\urldef\tempurl%
\url{https://doi.org/10.1109/FiCloud.2015.55}
\showDOI{\tempurl}


\bibitem[\protect\citeauthoryear{Kukade and Kale}{Kukade and Kale}{2015}]%
        {autoscaleindia}
\bibfield{author}{\bibinfo{person}{Priyanka~P. Kukade} {and}
  \bibinfo{person}{Prof.~Geetanjali Kale}.} \bibinfo{year}{2015}\natexlab{}.
\newblock \showarticletitle{Auto-Scaling of Micro-Services Using
  Containerization}.
\newblock \bibinfo{journal}{\emph{International Journal of Science and Research
  (IJSR)}} \bibinfo{volume}{4}, \bibinfo{number}{9} (\bibinfo{date}{sep}
  \bibinfo{year}{2015}).
\newblock


\bibitem[\protect\citeauthoryear{Kulkarni and Dwivedi}{Kulkarni and
  Dwivedi}{2008}]%
        {4578356}
\bibfield{author}{\bibinfo{person}{N. Kulkarni} {and} \bibinfo{person}{V.
  Dwivedi}.} \bibinfo{year}{2008}\natexlab{}.
\newblock \showarticletitle{The Role of Service Granularity in a Successful SOA
  Realization A Case Study}. In \bibinfo{booktitle}{\emph{2008 IEEE Congress on
  Services - Part I}}. \bibinfo{pages}{423--430}.
\newblock
\showISSN{2378-3818}
\urldef\tempurl%
\url{https://doi.org/10.1109/SERVICES-1.2008.86}
\showDOI{\tempurl}


\bibitem[\protect\citeauthoryear{Kwon and Tilevich}{Kwon and Tilevich}{2013}]%
        {cloudref}
\bibfield{author}{\bibinfo{person}{Young-Woo Kwon} {and} \bibinfo{person}{Eli
  Tilevich}.} \bibinfo{year}{2013}\natexlab{}.
\newblock \showarticletitle{Cloud refactoring: automated transitioning to
  cloud-based services}.
\newblock \bibinfo{journal}{\emph{Automated Software Engineering}}
  \bibinfo{volume}{21}, \bibinfo{number}{3} (\bibinfo{year}{2013}),
  \bibinfo{pages}{345--372}.
\newblock
\showISSN{1573-7535}
\urldef\tempurl%
\url{https://doi.org/10.1007/s10515-013-0136-9}
\showDOI{\tempurl}


\bibitem[\protect\citeauthoryear{Li}{Li}{2017}]%
        {8312516}
\bibfield{author}{\bibinfo{person}{S. Li}.} \bibinfo{year}{2017}\natexlab{}.
\newblock \showarticletitle{Understanding Quality Attributes in Microservice
  Architecture}. In \bibinfo{booktitle}{\emph{2017 24th Asia-Pacific Software
  Engineering Conference Workshops (APSECW)}}. \bibinfo{pages}{9--10}.
\newblock
\urldef\tempurl%
\url{https://doi.org/10.1109/APSECW.2017.33}
\showDOI{\tempurl}


\bibitem[\protect\citeauthoryear{Limthanmaphon and Zhang}{Limthanmaphon and
  Zhang}{2003}]%
        {case}
\bibfield{author}{\bibinfo{person}{Benchaphon Limthanmaphon} {and}
  \bibinfo{person}{Yanchun Zhang}.} \bibinfo{year}{2003}\natexlab{}.
\newblock \showarticletitle{Web Service Composition with Case-based Reasoning}.
  In \bibinfo{booktitle}{\emph{Proceedings of the 14th Australasian Database
  Conference - Volume 17}} \emph{(\bibinfo{series}{ADC '03})}.
  \bibinfo{publisher}{Australian Computer Society, Inc.},
  \bibinfo{address}{Darlinghurst, Australia, Australia},
  \bibinfo{pages}{201--208}.
\newblock
\showISBNx{0-909-92595-X}
\urldef\tempurl%
\url{http://dl.acm.org/citation.cfm?id=820085.820122}
\showURL{%
\tempurl}


\bibitem[\protect\citeauthoryear{Linthicum}{Linthicum}{2016}]%
        {7742277}
\bibfield{author}{\bibinfo{person}{D.~S. Linthicum}.}
  \bibinfo{year}{2016}\natexlab{}.
\newblock \showarticletitle{Practical Use of Microservices in Moving Workloads
  to the Cloud}.
\newblock \bibinfo{journal}{\emph{IEEE Cloud Computing}} \bibinfo{volume}{3},
  \bibinfo{number}{5} (\bibinfo{date}{Sept} \bibinfo{year}{2016}),
  \bibinfo{pages}{6--9}.
\newblock
\showISSN{2325-6095}
\urldef\tempurl%
\url{https://doi.org/10.1109/MCC.2016.114}
\showDOI{\tempurl}


\bibitem[\protect\citeauthoryear{Little}{Little}{2017}]%
        {differencesoa}
\bibfield{author}{\bibinfo{person}{Mark Little}.}
  \bibinfo{year}{2017}\natexlab{}.
\newblock \bibinfo{title}{The Difference between SOA and Microservices?}
\newblock
  \bibinfo{howpublished}{https://www.infoq.com/news/2017/07/soaandmicroservices}.
\newblock


\bibitem[\protect\citeauthoryear{LOFTIS}{LOFTIS}{2015}]%
        {micromatter}
\bibfield{author}{\bibinfo{person}{HUNTER LOFTIS}.}
  \bibinfo{year}{2015}\natexlab{}.
\newblock \bibinfo{title}{Why Microservices Matter}.
\newblock
  \bibinfo{howpublished}{https://blog.heroku.com/why\_microservices\_matter}.
\newblock


\bibitem[\protect\citeauthoryear{Losavio, Ordaz, and Esteller}{Losavio
  et~al\mbox{.}}{2015}]%
        {refactoring}
\bibfield{author}{\bibinfo{person}{Francisca Losavio}, \bibinfo{person}{Oscar
  Ordaz}, {and} \bibinfo{person}{Victor Esteller}.}
  \bibinfo{year}{2015}\natexlab{}.
\newblock \showarticletitle{Refactoring-Based Design of Reference
  Architecture.}
\newblock \bibinfo{journal}{\emph{Revista Antioque{\~n}a de las Ciencias
  Computacionales}} \bibinfo{volume}{5}, \bibinfo{number}{1}
  (\bibinfo{year}{2015}).
\newblock


\bibitem[\protect\citeauthoryear{MacKenzie, Laskey, McCabe, Brown, Metz, and
  Hamilton}{MacKenzie et~al\mbox{.}}{2006}]%
        {mackenzie2006reference}
\bibfield{author}{\bibinfo{person}{C~Matthew MacKenzie}, \bibinfo{person}{Ken
  Laskey}, \bibinfo{person}{Francis McCabe}, \bibinfo{person}{Peter~F Brown},
  \bibinfo{person}{Rebekah Metz}, {and} \bibinfo{person}{Booz~Allen Hamilton}.}
  \bibinfo{year}{2006}\natexlab{}.
\newblock \showarticletitle{Reference model for service oriented architecture
  1.0}.
\newblock \bibinfo{journal}{\emph{OASIS standard}}  \bibinfo{volume}{12}
  (\bibinfo{year}{2006}), \bibinfo{pages}{18}.
\newblock


\bibitem[\protect\citeauthoryear{Malek, Mikic-Rakic, and Medvidovic}{Malek
  et~al\mbox{.}}{2005}]%
        {dec3}
\bibfield{author}{\bibinfo{person}{Sam Malek}, \bibinfo{person}{Marija
  Mikic-Rakic}, {and} \bibinfo{person}{Nenad Medvidovic}.}
  \bibinfo{year}{2005}\natexlab{}.
\newblock \showarticletitle{A Decentralized Redeployment Algorithm for
  Improving the Availability of Distributed Systems}.
\newblock In \bibinfo{booktitle}{\emph{Component Deployment}},
  \bibfield{editor}{\bibinfo{person}{Alan Dearle} {and} \bibinfo{person}{Susan
  Eisenbach}} (Eds.). \bibinfo{series}{Lecture Notes in Computer Science},
  Vol.~\bibinfo{volume}{3798}. \bibinfo{publisher}{Springer Berlin Heidelberg},
  \bibinfo{pages}{99--114}.
\newblock
\showISBNx{978-3-540-30517-0}
\urldef\tempurl%
\url{https://doi.org/10.1007/11590712_8}
\showDOI{\tempurl}


\bibitem[\protect\citeauthoryear{Martin and Lewis}{Martin and Lewis}{2014}]%
        {microfow}
\bibfield{author}{\bibinfo{person}{Martin} {and} \bibinfo{person}{James
  Lewis}.} \bibinfo{year}{2014}\natexlab{}.
\newblock \bibinfo{title}{Microservices a definition of this new architectural
  term}.
\newblock
  \bibinfo{howpublished}{http://martinfowler.com/articles/microservices.html}.
\newblock


\bibitem[\protect\citeauthoryear{Maximilien and Singh}{Maximilien and
  Singh}{2004}]%
        {trust}
\bibfield{author}{\bibinfo{person}{E.~Michael Maximilien} {and}
  \bibinfo{person}{Munindar~P. Singh}.} \bibinfo{year}{2004}\natexlab{}.
\newblock \showarticletitle{Toward Autonomic Web Services Trust and Selection}.
  In \bibinfo{booktitle}{\emph{Proceedings of the 2Nd International Conference
  on Service Oriented Computing}} \emph{(\bibinfo{series}{ICSOC '04})}.
  \bibinfo{publisher}{ACM}, \bibinfo{address}{New York, NY, USA},
  \bibinfo{pages}{212--221}.
\newblock
\showISBNx{1-58113-871-7}
\urldef\tempurl%
\url{https://doi.org/10.1145/1035167.1035198}
\showDOI{\tempurl}


\bibitem[\protect\citeauthoryear{McIlraith and Son}{McIlraith and Son}{2002}]%
        {mci-son-kr02}
\bibfield{author}{\bibinfo{person}{S. McIlraith} {and} \bibinfo{person}{T.
  Son}.} \bibinfo{year}{2002}\natexlab{}.
\newblock \showarticletitle{Adapting Golog for Composition of Semantic Web
  Services}. \bibinfo{howpublished}{mci-son-kr02.pdf}. In
  \bibinfo{booktitle}{\emph{Proceedings of the Eighth International Conference
  on Knowledge Representation and Reasoning (KR2002)}}.
  \bibinfo{address}{Toulouse, France}, \bibinfo{pages}{482--493}.
\newblock


\bibitem[\protect\citeauthoryear{Medvidovic and Taylor}{Medvidovic and
  Taylor}{2000}]%
        {adl}
\bibfield{author}{\bibinfo{person}{Nenad Medvidovic} {and}
  \bibinfo{person}{Richard~N. Taylor}.} \bibinfo{year}{2000}\natexlab{}.
\newblock \showarticletitle{A Classification and Comparison Framework for
  Software Architecture Description Languages}.
\newblock \bibinfo{journal}{\emph{IEEE Trans. Softw. Eng.}}
  \bibinfo{volume}{26}, \bibinfo{number}{1} (\bibinfo{date}{Jan.}
  \bibinfo{year}{2000}), \bibinfo{pages}{70--93}.
\newblock
\showISSN{0098-5589}
\urldef\tempurl%
\url{https://doi.org/10.1109/32.825767}
\showDOI{\tempurl}


\bibitem[\protect\citeauthoryear{Miles}{Miles}{2016}]%
        {antifrag}
\bibfield{author}{\bibinfo{person}{Russ Miles}.}
  \bibinfo{year}{2016}\natexlab{}.
\newblock \bibinfo{booktitle}{\emph{Antifragile Software Building Adaptable
  Software with Microservices}}.
\newblock \bibinfo{publisher}{Leanpub}.
\newblock


\bibitem[\protect\citeauthoryear{Mo, Cai, Kazman, Xiao, and Feng}{Mo
  et~al\mbox{.}}{2016}]%
        {decouple}
\bibfield{author}{\bibinfo{person}{Ran Mo}, \bibinfo{person}{Yuanfang Cai},
  \bibinfo{person}{Rick Kazman}, \bibinfo{person}{Lu Xiao}, {and}
  \bibinfo{person}{Qiong Feng}.} \bibinfo{year}{2016}\natexlab{}.
\newblock \showarticletitle{Decoupling Level: A New Metric for Architectural
  Maintenance Complexity}. In \bibinfo{booktitle}{\emph{Proceedings of the 38th
  International Conference on Software Engineering}}
  \emph{(\bibinfo{series}{ICSE '16})}. \bibinfo{publisher}{ACM},
  \bibinfo{address}{New York, NY, USA}, \bibinfo{pages}{499--510}.
\newblock
\showISBNx{978-1-4503-3900-1}
\urldef\tempurl%
\url{https://doi.org/10.1145/2884781.2884825}
\showDOI{\tempurl}


\bibitem[\protect\citeauthoryear{Mongiello, Colucci, Vogli, Grieco, and
  Sciancalepore}{Mongiello et~al\mbox{.}}{2016}]%
        {Mongiello2016}
\bibfield{author}{\bibinfo{person}{Marina Mongiello}, \bibinfo{person}{Simona
  Colucci}, \bibinfo{person}{Elvis Vogli}, \bibinfo{person}{Luigi~Alfredo
  Grieco}, {and} \bibinfo{person}{Massimo Sciancalepore}.}
  \bibinfo{year}{2016}\natexlab{}.
\newblock \showarticletitle{Run-time architectural modeling for future internet
  applications}.
\newblock \bibinfo{journal}{\emph{Complex {\&} Intelligent Systems}}
  \bibinfo{volume}{2}, \bibinfo{number}{2} (\bibinfo{date}{01 Jun}
  \bibinfo{year}{2016}), \bibinfo{pages}{111--124}.
\newblock
\showISSN{2198-6053}
\urldef\tempurl%
\url{https://doi.org/10.1007/s40747-016-0020-x}
\showDOI{\tempurl}


\bibitem[\protect\citeauthoryear{Mont}{Mont}{2001}]%
        {b6b81df2-e27d-4bfe-81a4-9e44c7c0c6a3}
\bibfield{author}{\bibinfo{person}{Oksana Mont}.}
  \bibinfo{year}{2001}\natexlab{}.
\newblock \bibinfo{title}{Introducing and developing a Product-Service System
  (PSS) concept in Sweden}.
\newblock
\newblock
\showISSN{1650-1675}


\bibitem[\protect\citeauthoryear{Montesi and Weber}{Montesi and Weber}{2016}]%
        {DBLP:journals/corr/MontesiW16}
\bibfield{author}{\bibinfo{person}{Fabrizio Montesi} {and}
  \bibinfo{person}{Janine Weber}.} \bibinfo{year}{2016}\natexlab{}.
\newblock \showarticletitle{Circuit Breakers, Discovery, and {API} Gateways in
  Microservices}.
\newblock \bibinfo{journal}{\emph{CoRR}}  \bibinfo{volume}{abs/1609.05830}
  (\bibinfo{year}{2016}).
\newblock
\showeprint[arxiv]{1609.05830}
\urldef\tempurl%
\url{http://arxiv.org/abs/1609.05830}
\showURL{%
\tempurl}


\bibitem[\protect\citeauthoryear{Morin, Barais, Nain, and Jezequel}{Morin
  et~al\mbox{.}}{2009}]%
        {5070514}
\bibfield{author}{\bibinfo{person}{B. Morin}, \bibinfo{person}{O. Barais},
  \bibinfo{person}{G. Nain}, {and} \bibinfo{person}{J.~M. Jezequel}.}
  \bibinfo{year}{2009}\natexlab{}.
\newblock \showarticletitle{Taming Dynamically Adaptive Systems using models
  and aspects}. In \bibinfo{booktitle}{\emph{2009 IEEE 31st International
  Conference on Software Engineering}}. \bibinfo{pages}{122--132}.
\newblock
\showISSN{0270-5257}
\urldef\tempurl%
\url{https://doi.org/10.1109/ICSE.2009.5070514}
\showDOI{\tempurl}


\bibitem[\protect\citeauthoryear{Mozaffari}{Mozaffari}{2016}]%
        {redhat}
\bibfield{author}{\bibinfo{person}{Babak Mozaffari}.}
  \bibinfo{year}{2016}\natexlab{}.
\newblock \bibinfo{booktitle}{\emph{Microservice Architecture Building
  microservices with JBoss EAP 7} (\bibinfo{edition}{1.0} ed.)}.
\newblock RedHat.
\newblock


\bibitem[\protect\citeauthoryear{Müssig, Stricker, Lässig, and
  Heider}{Müssig et~al\mbox{.}}{2017}]%
        {iceis17}
\bibfield{author}{\bibinfo{person}{Daniel Müssig}, \bibinfo{person}{Robert
  Stricker}, \bibinfo{person}{Jörg Lässig}, {and} \bibinfo{person}{Jens
  Heider}.} \bibinfo{year}{2017}\natexlab{}.
\newblock \showarticletitle{Highly Scalable Microservice-based Enterprise
  Architecture for Smart Ecosystems in Hybrid Cloud Environments}. In
  \bibinfo{booktitle}{\emph{Proceedings of the 19th International Conference on
  Enterprise Information Systems - Volume 3: ICEIS,}}. INSTICC,
  \bibinfo{publisher}{SciTePress}, \bibinfo{pages}{454--459}.
\newblock
\showISBNx{978-989-758-249-3}
\urldef\tempurl%
\url{https://doi.org/10.5220/0006373304540459}
\showDOI{\tempurl}


\bibitem[\protect\citeauthoryear{Mustafa and Gómez}{Mustafa and
  Gómez}{2016}]%
        {fuzzycluster}
\bibfield{author}{\bibinfo{person}{Ola Mustafa} {and}
  \bibinfo{person}{Jorge~Marx Gómez}.} \bibinfo{year}{2016}\natexlab{}.
\newblock \showarticletitle{Using fuzzy clustering techniques to improve the
  design of microservices web applications}. In
  \bibinfo{booktitle}{\emph{Eureka International Virtual Meeting Eureka 2016
  OPTISAD}}.
\newblock


\bibitem[\protect\citeauthoryear{Mustafa and Gómez}{Mustafa and
  Gómez}{2017}]%
        {sustainecon}
\bibfield{author}{\bibinfo{person}{Ola Mustafa} {and}
  \bibinfo{person}{Jorge~Marx Gómez}.} \bibinfo{year}{2017}\natexlab{}.
\newblock \showarticletitle{Sustainable approach for improving microservices
  based web application}. In \bibinfo{booktitle}{\emph{Sustainability Dialogue:
  International Conference on Sustainability and Environmental Management}}.
\newblock


\bibitem[\protect\citeauthoryear{N.~Herzberg}{N.~Herzberg}{2018}]%
        {relpatterns}
\bibfield{author}{\bibinfo{person}{O.~Kopp J.~Lenhard N.~Herzberg,
  C.~Hochreiner}.} \bibinfo{year}{2018}\natexlab{}.
\newblock \showarticletitle{Analyzing the Relevance of SOA Patterns for
  Microservice-Based Systems}. In \bibinfo{booktitle}{\emph{10th ZEUS Workshop,
  ZEUS 2018}}.
\newblock


\bibitem[\protect\citeauthoryear{Naily, Setyautami, Muschevici, and
  Azurat}{Naily et~al\mbox{.}}{2018}]%
        {10.1007/978-3-319-74781-1_18}
\bibfield{author}{\bibinfo{person}{Moh.~Afifun Naily}, \bibinfo{person}{Maya
  Retno~Ayu Setyautami}, \bibinfo{person}{Radu Muschevici}, {and}
  \bibinfo{person}{Ade Azurat}.} \bibinfo{year}{2018}\natexlab{}.
\newblock \showarticletitle{A Framework for Modelling Variable Microservices as
  Software Product Lines}. In \bibinfo{booktitle}{\emph{Software Engineering
  and Formal Methods}}, \bibfield{editor}{\bibinfo{person}{Antonio Cerone}
  {and} \bibinfo{person}{Marco Roveri}} (Eds.). \bibinfo{publisher}{Springer
  International Publishing}, \bibinfo{address}{Cham},
  \bibinfo{pages}{246--261}.
\newblock
\showISBNx{978-3-319-74781-1}


\bibitem[\protect\citeauthoryear{Namiot and sneps sneppe}{Namiot and sneps
  sneppe}{2014}]%
        {article}
\bibfield{author}{\bibinfo{person}{Dmitry Namiot} {and}
  \bibinfo{person}{Manfred sneps sneppe}.} \bibinfo{year}{2014}\natexlab{}.
\newblock \showarticletitle{On Micro-services Architecture}.
\newblock   \bibinfo{volume}{2} (\bibinfo{date}{09} \bibinfo{year}{2014}),
  \bibinfo{pages}{24--27}.
\newblock


\bibitem[\protect\citeauthoryear{Newman}{Newman}{2014a}]%
        {practical1}
\bibfield{author}{\bibinfo{person}{Sam Newman}.}
  \bibinfo{year}{2014}\natexlab{a}.
\newblock \bibinfo{title}{Practical Considerations for Microservice
  Architectures. Part 1. (Sam Newman, UK)}.
\newblock \bibinfo{howpublished}{https://www.youtube.com/watch?v=12n9E5L6qgs}.
\newblock


\bibitem[\protect\citeauthoryear{Newman}{Newman}{2014b}]%
        {practical2}
\bibfield{author}{\bibinfo{person}{Sam Newman}.}
  \bibinfo{year}{2014}\natexlab{b}.
\newblock \bibinfo{title}{Practical Considerations for Microservice
  Architectures. Part 2. (Sam Newman, UK)}.
\newblock \bibinfo{howpublished}{https://www.youtube.com/watch?v=cU0J0w6sFoA}.
\newblock


\bibitem[\protect\citeauthoryear{Newman}{Newman}{2015a}]%
        {microbuild}
\bibfield{author}{\bibinfo{person}{Sam Newman}.}
  \bibinfo{year}{2015}\natexlab{a}.
\newblock \bibinfo{booktitle}{\emph{Building Microservices}
  (\bibinfo{edition}{first} ed.)}.
\newblock \bibinfo{publisher}{O'Reilly Media}.
\newblock


\bibitem[\protect\citeauthoryear{Newman}{Newman}{2015b}]%
        {samn}
\bibfield{author}{\bibinfo{person}{Sam Newman}.}
  \bibinfo{year}{2015}\natexlab{b}.
\newblock \bibinfo{title}{Microservices talk with Sam Newman}.
\newblock \bibinfo{howpublished}{https://www.youtube.com/watch?v=GDVcUM5wbxU}.
\newblock


\bibitem[\protect\citeauthoryear{Nguyen and Nahrstedt}{Nguyen and
  Nahrstedt}{2017}]%
        {8005348}
\bibfield{author}{\bibinfo{person}{P. Nguyen} {and} \bibinfo{person}{K.
  Nahrstedt}.} \bibinfo{year}{2017}\natexlab{}.
\newblock \showarticletitle{MONAD: Self-Adaptive Micro-Service Infrastructure
  for Heterogeneous Scientific Workflows}. In \bibinfo{booktitle}{\emph{2017
  IEEE International Conference on Autonomic Computing (ICAC)}}.
  \bibinfo{pages}{187--196}.
\newblock
\urldef\tempurl%
\url{https://doi.org/10.1109/ICAC.2017.38}
\showDOI{\tempurl}


\bibitem[\protect\citeauthoryear{Niu, Liu, and Li}{Niu et~al\mbox{.}}{2018}]%
        {infocom}
\bibfield{author}{\bibinfo{person}{Yipei Niu}, \bibinfo{person}{Fangming Liu},
  {and} \bibinfo{person}{Zongpeng Li}.} \bibinfo{year}{2018}\natexlab{}.
\newblock \showarticletitle{Load Balancing across Microservices}. In
  \bibinfo{booktitle}{\emph{IEEE International Conference on Computer
  Communications}}.
\newblock


\bibitem[\protect\citeauthoryear{Novatec}{Novatec}{2017}]%
        {cdc}
\bibfield{author}{\bibinfo{person}{Novatec}.} \bibinfo{year}{2017}\natexlab{}.
\newblock \bibinfo{title}{Introduction to Microservices Testing and Consumer
  Driven Contract Testing with PACT}.
\newblock
  \bibinfo{howpublished}{https://blog.novatec-gmbh.de/introduction-microservices-testing-consumer-driven-contract-testing-pact/}.
\newblock


\bibitem[\protect\citeauthoryear{Nygard}{Nygard}{2007}]%
        {release}
\bibfield{author}{\bibinfo{person}{Michael Nygard}.}
  \bibinfo{year}{2007}\natexlab{}.
\newblock \bibinfo{booktitle}{\emph{Release It! Design and Deploy
  Production-Ready Software}}.
\newblock \bibinfo{publisher}{Pragmatic Bookshelf}.
\newblock


\bibitem[\protect\citeauthoryear{OASIS}{OASIS}{2018}]%
        {amqp}
\bibfield{author}{\bibinfo{person}{OASIS}.} \bibinfo{year}{2018}\natexlab{}.
\newblock \bibinfo{title}{AMQP is the Internet Protocol for Business
  Messaging}.
\newblock \bibinfo{howpublished}{https://www.amqp.org/about/what}.
\newblock


\bibitem[\protect\citeauthoryear{O'Connor, Elger, and Clarke}{O'Connor
  et~al\mbox{.}}{[n. d.]}]%
        {doi:10.1002/smr.1866}
\bibfield{author}{\bibinfo{person}{Rory~V. O'Connor}, \bibinfo{person}{Peter
  Elger}, {and} \bibinfo{person}{Paul~M. Clarke}.} \bibinfo{year}{[n.
  d.]}\natexlab{}.
\newblock \showarticletitle{Continuous software engineering --- A microservices
  architecture perspective}.
\newblock \bibinfo{journal}{\emph{Journal of Software: Evolution and Process}}
  \bibinfo{volume}{29}, \bibinfo{number}{11} (\bibinfo{year}{[n. d.]}),
  \bibinfo{pages}{e1866}.
\newblock
\urldef\tempurl%
\url{https://doi.org/10.1002/smr.1866}
\showDOI{\tempurl}
\showeprint{https://onlinelibrary.wiley.com/doi/pdf/10.1002/smr.1866}


\bibitem[\protect\citeauthoryear{Oksa}{Oksa}{2016}]%
        {pop00190}
\bibfield{author}{\bibinfo{person}{M Oksa}.} \bibinfo{year}{2016}\natexlab{}.
\newblock \emph{\bibinfo{title}{Web API development and integration for
  microservice functionality in web applications}}.
\newblock \bibinfo{thesistype}{Master's\ thesis}. \bibinfo{school}{University
  of Jyväskylä}.
\newblock
\urldef\tempurl%
\url{http://urn.fi/URN:NBN:fi:jyu-201612215220}
\showURL{%
\tempurl}


\bibitem[\protect\citeauthoryear{Oksanych, Shevchenko, Shcherbak, and
  Shcherbak}{Oksanych et~al\mbox{.}}{2017}]%
        {pop00011}
\bibfield{author}{\bibinfo{person}{I Oksanych}, \bibinfo{person}{I Shevchenko},
  \bibinfo{person}{I Shcherbak}, {and} \bibinfo{person}{S. Shcherbak}.}
  \bibinfo{year}{2017}\natexlab{}.
\newblock \showarticletitle{Development of specialized services for predicting
  the business activity indicators based on micro-service architecture}.
\newblock \bibinfo{journal}{\emph{Information Technology}}
  (\bibinfo{year}{2017}).
\newblock


\bibitem[\protect\citeauthoryear{Olliffe}{Olliffe}{2015}]%
        {guts}
\bibfield{author}{\bibinfo{person}{Gary Olliffe}.}
  \bibinfo{year}{2015}\natexlab{}.
\newblock \bibinfo{title}{Microservices : Building Services with the Guts on
  the Outside}.
\newblock
  \bibinfo{howpublished}{https://blogs.gartner.com/gary-olliffe/2015/01/30/microservices-guts-on-the-outside/}.
\newblock


\bibitem[\protect\citeauthoryear{Oreizy}{Oreizy}{1998}]%
        {sysreconfig}
\bibfield{author}{\bibinfo{person}{P. Oreizy}.}
  \bibinfo{year}{1998}\natexlab{}.
\newblock \showarticletitle{On the role of software architectures in runtime
  system reconfiguration}.
\newblock \bibinfo{journal}{\emph{IEE Proceedings - Software}}
  \bibinfo{volume}{145} (\bibinfo{date}{October} \bibinfo{year}{1998}),
  \bibinfo{pages}{137--145(8)}.
\newblock
Issue 5.
\showISSN{1462-5970}
\urldef\tempurl%
\url{http://digital-library.theiet.org/content/journals/10.1049/ip-sen_19982296}
\showURL{%
\tempurl}


\bibitem[\protect\citeauthoryear{Oreizy, Medvidovic, and Taylor}{Oreizy
  et~al\mbox{.}}{1998}]%
        {Oreizy:1998:ARS:302163.302181}
\bibfield{author}{\bibinfo{person}{Peyman Oreizy}, \bibinfo{person}{Nenad
  Medvidovic}, {and} \bibinfo{person}{Richard~N. Taylor}.}
  \bibinfo{year}{1998}\natexlab{}.
\newblock \showarticletitle{Architecture-based Runtime Software Evolution}. In
  \bibinfo{booktitle}{\emph{Proceedings of the 20th International Conference on
  Software Engineering}} \emph{(\bibinfo{series}{ICSE '98})}.
  \bibinfo{publisher}{IEEE Computer Society}, \bibinfo{address}{Washington, DC,
  USA}, \bibinfo{pages}{177--186}.
\newblock
\showISBNx{0-8186-8368-6}
\urldef\tempurl%
\url{http://dl.acm.org/citation.cfm?id=302163.302181}
\showURL{%
\tempurl}


\bibitem[\protect\citeauthoryear{Ozkaya, Kazman, and Klein}{Ozkaya
  et~al\mbox{.}}{2007}]%
        {ipek}
\bibfield{author}{\bibinfo{person}{Ipek Ozkaya}, \bibinfo{person}{Rick Kazman},
  {and} \bibinfo{person}{Mark Klein}.} \bibinfo{year}{2007}\natexlab{}.
\newblock \bibinfo{booktitle}{\emph{Quality-Attribute-Based Economic Valuation
  of Architectural Patterns}}.
\newblock \bibinfo{type}{{T}echnical {R}eport} CMU/SEI-2007-TR-003.
  \bibinfo{institution}{Software Engineering Institute, Carnegie Mellon
  University}, \bibinfo{address}{Pittsburgh, PA}.
\newblock


\bibitem[\protect\citeauthoryear{Pahl and Jamshidi}{Pahl and Jamshidi}{2016}]%
        {Pahl:2016:MSM:3021834.3021846}
\bibfield{author}{\bibinfo{person}{Claus Pahl} {and} \bibinfo{person}{Pooyan
  Jamshidi}.} \bibinfo{year}{2016}\natexlab{}.
\newblock \showarticletitle{Microservices: A Systematic Mapping Study}. In
  \bibinfo{booktitle}{\emph{Proceedings of the 6th International Conference on
  Cloud Computing and Services Science - Volume 1 and 2}}
  \emph{(\bibinfo{series}{CLOSER 2016})}. \bibinfo{publisher}{SCITEPRESS -
  Science and Technology Publications, Lda}, \bibinfo{address}{Portugal},
  \bibinfo{pages}{137--146}.
\newblock
\showISBNx{978-989-758-182-3}
\urldef\tempurl%
\url{https://doi.org/10.5220/0005785501370146}
\showDOI{\tempurl}


\bibitem[\protect\citeauthoryear{Papatheocharous, Andersson, and
  Axelsson}{Papatheocharous et~al\mbox{.}}{2015}]%
        {10.1007/978-3-319-19593-3_7}
\bibfield{author}{\bibinfo{person}{Efi Papatheocharous},
  \bibinfo{person}{Jesper Andersson}, {and} \bibinfo{person}{Jakob Axelsson}.}
  \bibinfo{year}{2015}\natexlab{}.
\newblock \showarticletitle{Ecosystems and Open Innovation for Embedded
  Systems: A Systematic Mapping Study}. In \bibinfo{booktitle}{\emph{Software
  Business}}, \bibfield{editor}{\bibinfo{person}{Jo{\~a}o~M. Fernandes},
  \bibinfo{person}{Ricardo~J. Machado}, {and} \bibinfo{person}{Krzysztof Wnuk}}
  (Eds.). \bibinfo{publisher}{Springer International Publishing},
  \bibinfo{address}{Cham}, \bibinfo{pages}{81--95}.
\newblock
\showISBNx{978-3-319-19593-3}


\bibitem[\protect\citeauthoryear{Papazoglou, Traverso, Dustdar, Leymann, and
  Kr{\"a}mer}{Papazoglou et~al\mbox{.}}{2006}]%
        {Papazoglou2006ServiceOrientedCR}
\bibfield{author}{\bibinfo{person}{Michael Papazoglou}, \bibinfo{person}{Paolo
  Traverso}, \bibinfo{person}{Schahram Dustdar}, \bibinfo{person}{Frank
  Leymann}, {and} \bibinfo{person}{Bernd~J. Kr{\"a}mer}.}
  \bibinfo{year}{2006}\natexlab{}.
\newblock \showarticletitle{Service-Oriented Computing Research Roadmap}.
\newblock


\bibitem[\protect\citeauthoryear{Papazoglou}{Papazoglou}{2003}]%
        {1254461}
\bibfield{author}{\bibinfo{person}{M.~P. Papazoglou}.}
  \bibinfo{year}{2003}\natexlab{}.
\newblock \showarticletitle{Service-oriented computing: concepts,
  characteristics and directions}. In \bibinfo{booktitle}{\emph{Proceedings of
  the Fourth International Conference on Web Information Systems Engineering,
  2003. WISE 2003.}} \bibinfo{pages}{3--12}.
\newblock
\urldef\tempurl%
\url{https://doi.org/10.1109/WISE.2003.1254461}
\showDOI{\tempurl}


\bibitem[\protect\citeauthoryear{Papazoglou and Georgakopoulos}{Papazoglou and
  Georgakopoulos}{2003}]%
        {Papazoglou:2003:ISC:944217.944233}
\bibfield{author}{\bibinfo{person}{M.~P. Papazoglou} {and} \bibinfo{person}{D.
  Georgakopoulos}.} \bibinfo{year}{2003}\natexlab{}.
\newblock \showarticletitle{Introduction: Service-oriented Computing}.
\newblock \bibinfo{journal}{\emph{Commun. ACM}} \bibinfo{volume}{46},
  \bibinfo{number}{10} (\bibinfo{date}{Oct.} \bibinfo{year}{2003}),
  \bibinfo{pages}{24--28}.
\newblock
\showISSN{0001-0782}
\urldef\tempurl%
\url{https://doi.org/10.1145/944217.944233}
\showDOI{\tempurl}


\bibitem[\protect\citeauthoryear{Pautasso, Zimmermann, Amundsen, Lewis, and
  Josuttis}{Pautasso et~al\mbox{.}}{2017}]%
        {7819415}
\bibfield{author}{\bibinfo{person}{C. Pautasso}, \bibinfo{person}{O.
  Zimmermann}, \bibinfo{person}{M. Amundsen}, \bibinfo{person}{J. Lewis}, {and}
  \bibinfo{person}{N. Josuttis}.} \bibinfo{year}{2017}\natexlab{}.
\newblock \showarticletitle{Microservices in Practice, Part 1: Reality Check
  and Service Design}.
\newblock \bibinfo{journal}{\emph{IEEE Software}} \bibinfo{volume}{34},
  \bibinfo{number}{1} (\bibinfo{date}{Jan} \bibinfo{year}{2017}),
  \bibinfo{pages}{91--98}.
\newblock
\showISSN{0740-7459}
\urldef\tempurl%
\url{https://doi.org/10.1109/MS.2017.24}
\showDOI{\tempurl}


\bibitem[\protect\citeauthoryear{Penchikala}{Penchikala}{2017}]%
        {kaiser}
\bibfield{author}{\bibinfo{person}{Srini Penchikala}.}
  \bibinfo{year}{2017}\natexlab{}.
\newblock \bibinfo{title}{Susanne Kaiser on Microservices Journey from a
  Startup Perspective}.
\newblock
  \bibinfo{howpublished}{https://www.infoq.com/news/2017/07/kaiser-microservices-journey}.
\newblock


\bibitem[\protect\citeauthoryear{Perish}{Perish}{1999}]%
        {perish}
\bibfield{author}{\bibinfo{person}{Perish}.} \bibinfo{year}{1999}\natexlab{}.
\newblock \bibinfo{title}{Publish or Perish}.
\newblock
  \bibinfo{howpublished}{https://harzing.com/resources/publish-or-perish}.
\newblock


\bibitem[\protect\citeauthoryear{Perrey and Lycett}{Perrey and Lycett}{2003}]%
        {1210138}
\bibfield{author}{\bibinfo{person}{R. Perrey} {and} \bibinfo{person}{M.
  Lycett}.} \bibinfo{year}{2003}\natexlab{}.
\newblock \showarticletitle{Service-oriented architecture}. In
  \bibinfo{booktitle}{\emph{2003 Symposium on Applications and the Internet
  Workshops, 2003. Proceedings.}} \bibinfo{pages}{116--119}.
\newblock
\urldef\tempurl%
\url{https://doi.org/10.1109/SAINTW.2003.1210138}
\showDOI{\tempurl}


\bibitem[\protect\citeauthoryear{Perry, Porter, and Votta}{Perry
  et~al\mbox{.}}{2000}]%
        {Perry:2000:ESS:336512.336586}
\bibfield{author}{\bibinfo{person}{Dewayne~E. Perry}, \bibinfo{person}{Adam~A.
  Porter}, {and} \bibinfo{person}{Lawrence~G. Votta}.}
  \bibinfo{year}{2000}\natexlab{}.
\newblock \showarticletitle{Empirical Studies of Software Engineering: A
  Roadmap}. In \bibinfo{booktitle}{\emph{Proceedings of the Conference on The
  Future of Software Engineering}} \emph{(\bibinfo{series}{ICSE '00})}.
  \bibinfo{publisher}{ACM}, \bibinfo{address}{New York, NY, USA},
  \bibinfo{pages}{345--355}.
\newblock
\showISBNx{1-58113-253-0}
\urldef\tempurl%
\url{https://doi.org/10.1145/336512.336586}
\showDOI{\tempurl}


\bibitem[\protect\citeauthoryear{Petersen, Feldt, Mujtaba, and
  Mattsson}{Petersen et~al\mbox{.}}{2008}]%
        {Petersen:2008:SMS:2227115.2227123}
\bibfield{author}{\bibinfo{person}{Kai Petersen}, \bibinfo{person}{Robert
  Feldt}, \bibinfo{person}{Shahid Mujtaba}, {and} \bibinfo{person}{Michael
  Mattsson}.} \bibinfo{year}{2008}\natexlab{}.
\newblock \showarticletitle{Systematic Mapping Studies in Software
  Engineering}. In \bibinfo{booktitle}{\emph{Proceedings of the 12th
  International Conference on Evaluation and Assessment in Software
  Engineering}} \emph{(\bibinfo{series}{EASE'08})}. \bibinfo{publisher}{BCS
  Learning \& Development Ltd.}, \bibinfo{address}{Swindon, UK},
  \bibinfo{pages}{68--77}.
\newblock
\urldef\tempurl%
\url{http://dl.acm.org/citation.cfm?id=2227115.2227123}
\showURL{%
\tempurl}


\bibitem[\protect\citeauthoryear{Petersen, Vakkalanka, and Kuzniarz}{Petersen
  et~al\mbox{.}}{2015}]%
        {PETERSEN20151}
\bibfield{author}{\bibinfo{person}{Kai Petersen}, \bibinfo{person}{Sairam
  Vakkalanka}, {and} \bibinfo{person}{Ludwik Kuzniarz}.}
  \bibinfo{year}{2015}\natexlab{}.
\newblock \showarticletitle{Guidelines for conducting systematic mapping
  studies in software engineering: An update}.
\newblock \bibinfo{journal}{\emph{Information and Software Technology}}
  \bibinfo{volume}{64} (\bibinfo{year}{2015}), \bibinfo{pages}{1 -- 18}.
\newblock
\showISSN{0950-5849}
\urldef\tempurl%
\url{https://doi.org/10.1016/j.infsof.2015.03.007}
\showDOI{\tempurl}


\bibitem[\protect\citeauthoryear{Porrmann, Essmann, and Colombo}{Porrmann
  et~al\mbox{.}}{2017}]%
        {8216583}
\bibfield{author}{\bibinfo{person}{T. Porrmann}, \bibinfo{person}{R. Essmann},
  {and} \bibinfo{person}{A.~W. Colombo}.} \bibinfo{year}{2017}\natexlab{}.
\newblock \showarticletitle{Development of an event-oriented, cloud-based SCADA
  system using a microservice architecture under the RAMI4.0 specification:
  Lessons learned}. In \bibinfo{booktitle}{\emph{IECON 2017 - 43rd Annual
  Conference of the IEEE Industrial Electronics Society}}.
  \bibinfo{pages}{3441--3448}.
\newblock
\urldef\tempurl%
\url{https://doi.org/10.1109/IECON.2017.8216583}
\showDOI{\tempurl}


\bibitem[\protect\citeauthoryear{Posta}{Posta}{2017a}]%
        {hardest}
\bibfield{author}{\bibinfo{person}{Christian Posta}.}
  \bibinfo{year}{2017}\natexlab{a}.
\newblock \bibinfo{title}{The Hardest Part of Microservices: Calling Your
  Services}.
\newblock
  \bibinfo{howpublished}{http://blog.christianposta.com/microservices/the-hardest-part-of-microservices-calling-your-services/}.
\newblock


\bibitem[\protect\citeauthoryear{Posta}{Posta}{2017b}]%
        {micropart2}
\bibfield{author}{\bibinfo{person}{Christian Posta}.}
  \bibinfo{year}{2017}\natexlab{b}.
\newblock \bibinfo{title}{Low-risk Monolith to Microservice Evolution Part II}.
\newblock
  \bibinfo{howpublished}{http://blog.christianposta.com/microservices/low-risk-monolith-to-microservice-evolution-part-ii/}.
\newblock


\bibitem[\protect\citeauthoryear{Probst and Becker}{Probst and Becker}{2016}]%
        {netflixeng}
\bibfield{author}{\bibinfo{person}{Katharina Probst} {and}
  \bibinfo{person}{Justin Becker}.} \bibinfo{year}{2016}\natexlab{}.
\newblock \bibinfo{title}{Engineering Trade-Offs and The Netflix API
  Re-Architecture}.
\newblock
  \bibinfo{howpublished}{https://medium.com/netflix-techblog/engineering-trade-offs-and-the-netflix-api-re-architecture-64f122b277dd}.
\newblock


\bibitem[\protect\citeauthoryear{Rademacher, Sachweh, and Zündorf}{Rademacher
  et~al\mbox{.}}{2017}]%
        {7958454}
\bibfield{author}{\bibinfo{person}{F. Rademacher}, \bibinfo{person}{S.
  Sachweh}, {and} \bibinfo{person}{A. Zündorf}.}
  \bibinfo{year}{2017}\natexlab{}.
\newblock \showarticletitle{Differences between Model-Driven Development of
  Service-Oriented and Microservice Architecture}. In
  \bibinfo{booktitle}{\emph{2017 IEEE International Conference on Software
  Architecture Workshops (ICSAW)}}. \bibinfo{pages}{38--45}.
\newblock
\urldef\tempurl%
\url{https://doi.org/10.1109/ICSAW.2017.32}
\showDOI{\tempurl}


\bibitem[\protect\citeauthoryear{Rademacher, Sachweh, and
  Z{\"u}ndorf}{Rademacher et~al\mbox{.}}{2018}]%
        {10.1007/978-3-319-74781-1_17}
\bibfield{author}{\bibinfo{person}{Florian Rademacher}, \bibinfo{person}{Sabine
  Sachweh}, {and} \bibinfo{person}{Albert Z{\"u}ndorf}.}
  \bibinfo{year}{2018}\natexlab{}.
\newblock \showarticletitle{Towards a UML Profile for Domain-Driven Design of
  Microservice Architectures}. In \bibinfo{booktitle}{\emph{Software
  Engineering and Formal Methods}}, \bibfield{editor}{\bibinfo{person}{Antonio
  Cerone} {and} \bibinfo{person}{Marco Roveri}} (Eds.).
  \bibinfo{publisher}{Springer International Publishing},
  \bibinfo{address}{Cham}, \bibinfo{pages}{230--245}.
\newblock
\showISBNx{978-3-319-74781-1}


\bibitem[\protect\citeauthoryear{Rajagopalan and Jamjoom}{Rajagopalan and
  Jamjoom}{2015}]%
        {190613}
\bibfield{author}{\bibinfo{person}{Shriram Rajagopalan} {and}
  \bibinfo{person}{Hani Jamjoom}.} \bibinfo{year}{2015}\natexlab{}.
\newblock \showarticletitle{App{\textendash}Bisect: Autonomous Healing for
  Microservice-Based Apps}. In \bibinfo{booktitle}{\emph{7th {USENIX} Workshop
  on Hot Topics in Cloud Computing (HotCloud 15)}}.
  \bibinfo{publisher}{{USENIX} Association}, \bibinfo{address}{Santa Clara,
  CA}.
\newblock
\urldef\tempurl%
\url{https://www.usenix.org/conference/hotcloud15/workshop-program/presentation/rajagopalan}
\showURL{%
\tempurl}


\bibitem[\protect\citeauthoryear{Rajasekar, Terrell Russell~and, de~Torcy, Xu,
  Wan, Moore, Schroeder, Chen, Conway, and Ward}{Rajasekar
  et~al\mbox{.}}{2015}]%
        {RODS}
\bibfield{author}{\bibinfo{person}{Arcot Rajasekar},
  \bibinfo{person}{Jason~Coposky Terrell Russell~and}, \bibinfo{person}{Antoine
  de Torcy}, \bibinfo{person}{Hao Xu}, \bibinfo{person}{Michael Wan},
  \bibinfo{person}{Reagan~W. Moore}, \bibinfo{person}{Wayne Schroeder},
  \bibinfo{person}{Sheau-Yen Chen}, \bibinfo{person}{Mike Conway}, {and}
  \bibinfo{person}{Jewel~H. Ward}.} \bibinfo{year}{2015}\natexlab{}.
\newblock \bibinfo{booktitle}{\emph{The integrated Rule-Oriented Data System
  (iRODS 4.0) Microservice Workbook} (\bibinfo{edition}{first} ed.)}.
\newblock \bibinfo{publisher}{CreateSpace Independent Publishing Platform}.
\newblock


\bibitem[\protect\citeauthoryear{Reinhold}{Reinhold}{2016}]%
        {uber}
\bibfield{author}{\bibinfo{person}{Emily Reinhold}.}
  \bibinfo{year}{2016}\natexlab{}.
\newblock \bibinfo{title}{Lessons Learned on Uber's Journey into
  Microservices}.
\newblock
  \bibinfo{howpublished}{https://www.infoq.com/presentations/uber-darwin}.
\newblock


\bibitem[\protect\citeauthoryear{Richards}{Richards}{2016}]%
        {markrich}
\bibfield{author}{\bibinfo{person}{Mark Richards}.}
  \bibinfo{year}{2016}\natexlab{}.
\newblock \bibinfo{booktitle}{\emph{Microservices vs. Service-Oriented
  Architecture}}.
\newblock \bibinfo{publisher}{O Reilly Media}.
\newblock


\bibitem[\protect\citeauthoryear{Richardson}{Richardson}{2014}]%
        {chris}
\bibfield{author}{\bibinfo{person}{Chris Richardson}.}
  \bibinfo{year}{2014}\natexlab{}.
\newblock \bibinfo{title}{A pattern language for microservices}.
\newblock \bibinfo{howpublished}{http://microservices.io/patterns/index.html}.
\newblock


\bibitem[\protect\citeauthoryear{Richardson}{Richardson}{2018}]%
        {richardson2018microservice}
\bibfield{author}{\bibinfo{person}{C. Richardson}.}
  \bibinfo{year}{2018}\natexlab{}.
\newblock \bibinfo{booktitle}{\emph{Microservice Patterns}}.
\newblock \bibinfo{publisher}{Manning Publications Company}.
\newblock
\showISBNx{9781617294549}
\urldef\tempurl%
\url{https://books.google.co.uk/books?id=UeK1swEACAAJ}
\showURL{%
\tempurl}


\bibitem[\protect\citeauthoryear{Richardson and Smith}{Richardson and
  Smith}{2016}]%
        {nginx}
\bibfield{author}{\bibinfo{person}{Chris Richardson} {and}
  \bibinfo{person}{Floyd Smith}.} \bibinfo{year}{2016}\natexlab{}.
\newblock \bibinfo{title}{Microservices: From Design to Deployment}.
\newblock


\bibitem[\protect\citeauthoryear{Richardson, Wolff, Miles, Kaiser, Newman, and
  Tilkov}{Richardson et~al\mbox{.}}{2016}]%
        {panel}
\bibfield{author}{\bibinfo{person}{C. Richardson}, \bibinfo{person}{E. Wolff},
  \bibinfo{person}{R. Miles}, \bibinfo{person}{S. Kaiser}, \bibinfo{person}{S.
  Newman}, {and} \bibinfo{person}{S. Tilkov}.} \bibinfo{year}{2016}\natexlab{}.
\newblock \bibinfo{title}{microXchg 2016 - C. Richardson, E. Wolff, R. Miles,
  S. Kaiser, S. Newman, S. Tilkov}.
\newblock
\newblock
\urldef\tempurl%
\url{https://www.youtube.com/watch?v=wHuI7C3-Eis\&list=PLx2By31njbhrs8caX08BEusyD_fBDu-XG\&feature=player_detailpage}
\showURL{%
\tempurl}


\bibitem[\protect\citeauthoryear{Richter, Konrad, Utecht, and Polze}{Richter
  et~al\mbox{.}}{2017}]%
        {8004304}
\bibfield{author}{\bibinfo{person}{D. Richter}, \bibinfo{person}{M. Konrad},
  \bibinfo{person}{K. Utecht}, {and} \bibinfo{person}{A. Polze}.}
  \bibinfo{year}{2017}\natexlab{}.
\newblock \showarticletitle{Highly-Available Applications on Unreliable
  Infrastructure: Microservice Architectures in Practice}. In
  \bibinfo{booktitle}{\emph{2017 IEEE International Conference on Software
  Quality, Reliability and Security Companion (QRS-C)}}.
  \bibinfo{pages}{130--137}.
\newblock
\urldef\tempurl%
\url{https://doi.org/10.1109/QRS-C.2017.28}
\showDOI{\tempurl}


\bibitem[\protect\citeauthoryear{Riegg~Cellini and Edwin~Kee}{Riegg~Cellini and
  Edwin~Kee}{2015}]%
        {costeff}
\bibfield{author}{\bibinfo{person}{Stephanie Riegg~Cellini} {and}
  \bibinfo{person}{James Edwin~Kee}.} \bibinfo{year}{2015}\natexlab{}.
\newblock \bibinfo{booktitle}{\emph{Cost-Effectiveness and Cost-Benefit
  Analysis}}.
\newblock \bibinfo{publisher}{JohnWileySons}, \bibinfo{pages}{636--672}.
\newblock
\showISBNx{9781119171386}
\urldef\tempurl%
\url{https://doi.org/10.1002/9781119171386.ch24}
\showDOI{\tempurl}


\bibitem[\protect\citeauthoryear{Rodger}{Rodger}{2017}]%
        {rodger2017tao}
\bibfield{author}{\bibinfo{person}{R. Rodger}.}
  \bibinfo{year}{2017}\natexlab{}.
\newblock \bibinfo{booktitle}{\emph{The Tao of Microservices}}.
\newblock \bibinfo{publisher}{Manning Publications Company}.
\newblock
\showISBNx{9781617293146}
\urldef\tempurl%
\url{https://books.google.co.uk/books?id=uosOkAEACAAJ}
\showURL{%
\tempurl}


\bibitem[\protect\citeauthoryear{Rombach}{Rombach}{1987}]%
        {1702220}
\bibfield{author}{\bibinfo{person}{H.D. Rombach}.}
  \bibinfo{year}{1987}\natexlab{}.
\newblock \showarticletitle{A Controlled Expeniment on the Impact of Software
  Structure on Maintainability}.
\newblock \bibinfo{journal}{\emph{Software Engineering, IEEE Transactions on}}
  \bibinfo{volume}{SE-13}, \bibinfo{number}{3} (\bibinfo{date}{March}
  \bibinfo{year}{1987}), \bibinfo{pages}{344--354}.
\newblock
\showISSN{0098-5589}
\urldef\tempurl%
\url{https://doi.org/10.1109/TSE.1987.233165}
\showDOI{\tempurl}


\bibitem[\protect\citeauthoryear{Rosen, Lublinsky, Smith, and Balcer}{Rosen
  et~al\mbox{.}}{2012}]%
        {rosen2012applied}
\bibfield{author}{\bibinfo{person}{M. Rosen}, \bibinfo{person}{B. Lublinsky},
  \bibinfo{person}{K.T. Smith}, {and} \bibinfo{person}{M.J. Balcer}.}
  \bibinfo{year}{2012}\natexlab{}.
\newblock \bibinfo{booktitle}{\emph{Applied SOA: Service-Oriented Architecture
  and Design Strategies}}.
\newblock \bibinfo{publisher}{Wiley}.
\newblock
\showISBNx{9781118079799}
\urldef\tempurl%
\url{https://books.google.co.uk/books?id=GFL9lWKojFYC}
\showURL{%
\tempurl}


\bibitem[\protect\citeauthoryear{Sacolick}{Sacolick}{2018}]%
        {cicdterms}
\bibfield{author}{\bibinfo{person}{Isaac Sacolick}.}
  \bibinfo{year}{2018}\natexlab{}.
\newblock \bibinfo{title}{What is CI/CD? Continuous integration and continuous
  delivery explained}.
\newblock
  \bibinfo{howpublished}{https://www.infoworld.com/article/3271126/ci-cd/what-is-cicd-continuous-integration-and-continuous-delivery-explained.html}.
\newblock


\bibitem[\protect\citeauthoryear{Sader}{Sader}{2015}]%
        {refarch}
\bibfield{author}{\bibinfo{person}{Yamen Sader}.}
  \bibinfo{year}{2015}\natexlab{}.
\newblock \bibinfo{title}{A Microservices Reference Architecture}.
\newblock \bibinfo{howpublished}{https://www.youtube.com/watch?v=KHqMPRA6jVI}.
\newblock


\bibitem[\protect\citeauthoryear{Salah, Zemerly, Yeun, Al-Qutayri, and
  Al-Hammadi}{Salah et~al\mbox{.}}{2016}]%
        {7856721}
\bibfield{author}{\bibinfo{person}{T. Salah}, \bibinfo{person}{M.~Jamal
  Zemerly}, \bibinfo{person}{Chan~Yeob Yeun}, \bibinfo{person}{M. Al-Qutayri},
  {and} \bibinfo{person}{Y. Al-Hammadi}.} \bibinfo{year}{2016}\natexlab{}.
\newblock \showarticletitle{The evolution of distributed systems towards
  microservices architecture}. In \bibinfo{booktitle}{\emph{2016 11th
  International Conference for Internet Technology and Secured Transactions
  (ICITST)}}. \bibinfo{pages}{318--325}.
\newblock
\urldef\tempurl%
\url{https://doi.org/10.1109/ICITST.2016.7856721}
\showDOI{\tempurl}


\bibitem[\protect\citeauthoryear{Salvadori, Oliveira, Huf, Inacio, and
  Siqueira}{Salvadori et~al\mbox{.}}{2017}]%
        {Salvadori:2017:OAF:3151759.3151793}
\bibfield{author}{\bibinfo{person}{Ivan Salvadori}, \bibinfo{person}{Bruno
  C.~N. Oliveira}, \bibinfo{person}{Alexis Huf}, \bibinfo{person}{Eduardo~C.
  Inacio}, {and} \bibinfo{person}{Frank Siqueira}.}
  \bibinfo{year}{2017}\natexlab{}.
\newblock \showarticletitle{An Ontology Alignment Framework for Data-driven
  Microservices}. In \bibinfo{booktitle}{\emph{Proceedings of the 19th
  International Conference on Information Integration and Web-based
  Applications \& Services}} \emph{(\bibinfo{series}{iiWAS '17})}.
  \bibinfo{publisher}{ACM}, \bibinfo{address}{New York, NY, USA},
  \bibinfo{pages}{425--433}.
\newblock
\showISBNx{978-1-4503-5299-4}
\urldef\tempurl%
\url{https://doi.org/10.1145/3151759.3151793}
\showDOI{\tempurl}


\bibitem[\protect\citeauthoryear{Sama, Elbaum, Raimondi, Rosenblum, and
  Wang}{Sama et~al\mbox{.}}{2010}]%
        {caaa}
\bibfield{author}{\bibinfo{person}{Michele Sama}, \bibinfo{person}{Sebastian
  Elbaum}, \bibinfo{person}{Franco Raimondi}, \bibinfo{person}{David~S.
  Rosenblum}, {and} \bibinfo{person}{Zhimin Wang}.}
  \bibinfo{year}{2010}\natexlab{}.
\newblock \bibinfo{title}{Context-Aware Adaptive Applications: Fault Patterns
  and Their Automated Identification}.
\newblock , \bibinfo{numpages}{18}~pages.
\newblock
\showISSN{0098-5589}
\urldef\tempurl%
\url{https://doi.org/10.1109/TSE.2010.35}
\showDOI{\tempurl}


\bibitem[\protect\citeauthoryear{Sama, Rosenblum, Wang, and Elbaum}{Sama
  et~al\mbox{.}}{2008}]%
        {layers}
\bibfield{author}{\bibinfo{person}{Michele Sama}, \bibinfo{person}{David~S.
  Rosenblum}, \bibinfo{person}{Zhimin Wang}, {and} \bibinfo{person}{Sebastian
  Elbaum}.} \bibinfo{year}{2008}\natexlab{}.
\newblock \showarticletitle{Multi-layer Faults in the Architectures of Mobile,
  Context-aware Adaptive Applications: A Position Paper}. In
  \bibinfo{booktitle}{\emph{Proceedings of the 1st International Workshop on
  Software Architectures and Mobility}} \emph{(\bibinfo{series}{SAM '08})}.
  \bibinfo{publisher}{ACM}, \bibinfo{address}{New York, NY, USA},
  \bibinfo{pages}{47--49}.
\newblock
\showISBNx{978-1-60558-022-7}
\urldef\tempurl%
\url{https://doi.org/10.1145/1370888.1370901}
\showDOI{\tempurl}


\bibitem[\protect\citeauthoryear{Sampaio, Kadiyala, Hu, Steinbacher, Erwin,
  Rosa, Beschastnikh, and Rubin}{Sampaio et~al\mbox{.}}{2017}]%
        {Sampaio2017SupportingME}
\bibfield{author}{\bibinfo{person}{Adalberto~R. Sampaio},
  \bibinfo{person}{Harshavardhan Kadiyala}, \bibinfo{person}{Bo Hu},
  \bibinfo{person}{John Steinbacher}, \bibinfo{person}{Tony Erwin},
  \bibinfo{person}{Nelson Rosa}, \bibinfo{person}{Ivan Beschastnikh}, {and}
  \bibinfo{person}{Julia Rubin}.} \bibinfo{year}{2017}\natexlab{}.
\newblock \showarticletitle{Supporting Microservice Evolution}.
\newblock \bibinfo{journal}{\emph{2017 IEEE International Conference on
  Software Maintenance and Evolution (ICSME)}} (\bibinfo{year}{2017}),
  \bibinfo{pages}{539--543}.
\newblock


\bibitem[\protect\citeauthoryear{Sanjeev~Shareema}{Sanjeev~Shareema}{[n. d.]}]%
        {ibmdevops}
\bibfield{author}{\bibinfo{person}{Bernie~Coyne Sanjeev~Shareema}.}
  \bibinfo{year}{[n. d.]}\natexlab{}.
\newblock \bibinfo{title}{DevOps for Dummies - 3rd IBM Limited Edition}.
\newblock
  \bibinfo{howpublished}{https://www.ibm.com/ibm/devops/us/en/resources/dummiesbooks/}.
\newblock


\bibitem[\protect\citeauthoryear{Schaefer}{Schaefer}{2016}]%
        {zalando}
\bibfield{author}{\bibinfo{person}{Rodrigue Schaefer}.}
  \bibinfo{year}{2016}\natexlab{}.
\newblock \bibinfo{title}{GOTO 2016 - From Monolith to Microservices at Zalando
  - Rodrigue Schaefer}.
\newblock \bibinfo{howpublished}{https://youtu.be/gEeHZwjwehs}.
\newblock


\bibitem[\protect\citeauthoryear{Schaevitz}{Schaevitz}{2017}]%
        {205506}
\bibfield{author}{\bibinfo{person}{Samantha Schaevitz}.}
  \bibinfo{year}{2017}\natexlab{}.
\newblock \bibinfo{title}{Deploying Changes to Production in the Age of the
  Microservice}.
\newblock
  \bibinfo{howpublished}{https://www.usenix.org/conference/srecon17europe/program/presentation/schaevitz}.
\newblock


\bibitem[\protect\citeauthoryear{Schmidt, MacDonell, and Connor}{Schmidt
  et~al\mbox{.}}{2012}]%
        {Schmidt2012}
\bibfield{author}{\bibinfo{person}{Frederik Schmidt},
  \bibinfo{person}{Stephen~G. MacDonell}, {and} \bibinfo{person}{Andrew~M.
  Connor}.} \bibinfo{year}{2012}\natexlab{}.
\newblock \bibinfo{booktitle}{\emph{An Automatic Architecture Reconstruction
  and Refactoring Framework}}.
\newblock \bibinfo{publisher}{Springer Berlin Heidelberg},
  \bibinfo{address}{Berlin, Heidelberg}, \bibinfo{pages}{95--111}.
\newblock
\showISBNx{978-3-642-23202-2}
\urldef\tempurl%
\url{https://doi.org/10.1007/978-3-642-23202-2_7}
\showDOI{\tempurl}


\bibitem[\protect\citeauthoryear{Shahir~Daya}{Shahir~Daya}{2016}]%
        {ibmredbooks}
\bibfield{author}{\bibinfo{person}{Kameswara Eati Carlos M Ferreira Dejan
  Glozic Vasfi Gucer Manav Gupta Sunil Joshi Valerie Lampkin Marcelo Martins
  Shishir Narain Ramratan~Vennam Shahir~Daya, Nguyen Van~Duy}.}
  \bibinfo{year}{2016}\natexlab{}.
\newblock \bibinfo{title}{Microservices from Theory to Practice: Creating
  Applications in IBM Bluemix Using the Microservices Approach}.
\newblock
  \bibinfo{howpublished}{http://www.redbooks.ibm.com/abstracts/sg248275.html?Open}.
\newblock


\bibitem[\protect\citeauthoryear{Sharma}{Sharma}{2017}]%
        {sharma2017mastering}
\bibfield{author}{\bibinfo{person}{S. Sharma}.}
  \bibinfo{year}{2017}\natexlab{}.
\newblock \bibinfo{booktitle}{\emph{Mastering Microservices with Java 9: Build
  domain-driven microservice-based applications with Spring, Spring Cloud, and
  Angular}}.
\newblock \bibinfo{publisher}{Packt Publishing}.
\newblock
\showISBNx{9781787282414}
\urldef\tempurl%
\url{https://books.google.co.uk/books?id=WsxPDwAAQBAJ}
\showURL{%
\tempurl}


\bibitem[\protect\citeauthoryear{Sheppard}{Sheppard}{1990}]%
        {42969}
\bibfield{author}{\bibinfo{person}{M. Sheppard}.}
  \bibinfo{year}{1990}\natexlab{}.
\newblock \showarticletitle{Design metrics: an empirical analysis}.
\newblock \bibinfo{journal}{\emph{Software Engineering Journal}}
  \bibinfo{volume}{5}, \bibinfo{number}{1} (\bibinfo{date}{Jan}
  \bibinfo{year}{1990}), \bibinfo{pages}{3--10}.
\newblock
\showISSN{0268-6961}


\bibitem[\protect\citeauthoryear{Shoumik, Talukder, Jami, Protik, and
  Hoque}{Shoumik et~al\mbox{.}}{2017}]%
        {8281846}
\bibfield{author}{\bibinfo{person}{F.~S. Shoumik}, \bibinfo{person}{M.~I. M.~M.
  Talukder}, \bibinfo{person}{A.~I. Jami}, \bibinfo{person}{N.~W. Protik},
  {and} \bibinfo{person}{M.~M. Hoque}.} \bibinfo{year}{2017}\natexlab{}.
\newblock \showarticletitle{Scalable micro-service based approach to FHIR
  server with golang and No-SQL}. In \bibinfo{booktitle}{\emph{2017 20th
  International Conference of Computer and Information Technology (ICCIT)}}.
  \bibinfo{pages}{1--6}.
\newblock
\urldef\tempurl%
\url{https://doi.org/10.1109/ICCITECHN.2017.8281846}
\showDOI{\tempurl}


\bibitem[\protect\citeauthoryear{Shoup}{Shoup}{2016}]%
        {randy}
\bibfield{author}{\bibinfo{person}{Randy Shoup}.}
  \bibinfo{year}{2016}\natexlab{}.
\newblock \bibinfo{title}{GOTO 2016 - Pragmatic Microservices - Randy Shoup}.
\newblock \bibinfo{howpublished}{https://youtu.be/9vS7TbgirgY}.
\newblock


\bibitem[\protect\citeauthoryear{Simons}{Simons}{2016}]%
        {decoupled}
\bibfield{author}{\bibinfo{person}{David Simons}.}
  \bibinfo{year}{2016}\natexlab{}.
\newblock \bibinfo{title}{Decoupled APIs through Microservices}.
\newblock
  \bibinfo{howpublished}{https://www.infoq.com/presentations/api-microservices-tools}.
\newblock


\bibitem[\protect\citeauthoryear{Slack}{Slack}{2005}]%
        {SLACK2005}
\bibfield{author}{\bibinfo{person}{Nigel Slack}.}
  \bibinfo{year}{2005}\natexlab{}.
\newblock \showarticletitle{{Operations strategy: will it ever realize its
  potential?}}
\newblock \bibinfo{journal}{\emph{{Gestao \& Producao}}}  \bibinfo{volume}{12}
  (\bibinfo{date}{12} \bibinfo{year}{2005}), \bibinfo{pages}{323 -- 332}.
\newblock
\showISSN{0104-530X}
\urldef\tempurl%
\url{http://www.scielo.br/scielo.php?script=sci_arttext&pid=S0104-530X2005000300004&nrm=iso}
\showURL{%
\tempurl}


\bibitem[\protect\citeauthoryear{Soenen, Tavernier, Colle, and Pickavet}{Soenen
  et~al\mbox{.}}{2017}]%
        {8093034}
\bibfield{author}{\bibinfo{person}{T. Soenen}, \bibinfo{person}{W. Tavernier},
  \bibinfo{person}{D. Colle}, {and} \bibinfo{person}{M. Pickavet}.}
  \bibinfo{year}{2017}\natexlab{}.
\newblock \showarticletitle{Optimising microservice-based reliable NFV
  management orchestration architectures}. In \bibinfo{booktitle}{\emph{2017
  9th International Workshop on Resilient Networks Design and Modeling
  (RNDM)}}. \bibinfo{pages}{1--7}.
\newblock
\urldef\tempurl%
\url{https://doi.org/10.1109/RNDM.2017.8093034}
\showDOI{\tempurl}


\bibitem[\protect\citeauthoryear{Soldani, Tamburri, and Heuvel}{Soldani
  et~al\mbox{.}}{2018}]%
        {SOLDANI2018215}
\bibfield{author}{\bibinfo{person}{Jacopo Soldani},
  \bibinfo{person}{Damian~Andrew Tamburri}, {and} \bibinfo{person}{Willem-Jan
  Van~Den Heuvel}.} \bibinfo{year}{2018}\natexlab{}.
\newblock \showarticletitle{The pains and gains of microservices: A Systematic
  grey literature review}.
\newblock \bibinfo{journal}{\emph{Journal of Systems and Software}}
  \bibinfo{volume}{146} (\bibinfo{year}{2018}), \bibinfo{pages}{215 -- 232}.
\newblock
\showISSN{0164-1212}
\urldef\tempurl%
\url{https://doi.org/10.1016/j.jss.2018.09.082}
\showDOI{\tempurl}


\bibitem[\protect\citeauthoryear{Sorgalla}{Sorgalla}{[n. d.]}]%
        {pop00081}
\bibfield{author}{\bibinfo{person}{J Sorgalla}.} \bibinfo{year}{[n.
  d.]}\natexlab{}.
\newblock \showarticletitle{AjiL: A Graphical Modeling Language for the
  Development of Microservice Architectures}.
  \bibinfo{howpublished}{http://conf-micro.services/papers/Sorgalla.pdf}. In
  \bibinfo{booktitle}{\emph{Microservices}}.
\newblock


\bibitem[\protect\citeauthoryear{Stal}{Stal}{2014}]%
        {Stal201463}
\bibfield{author}{\bibinfo{person}{Michael Stal}.}
  \bibinfo{year}{2014}\natexlab{}.
\newblock \showarticletitle{Chapter 3 - Refactoring Software Architectures}.
\newblock In \bibinfo{booktitle}{\emph{Agile Software Architecture}},
  \bibfield{editor}{\bibinfo{person}{Muhammad~Ali Babar}, \bibinfo{person}{},
  \bibinfo{person}{Alan~W. Brown}, \bibinfo{person}{}, {and}
  \bibinfo{person}{Ivan Mistrik}} (Eds.). \bibinfo{publisher}{Morgan Kaufmann},
  \bibinfo{address}{Boston}, \bibinfo{pages}{63 -- 82}.
\newblock
\showISBNx{978-0-12-407772-0}
\urldef\tempurl%
\url{https://doi.org/10.1016/B978-0-12-407772-0.00003-4}
\showDOI{\tempurl}


\bibitem[\protect\citeauthoryear{Steinacker}{Steinacker}{2015}]%
        {ottodev}
\bibfield{author}{\bibinfo{person}{Guido Steinacker}.}
  \bibinfo{year}{2015}\natexlab{}.
\newblock \bibinfo{title}{On Monoliths and Microservices}.
\newblock
  \bibinfo{howpublished}{https://dev.otto.de/2015/09/30/on-monoliths-and-microservices/}.
\newblock


\bibitem[\protect\citeauthoryear{Stenberg}{Stenberg}{2014}]%
        {hexagon}
\bibfield{author}{\bibinfo{person}{Jan Stenberg}.}
  \bibinfo{year}{2014}\natexlab{}.
\newblock \bibinfo{title}{Exploring the Hexagonal Architecture}.
\newblock
  \bibinfo{howpublished}{https://www.infoq.com/news/2014/10/exploring-hexagonal-architecture}.
\newblock


\bibitem[\protect\citeauthoryear{Stenberg}{Stenberg}{2015}]%
        {amazonmicro}
\bibfield{author}{\bibinfo{person}{Jan Stenberg}.}
  \bibinfo{year}{2015}\natexlab{}.
\newblock \bibinfo{title}{Microservices and Teams at Amazon}.
\newblock
  \bibinfo{howpublished}{https://www.infoq.com/news/2015/12/microservices-amazon}.
\newblock


\bibitem[\protect\citeauthoryear{Stenberg}{Stenberg}{2017}]%
        {heritage}
\bibfield{author}{\bibinfo{person}{Jan Stenberg}.}
  \bibinfo{year}{2017}\natexlab{}.
\newblock \bibinfo{title}{About the SOA Heritage Impact on Microservices}.
\newblock
  \bibinfo{howpublished}{https://www.infoq.com/news/2017/11/soa-impact-microservices}.
\newblock


\bibitem[\protect\citeauthoryear{Stenberg}{Stenberg}{2018}]%
        {stratdecomp}
\bibfield{author}{\bibinfo{person}{Jan Stenberg}.}
  \bibinfo{year}{2018}\natexlab{}.
\newblock \bibinfo{title}{Strategies for Decomposing a System into
  Microservices}.
\newblock
  \bibinfo{howpublished}{https://www.infoq.com/news/2018/06/decomposing-system-microservices}.
\newblock


\bibitem[\protect\citeauthoryear{Sullivan, Chalasani, Jha, and
  Sazawal}{Sullivan et~al\mbox{.}}{1999}]%
        {kevin}
\bibfield{author}{\bibinfo{person}{Kevin~J Sullivan}, \bibinfo{person}{Prasad
  Chalasani}, \bibinfo{person}{Somesh Jha}, {and} \bibinfo{person}{Vibha
  Sazawal}.} \bibinfo{year}{1999}\natexlab{}.
\newblock \showarticletitle{Software Design as an Investment Activity: A Real
  Options Perspective}.
\newblock \bibinfo{journal}{\emph{Real options and business strategy:
  Applications to decision making}} \bibinfo{number}{10}
  (\bibinfo{year}{1999}), \bibinfo{pages}{215 --- 262}.
\newblock


\bibitem[\protect\citeauthoryear{Sullivan, Griswold, Cai, and Hallen}{Sullivan
  et~al\mbox{.}}{2001}]%
        {module}
\bibfield{author}{\bibinfo{person}{Kevin~J. Sullivan},
  \bibinfo{person}{William~G. Griswold}, \bibinfo{person}{Yuanfang Cai}, {and}
  \bibinfo{person}{Ben Hallen}.} \bibinfo{year}{2001}\natexlab{}.
\newblock \showarticletitle{The Structure and Value of Modularity in Software
  Design}. In \bibinfo{booktitle}{\emph{Proceedings of the 8th European
  Software Engineering Conference Held Jointly with 9th ACM SIGSOFT
  International Symposium on Foundations of Software Engineering}}
  \emph{(\bibinfo{series}{ESEC/FSE-9})}. \bibinfo{publisher}{ACM},
  \bibinfo{address}{New York, NY, USA}, \bibinfo{pages}{99--108}.
\newblock
\showISBNx{1-58113-390-1}
\urldef\tempurl%
\url{https://doi.org/10.1145/503209.503224}
\showDOI{\tempurl}


\bibitem[\protect\citeauthoryear{Terzić, Dimitrieski, Kordić~(Aleksić,
  Milosavljevic, and Luković}{Terzić et~al\mbox{.}}{2017}]%
        {microbuilder}
\bibfield{author}{\bibinfo{person}{Branko Terzić}, \bibinfo{person}{Vladimir
  Dimitrieski}, \bibinfo{person}{Slavica Kordić~(Aleksić},
  \bibinfo{person}{Gordana Milosavljevic}, {and} \bibinfo{person}{Ivan
  Luković}.} \bibinfo{year}{2017}\natexlab{}.
\newblock \showarticletitle{MicroBuilder: A Model-Driven Tool for the
  Specification of REST Microservice Architectures}.
\newblock  (\bibinfo{date}{03} \bibinfo{year}{2017}).
\newblock


\bibitem[\protect\citeauthoryear{ThoughtWorks}{ThoughtWorks}{2016}]%
        {otto}
\bibfield{author}{\bibinfo{person}{ThoughtWorks}.}
  \bibinfo{year}{2016}\natexlab{}.
\newblock \bibinfo{title}{Otto | From Legacy Systems to Fast and Flexible
  Platforms}.
\newblock \bibinfo{howpublished}{https://youtu.be/bSvjZi3WKZQ}.
\newblock


\bibitem[\protect\citeauthoryear{Tilkov}{Tilkov}{2016}]%
        {stefanpresent}
\bibfield{author}{\bibinfo{person}{Stefan Tilkov}.}
  \bibinfo{year}{2016}\natexlab{}.
\newblock \bibinfo{title}{One size does not fit all}.
\newblock
  \bibinfo{howpublished}{https://gotocon.com/dl/goto-london-2016/slides/Stefan\_Tilkov-OneSizeDoesNotFitAllGOTOLondon16.pdf}.
\newblock


\bibitem[\protect\citeauthoryear{Toffetti, Brunner, Bl\"{o}chlinger, Dudouet,
  and Edmonds}{Toffetti et~al\mbox{.}}{2015}]%
        {Toffetti:2015:ASM:2747470.2747474}
\bibfield{author}{\bibinfo{person}{Giovanni Toffetti}, \bibinfo{person}{Sandro
  Brunner}, \bibinfo{person}{Martin Bl\"{o}chlinger}, \bibinfo{person}{Florian
  Dudouet}, {and} \bibinfo{person}{Andrew Edmonds}.}
  \bibinfo{year}{2015}\natexlab{}.
\newblock \showarticletitle{An Architecture for Self-managing Microservices}.
  In \bibinfo{booktitle}{\emph{Proceedings of the 1st International Workshop on
  Automated Incident Management in Cloud}} \emph{(\bibinfo{series}{AIMC '15})}.
  \bibinfo{publisher}{ACM}, \bibinfo{address}{New York, NY, USA},
  \bibinfo{pages}{19--24}.
\newblock
\showISBNx{978-1-4503-3476-1}
\urldef\tempurl%
\url{https://doi.org/10.1145/2747470.2747474}
\showDOI{\tempurl}


\bibitem[\protect\citeauthoryear{Tonse}{Tonse}{2015}]%
        {netflixipc}
\bibfield{author}{\bibinfo{person}{Sudhir Tonse}.}
  \bibinfo{year}{2015}\natexlab{}.
\newblock \bibinfo{title}{Scalable Microservices at Netflix. Challenges and
  Tools of the Trade}.
\newblock
  \bibinfo{howpublished}{http://www.infoq.com/presentations/netflix-ipc}.
\newblock


\bibitem[\protect\citeauthoryear{Trends}{Trends}{[n. d.]}]%
        {googletrend}
\bibfield{author}{\bibinfo{person}{Google Trends}.} \bibinfo{year}{[n.
  d.]}\natexlab{}.
\newblock \bibinfo{title}{Microservices - Explore}.
\newblock
  \bibinfo{howpublished}{https://trends.google.com/trends/explore?date=2013-01-01\&q=Microservices}.
\newblock


\bibitem[\protect\citeauthoryear{Uhle}{Uhle}{2014}]%
        {depmodel}
\bibfield{author}{\bibinfo{person}{Johan Uhle}.}
  \bibinfo{year}{2014}\natexlab{}.
\newblock \emph{\bibinfo{title}{On Dependability Modeling in a Deployed
  Microservice Architecture}}.
\newblock \bibinfo{thesistype}{Master's\ thesis}. \bibinfo{school}{Universitat
  Potsdam}.
\newblock


\bibitem[\protect\citeauthoryear{Ulander}{Ulander}{2017}]%
        {pop00150}
\bibfield{author}{\bibinfo{person}{D Ulander}.}
  \bibinfo{year}{2017}\natexlab{}.
\newblock \emph{\bibinfo{title}{Software Architectural Metrics for the Scania
  Internet of Things Platform: From a Microservice Perspectiv}}.
\newblock \bibinfo{thesistype}{Master's\ thesis}. \bibinfo{school}{Uppsala
  University}.
\newblock
\urldef\tempurl%
\url{http://www.diva-portal.org/smash/record.jsf?pid=diva2:1115342}
\showURL{%
\tempurl}


\bibitem[\protect\citeauthoryear{van Engelen}{van Engelen}{2004}]%
        {gsoap}
\bibfield{author}{\bibinfo{person}{Robert van Engelen}.}
  \bibinfo{year}{2004}\natexlab{}.
\newblock \showarticletitle{Code Generation Techniques for Developing
  Light-weight XML Web Services for Embedded Devices}.
  \bibinfo{howpublished}{http://doi.acm.org/10.1145/967900.968075}. In
  \bibinfo{booktitle}{\emph{Proceedings of the 2004 ACM Symposium on Applied
  Computing}} \emph{(\bibinfo{series}{SAC '04})}. \bibinfo{publisher}{ACM},
  \bibinfo{address}{New York, NY, USA}, \bibinfo{pages}{854--861}.
\newblock
\showISBNx{1-58113-812-1}
\urldef\tempurl%
\url{https://doi.org/10.1145/967900.968075}
\showDOI{\tempurl}


\bibitem[\protect\citeauthoryear{Vandermerwe and Rada}{Vandermerwe and
  Rada}{1988}]%
        {VANDERMERWE1988314}
\bibfield{author}{\bibinfo{person}{Sandra Vandermerwe} {and}
  \bibinfo{person}{Juan Rada}.} \bibinfo{year}{1988}\natexlab{}.
\newblock \showarticletitle{Servitization of business: Adding value by adding
  services}.
\newblock \bibinfo{journal}{\emph{European Management Journal}}
  \bibinfo{volume}{6}, \bibinfo{number}{4} (\bibinfo{year}{1988}),
  \bibinfo{pages}{314 -- 324}.
\newblock
\showISSN{0263-2373}
\urldef\tempurl%
\url{https://doi.org/10.1016/0263-2373(88)90033-3}
\showDOI{\tempurl}


\bibitem[\protect\citeauthoryear{Villamizar, Garcés, Ochoa, Castro, Salamanca,
  Verano, Casallas, Gil, Valencia, Zambrano, and Lang}{Villamizar
  et~al\mbox{.}}{2016}]%
        {7515686}
\bibfield{author}{\bibinfo{person}{M. Villamizar}, \bibinfo{person}{O.
  Garcés}, \bibinfo{person}{L. Ochoa}, \bibinfo{person}{H. Castro},
  \bibinfo{person}{L. Salamanca}, \bibinfo{person}{M. Verano},
  \bibinfo{person}{R. Casallas}, \bibinfo{person}{S. Gil}, \bibinfo{person}{C.
  Valencia}, \bibinfo{person}{A. Zambrano}, {and} \bibinfo{person}{M. Lang}.}
  \bibinfo{year}{2016}\natexlab{}.
\newblock \showarticletitle{Infrastructure Cost Comparison of Running Web
  Applications in the Cloud Using AWS Lambda and Monolithic and Microservice
  Architectures}. In \bibinfo{booktitle}{\emph{2016 16th IEEE/ACM International
  Symposium on Cluster, Cloud and Grid Computing (CCGrid)}}.
  \bibinfo{pages}{179--182}.
\newblock
\urldef\tempurl%
\url{https://doi.org/10.1109/CCGrid.2016.37}
\showDOI{\tempurl}


\bibitem[\protect\citeauthoryear{Vlaovic, Pilani, Parulekar, and Handa}{Vlaovic
  et~al\mbox{.}}{2016}]%
        {netflixaws}
\bibfield{author}{\bibinfo{person}{Stevan Vlaovic}, \bibinfo{person}{Rahul
  Pilani}, \bibinfo{person}{Subir Parulekar}, {and} \bibinfo{person}{Sangeeta
  Handa}.} \bibinfo{year}{2016}\natexlab{}.
\newblock \bibinfo{title}{Netflix Billing Migration to AWS}.
\newblock
  \bibinfo{howpublished}{https://medium.com/netflix-techblog/netflix-billing-migration-to-aws-451fba085a4}.
\newblock


\bibitem[\protect\citeauthoryear{Vogel and Giese}{Vogel and Giese}{2014}]%
        {Vogel:2014:MES:2578044.2555612}
\bibfield{author}{\bibinfo{person}{Thomas Vogel} {and} \bibinfo{person}{Holger
  Giese}.} \bibinfo{year}{2014}\natexlab{}.
\newblock \showarticletitle{Model-Driven Engineering of Self-Adaptive Software
  with EUREMA}.
\newblock \bibinfo{journal}{\emph{ACM Trans. Auton. Adapt. Syst.}}
  \bibinfo{volume}{8}, \bibinfo{number}{4}, Article \bibinfo{articleno}{18}
  (\bibinfo{date}{Jan.} \bibinfo{year}{2014}), \bibinfo{numpages}{33}~pages.
\newblock
\showISSN{1556-4665}
\urldef\tempurl%
\url{https://doi.org/10.1145/2555612}
\showDOI{\tempurl}


\bibitem[\protect\citeauthoryear{Vromant, Weyns, Malek, and Andersson}{Vromant
  et~al\mbox{.}}{2011}]%
        {dec4}
\bibfield{author}{\bibinfo{person}{Pieter Vromant}, \bibinfo{person}{Danny
  Weyns}, \bibinfo{person}{Sam Malek}, {and} \bibinfo{person}{Jesper
  Andersson}.} \bibinfo{year}{2011}\natexlab{}.
\newblock \showarticletitle{On Interacting Control Loops in Self-adaptive
  Systems}. In \bibinfo{booktitle}{\emph{Proceedings of the 6th International
  Symposium on Software Engineering for Adaptive and Self-Managing Systems}}
  \emph{(\bibinfo{series}{SEAMS '11})}. \bibinfo{publisher}{ACM},
  \bibinfo{address}{New York, NY, USA}, \bibinfo{pages}{202--207}.
\newblock
\showISBNx{978-1-4503-0575-4}
\urldef\tempurl%
\url{https://doi.org/10.1145/1988008.1988037}
\showDOI{\tempurl}


\bibitem[\protect\citeauthoryear{Vukovic and Robinson}{Vukovic and
  Robinson}{2005}]%
        {proser}
\bibfield{author}{\bibinfo{person}{Maja Vukovic} {and} \bibinfo{person}{Peter
  Robinson}.} \bibinfo{year}{2005}\natexlab{}.
\newblock \showarticletitle{SHOP2 and TLPlan for proactive service
  composition}. In \bibinfo{booktitle}{\emph{UK-Russia Workshop on Proactive
  Computing}}.
\newblock


\bibitem[\protect\citeauthoryear{Vural, Koyuncu, and Guney}{Vural
  et~al\mbox{.}}{2017}]%
        {10.1007/978-3-319-62407-5_14}
\bibfield{author}{\bibinfo{person}{Hulya Vural}, \bibinfo{person}{Murat
  Koyuncu}, {and} \bibinfo{person}{Sinem Guney}.}
  \bibinfo{year}{2017}\natexlab{}.
\newblock \showarticletitle{A Systematic Literature Review on Microservices}.
  In \bibinfo{booktitle}{\emph{Computational Science and Its Applications --
  ICCSA 2017}}, \bibfield{editor}{\bibinfo{person}{Osvaldo Gervasi},
  \bibinfo{person}{Beniamino Murgante}, \bibinfo{person}{Sanjay Misra},
  \bibinfo{person}{Giuseppe Borruso}, \bibinfo{person}{Carmelo~M. Torre},
  \bibinfo{person}{Ana Maria~A.C. Rocha}, \bibinfo{person}{David Taniar},
  \bibinfo{person}{Bernady~O. Apduhan}, \bibinfo{person}{Elena Stankova}, {and}
  \bibinfo{person}{Alfredo Cuzzocrea}} (Eds.). \bibinfo{publisher}{Springer
  International Publishing}, \bibinfo{address}{Cham},
  \bibinfo{pages}{203--217}.
\newblock
\showISBNx{978-3-319-62407-5}


\bibitem[\protect\citeauthoryear{Wagner}{Wagner}{2015}]%
        {aws}
\bibfield{author}{\bibinfo{person}{Tim Wagner}.}
  \bibinfo{year}{2015}\natexlab{}.
\newblock \bibinfo{title}{Microservices without the Servers}.
\newblock
\newblock
\urldef\tempurl%
\url{https://aws.amazon.com/blogs/compute/microservices-without-the-servers/}
\showURL{%
\tempurl}


\bibitem[\protect\citeauthoryear{Wang, Chao, Lo, Farmer, and Kuo}{Wang
  et~al\mbox{.}}{2009}]%
        {reput}
\bibfield{author}{\bibinfo{person}{Ping Wang}, \bibinfo{person}{Kuo-Ming Chao},
  \bibinfo{person}{Chi-Chun Lo}, \bibinfo{person}{R. Farmer}, {and}
  \bibinfo{person}{Pu-Tsun Kuo}.} \bibinfo{year}{2009}\natexlab{}.
\newblock \showarticletitle{A Reputation-Based Service Selection Scheme}. In
  \bibinfo{booktitle}{\emph{e-Business Engineering, 2009. ICEBE '09. IEEE
  International Conference on}}. \bibinfo{pages}{501--506}.
\newblock
\urldef\tempurl%
\url{https://doi.org/10.1109/ICEBE.2009.80}
\showDOI{\tempurl}


\bibitem[\protect\citeauthoryear{Wasson and Tam}{Wasson and Tam}{2018}]%
        {cicdmicro}
\bibfield{author}{\bibinfo{person}{Mike Wasson} {and} \bibinfo{person}{Benjamin
  Tam}.} \bibinfo{year}{2018}\natexlab{}.
\newblock \bibinfo{title}{Designing microservices: Continuous integration}.
\newblock
  \bibinfo{howpublished}{https://docs.microsoft.com/en-us/azure/architecture/microservices/ci-cd}.
\newblock


\bibitem[\protect\citeauthoryear{Weyns, Malek, and Andersson}{Weyns
  et~al\mbox{.}}{2010}]%
        {dec5}
\bibfield{author}{\bibinfo{person}{Danny Weyns}, \bibinfo{person}{Sam Malek},
  {and} \bibinfo{person}{Jesper Andersson}.} \bibinfo{year}{2010}\natexlab{}.
\newblock \showarticletitle{On Decentralized Self-adaptation: Lessons from the
  Trenches and Challenges for the Future}. In
  \bibinfo{booktitle}{\emph{Proceedings of the 2010 ICSE Workshop on Software
  Engineering for Adaptive and Self-Managing Systems}}
  \emph{(\bibinfo{series}{SEAMS '10})}. \bibinfo{publisher}{ACM},
  \bibinfo{address}{New York, NY, USA}, \bibinfo{pages}{84--93}.
\newblock
\showISBNx{978-1-60558-971-8}
\urldef\tempurl%
\url{https://doi.org/10.1145/1808984.1808994}
\showDOI{\tempurl}


\bibitem[\protect\citeauthoryear{White, Strowd, and Schmidt}{White
  et~al\mbox{.}}{2008}]%
        {white08creatingself-healing}
\bibfield{author}{\bibinfo{person}{Jules White}, \bibinfo{person}{Harrison~D.
  Strowd}, {and} \bibinfo{person}{Douglas~C. Schmidt}.}
  \bibinfo{year}{2008}\natexlab{}.
\newblock \showarticletitle{Creating Self-healing Service Compositions with
  Feature Modeling and Microrebooting}. In \bibinfo{booktitle}{\emph{The
  International Journal of Business Process Integration and Management
  (IJBPIM), Special issue on Model-Driven Service-Oriented Architectures}}.
\newblock


\bibitem[\protect\citeauthoryear{Wieringa, Maiden, Mead, and Rolland}{Wieringa
  et~al\mbox{.}}{2005}]%
        {Wieringa:2005:REP:1107677.1107683}
\bibfield{author}{\bibinfo{person}{Roel Wieringa}, \bibinfo{person}{Neil
  Maiden}, \bibinfo{person}{Nancy Mead}, {and} \bibinfo{person}{Colette
  Rolland}.} \bibinfo{year}{2005}\natexlab{}.
\newblock \showarticletitle{Requirements Engineering Paper Classification and
  Evaluation Criteria: A Proposal and a Discussion}.
\newblock \bibinfo{journal}{\emph{Requir. Eng.}} \bibinfo{volume}{11},
  \bibinfo{number}{1} (\bibinfo{date}{Dec.} \bibinfo{year}{2005}),
  \bibinfo{pages}{102--107}.
\newblock
\showISSN{0947-3602}
\urldef\tempurl%
\url{https://doi.org/10.1007/s00766-005-0021-6}
\showDOI{\tempurl}


\bibitem[\protect\citeauthoryear{Wiggins}{Wiggins}{2017}]%
        {12fact}
\bibfield{author}{\bibinfo{person}{Adam Wiggins}.}
  \bibinfo{year}{2017}\natexlab{}.
\newblock \bibinfo{booktitle}{\emph{The Twelve-Factor App}}.
\newblock


\bibitem[\protect\citeauthoryear{Wizenty, Sorgalla, Rademacher, and
  Sachweh}{Wizenty et~al\mbox{.}}{2017}]%
        {Wizenty:2017:MBM:3129790.3129821}
\bibfield{author}{\bibinfo{person}{Philip Wizenty}, \bibinfo{person}{Jonas
  Sorgalla}, \bibinfo{person}{Florian Rademacher}, {and}
  \bibinfo{person}{Sabine Sachweh}.} \bibinfo{year}{2017}\natexlab{}.
\newblock \showarticletitle{MAGMA: Build Management-based Generation of
  Microservice Infrastructures}. In \bibinfo{booktitle}{\emph{Proceedings of
  the 11th European Conference on Software Architecture: Companion
  Proceedings}} \emph{(\bibinfo{series}{ECSA '17})}. \bibinfo{publisher}{ACM},
  \bibinfo{address}{New York, NY, USA}, \bibinfo{pages}{61--65}.
\newblock
\showISBNx{978-1-4503-5217-8}
\urldef\tempurl%
\url{https://doi.org/10.1145/3129790.3129821}
\showDOI{\tempurl}


\bibitem[\protect\citeauthoryear{Wolff}{Wolff}{2016}]%
        {wolff2016microservices}
\bibfield{author}{\bibinfo{person}{E. Wolff}.} \bibinfo{year}{2016}\natexlab{}.
\newblock \bibinfo{booktitle}{\emph{Microservices: Flexible Software
  Architecture}}.
\newblock \bibinfo{publisher}{Pearson Education}.
\newblock
\showISBNx{9780134650401}
\urldef\tempurl%
\url{https://books.google.co.uk/books?id=zucwDQAAQBAJ}
\showURL{%
\tempurl}


\bibitem[\protect\citeauthoryear{Xiang~Zhou}{Xiang~Zhou}{2018}]%
        {benchmicro}
\bibfield{author}{\bibinfo{person}{Tao Xie Jun Sun Chenjie Xu Chao Ji
  Wenyun~Zhao Xiang~Zhou, Xin~Peng}.} \bibinfo{year}{2018}\natexlab{}.
\newblock \showarticletitle{Benchmarking Microservice Systems for Software
  Engineering Research}. In \bibinfo{booktitle}{\emph{ICSE '18 Companion}}.
\newblock


\bibitem[\protect\citeauthoryear{Yu, Silveira, and Sundaram}{Yu
  et~al\mbox{.}}{2016}]%
        {7867539}
\bibfield{author}{\bibinfo{person}{Yale Yu}, \bibinfo{person}{H. Silveira},
  {and} \bibinfo{person}{M. Sundaram}.} \bibinfo{year}{2016}\natexlab{}.
\newblock \showarticletitle{A microservice based reference architecture model
  in the context of enterprise architecture}. In \bibinfo{booktitle}{\emph{2016
  IEEE Advanced Information Management, Communicates, Electronic and Automation
  Control Conference (IMCEC)}}. \bibinfo{pages}{1856--1860}.
\newblock
\urldef\tempurl%
\url{https://doi.org/10.1109/IMCEC.2016.7867539}
\showDOI{\tempurl}


\bibitem[\protect\citeauthoryear{Zachman}{Zachman}{2003}]%
        {zachman}
\bibfield{author}{\bibinfo{person}{John~A. Zachman}.}
  \bibinfo{year}{2003}\natexlab{}.
\newblock \bibinfo{booktitle}{\emph{The Zachman Framework For Enterprise
  Architecture: Primer for Enterprise Engineering and Manufacturing}}.
\newblock \bibinfo{publisher}{Zachman International}.
\newblock


\bibitem[\protect\citeauthoryear{Zeiner, Goller, Exp{\'o}sito~Jim{\'e}nez,
  Salmhofer, and Haas}{Zeiner et~al\mbox{.}}{2016}]%
        {Zeiner2016}
\bibfield{author}{\bibinfo{person}{Herwig Zeiner}, \bibinfo{person}{Michael
  Goller}, \bibinfo{person}{V{\'i}ctor~Juan Exp{\'o}sito~Jim{\'e}nez},
  \bibinfo{person}{Florian Salmhofer}, {and} \bibinfo{person}{Werner Haas}.}
  \bibinfo{year}{2016}\natexlab{}.
\newblock \showarticletitle{SeCoS: Web of Things platform based on a
  microservices architecture and support of time-awareness}.
\newblock \bibinfo{journal}{\emph{e {\&} i Elektrotechnik und
  Informationstechnik}} \bibinfo{volume}{133}, \bibinfo{number}{3}
  (\bibinfo{date}{01 Jun} \bibinfo{year}{2016}), \bibinfo{pages}{158--162}.
\newblock
\showISSN{1613-7620}
\urldef\tempurl%
\url{https://doi.org/10.1007/s00502-016-0404-z}
\showDOI{\tempurl}


\bibitem[\protect\citeauthoryear{Zhang and Cheng}{Zhang and Cheng}{2006}]%
        {Zhang:2006:MDD:1134285.1134337}
\bibfield{author}{\bibinfo{person}{Ji Zhang} {and} \bibinfo{person}{Betty H.~C.
  Cheng}.} \bibinfo{year}{2006}\natexlab{}.
\newblock \showarticletitle{Model-based Development of Dynamically Adaptive
  Software}. In \bibinfo{booktitle}{\emph{Proceedings of the 28th International
  Conference on Software Engineering}} \emph{(\bibinfo{series}{ICSE '06})}.
  \bibinfo{publisher}{ACM}, \bibinfo{address}{New York, NY, USA},
  \bibinfo{pages}{371--380}.
\newblock
\showISBNx{1-59593-375-1}
\urldef\tempurl%
\url{https://doi.org/10.1145/1134285.1134337}
\showDOI{\tempurl}


\bibitem[\protect\citeauthoryear{Zheng, Zhang, Zheng, Fu, and Liu}{Zheng
  et~al\mbox{.}}{2017}]%
        {8026911}
\bibfield{author}{\bibinfo{person}{T. Zheng}, \bibinfo{person}{Y. Zhang},
  \bibinfo{person}{X. Zheng}, \bibinfo{person}{M. Fu}, {and}
  \bibinfo{person}{X. Liu}.} \bibinfo{year}{2017}\natexlab{}.
\newblock \showarticletitle{BigVM: A Multi-Layer-Microservice-Based Platform
  for Deploying SaaS}. In \bibinfo{booktitle}{\emph{2017 Fifth International
  Conference on Advanced Cloud and Big Data (CBD)}}. \bibinfo{pages}{45--50}.
\newblock
\urldef\tempurl%
\url{https://doi.org/10.1109/CBD.2017.16}
\showDOI{\tempurl}


\bibitem[\protect\citeauthoryear{Zimmermann}{Zimmermann}{2012}]%
        {Zimmermann:2012:ADI:2361999.2362021}
\bibfield{author}{\bibinfo{person}{Olaf Zimmermann}.}
  \bibinfo{year}{2012}\natexlab{}.
\newblock \showarticletitle{Architectural Decision Identification in
  Architectural Patterns}. In \bibinfo{booktitle}{\emph{Proceedings of the
  WICSA/ECSA 2012 Companion Volume}} \emph{(\bibinfo{series}{WICSA/ECSA '12})}.
  \bibinfo{publisher}{ACM}, \bibinfo{address}{New York, NY, USA},
  \bibinfo{pages}{96--103}.
\newblock
\showISBNx{978-1-4503-1568-5}
\urldef\tempurl%
\url{https://doi.org/10.1145/2361999.2362021}
\showDOI{\tempurl}


\bibitem[\protect\citeauthoryear{Zimmermann}{Zimmermann}{2017a}]%
        {archzimmer}
\bibfield{author}{\bibinfo{person}{Olaf Zimmermann}.}
  \bibinfo{year}{2017}\natexlab{a}.
\newblock \showarticletitle{Architectural refactoring for the cloud: a
  decision-centric view on cloud migration}.
\newblock \bibinfo{journal}{\emph{Computing}} \bibinfo{volume}{99},
  \bibinfo{number}{2} (\bibinfo{date}{01 Feb} \bibinfo{year}{2017}),
  \bibinfo{pages}{129--145}.
\newblock
\showISSN{1436-5057}
\urldef\tempurl%
\url{https://doi.org/10.1007/s00607-016-0520-y}
\showDOI{\tempurl}


\bibitem[\protect\citeauthoryear{Zimmermann}{Zimmermann}{2017b}]%
        {Zimmermann2017}
\bibfield{author}{\bibinfo{person}{Olaf Zimmermann}.}
  \bibinfo{year}{2017}\natexlab{b}.
\newblock \showarticletitle{Microservices tenets}.
\newblock \bibinfo{journal}{\emph{Computer Science - Research and Development}}
  \bibinfo{volume}{32}, \bibinfo{number}{3} (\bibinfo{date}{01 Jul}
  \bibinfo{year}{2017}), \bibinfo{pages}{301--310}.
\newblock
\showISSN{1865-2042}
\urldef\tempurl%
\url{https://doi.org/10.1007/s00450-016-0337-0}
\showDOI{\tempurl}


\bibitem[\protect\citeauthoryear{Zimmermann, Doubrovski, Grundler, and
  Hogg}{Zimmermann et~al\mbox{.}}{2005}]%
        {Zimmermann:2005:SAB:1094855.1094965}
\bibfield{author}{\bibinfo{person}{Olaf Zimmermann}, \bibinfo{person}{Vadim
  Doubrovski}, \bibinfo{person}{Jonas Grundler}, {and} \bibinfo{person}{Kerard
  Hogg}.} \bibinfo{year}{2005}\natexlab{}.
\newblock \showarticletitle{Service-oriented Architecture and Business Process
  Choreography in an Order Management Scenario: Rationale, Concepts, Lessons
  Learned}. In \bibinfo{booktitle}{\emph{Companion to the 20th Annual ACM
  SIGPLAN Conference on Object-oriented Programming, Systems, Languages, and
  Applications}} \emph{(\bibinfo{series}{OOPSLA '05})}.
  \bibinfo{publisher}{ACM}, \bibinfo{address}{New York, NY, USA},
  \bibinfo{pages}{301--312}.
\newblock
\showISBNx{1-59593-193-7}
\urldef\tempurl%
\url{https://doi.org/10.1145/1094855.1094965}
\showDOI{\tempurl}


\end{thebibliography}
	
\end{document}